\documentclass[fleqn,usenatbib]{mnras}
\usepackage[T1]{fontenc}
\usepackage{ae,aecompl}
\pdfoutput=1
\usepackage{graphicx}	
\usepackage{amsmath}	
\usepackage{amssymb}	
\usepackage {pdflscape}
\usepackage{hyperref}

\title[ \textit{PARSEC} SSP for  \textit{GRASIL}]{Modelling the UV to radio SEDs of nearby star-forming galaxies: new  \textit{PARSEC} SSP for   \textit{GRASIL} }
\author[Obi]{I.A. Obi$^{1}$\thanks{E-mail: iaobi@sissa.it},
A. Bressan$^{1}$,
F. Perrotta$^{1}$,
L. Silva$^{2}$
O. Vega$^{3}$,
Y. Chen$^{4}$,
A. Lapi$^{1}$,\and  
C. Mancuso$^{1}$,
L. Girardi$^{5}$,
G.L. Granato$^{2}$,
P. Marigo$^{4}$,
and A. Slemer$^{4}$.
\\
$^{1}$SISSA, via Bonomea 265, I-34136 Trieste, Italy\\
$^{2}$INAF - Osservatorio Astronomico di Trieste, via Tiepolo 11, 34131
Trieste, Italy\\
$^{3}$INAOE, Luis Enrique Erro 1, 72840 Tonantzintla, Puebla, Mexico\\
$^{4}$Dipartimento di Fisica e Astronomia Galileo Galilei,
 Universit\'a di Padova, Vicolo dell'Osservatorio 3, I-35122 Padova, Italy\\
$^{5}$INAF - Osservatorio Astronomico di Padova, Vicolo dell'Osservatorio 5, I-35122 Padova, Italy\\
}
\date{Accepted XXX. Received YYY; in original form ZZZ}
\pubyear{2017}

\begin{document}
\label{firstpage}
\maketitle

\begin{abstract}
By means of the updated \textit{\small{PARSEC}} database of evolutionary tracks of massive stars, we compute  the integrated stellar light, the ionizing photon budget and the supernova rates of young simple stellar populations (SSPs), for different metallicities and IMF 
upper mass limits.
Using \textit{\small{CLOUDY}} we compute and include in the SSP spectra the nebular emission contribution. We also revisit the thermal and  non-thermal radio emission contribution from young stars.  Using \textit{\small{GRASIL}} we can  thus predict the panchromatic spectrum and the main recombination lines  of any type of star-forming galaxy,
including the effects of dust absorption and re-emission. 
We check the new models against the spectral energy distributions (SEDs) of selected well-observed nearby galaxies.
From the best-fit models we obtain  
a consistent set of star formation rate (SFR) calibrations at wavelengths ranging from ultraviolet (UV) to radio.
We also provide analytical calibrations that take into account the  dependence on  metallcity and IMF upper mass limit of the SSPs. We show that the latter limit can be well constrained by combining information from the observed far infrared, 24~$\micron$,   33~GHz and H$\alpha$ luminosities. Another interesting property derived from the fits is that, while in a normal galaxy  the attenuation in the lines is significantly higher than that in the nearby continuum,  in individual star bursting regions they are similar, supporting the notion that this effect is due to an age selective extinction. Since in these conditions the Balmer decrement method may not be accurate, we provide relations to estimate the  attenuation from the  observed 24~$\micron$  or 33~GHz fluxes. These relations can be useful for the analysis of young high redshift galaxies.
\end{abstract}

\section{Introduction}
\label{sec:intro}
Modelling  the SEDs of galaxies has proven to be a very powerful tool in our current understanding  of the different physical processes that come into play in the formation and evolution of galaxies. Various physical properties of galaxies like stellar, metal and dust content, star formation rate, dust obscuration, etc are estimated by fitting  the theoretical SEDs to the observed ones.
In the UV to  infrared (IR) spectral regions of a SED, dust
plays an important role. 
It absorbs (as well as  scatters) the UV-near-infrared (NIR) light  and re-radiates it in the IR. This IR emission (3 - 1000 $\micron$)  may arise from (a) the
emission from dust heated by young OB stars (neglecting heating from AGN), (b) the emission from the  circumstellar envelopes of evolved stars  and (c) the cirrus emission from dust distributed throughout in the  ISM and heated by the general interstellar  radiation field.
Point (b) leads to  wrong  estimates of the physical properties related to  star formation.
The radio band which is not sensitive to dust attenuation
is  usually used to check and complement the interpretations arrived by using the optical/IR bands. Radio emission from normal star-forming galaxies is usually dominated by the non-thermal component (up to $\approx$ 90 per~cent of the radio flux) which is believed to be due to the synchrotron emission from relativistic electrons accelerated into the shocked ISM by
core-collapse supernovae (CCSN) explosions from massive young stars (above ~$\approx 8~\rm{M}_{\sun}$) ending their lifes \citep{Condon1992}. 
Recent advances in hydrodynamical simulations of CCSN  have  indicated a range of stellar masses where the stars fail to explode but  rather end up directly  as a black hole.
This raises concern about the widely accepted notion that   all stars more massive than $\approx 8~\rm{M}_{\sun}$  end up as a  CCSN.

Over the years, great progress has been made both in the development of tools and models use to extract the  information encoded in the SEDs and in multi-wavelength surveys that sample  the UV to radio SED of local and high redshift galaxies. Furthermore, new observing facilities with unprecedented resolution  and sensivity (e.g. SKA, \textit{JWST}, EVLT etc) will be put in place in the nearest feature. At the same time, semi-analytic models of galaxy formation, in the context of the current cosmological standard model, are appearing and   providing realistic predictions of  the physical properties of galaxies with  which the results of the SED fitting can be compared.

Stellar population synthesis (SPS) still remains the basis of SED modelling.
The most common method  used in computing the SEDs of SSPs is that  of isochrone synthesis
\citep{Chiosi1988, Maeder1988, Charlot1991} which uses the locus of stars in an isochrone to  integrate the spectra of all stars along an isochrone to get the total flux. This involves computing first the  stellar evolutionary tracks  for different masses and metal contents and  building the isochrones from the tracks.
With the isochrones, the  resulting stellar spectrum is  computed using  stellar atmosphere libraries. 
In recent years, significant  efforts from different research groups have been put into  providing homogeneous sets of evolutionary tracks
and improving the stellar libraries. As a result of these improvements, SPS models can remarkably reproduce well  the UV-NIR SEDs  and high-resolution spectra in the optical wavelength band. 
However, despite these improvements,  there are still challenges  in the field, especially  in the treatment of the phases  of stellar evolution that are  weakly understood.
The most important being the  short lived phases: massive stars, thermally pulsing asymptotic giant branch (TP-AGB) stars, blue stragglers
and extreme horizontal branch stars.
The  TP-AGB stars have however  increasingly received attention leading to  a rapid  progress in their  modelling \citep{Maraston2005, Marigo2007,Marigo2017}. 

The stellar evolution code used in Padova to compute sets of stellar evolutionary tracks that are of  wide useage in the astronomical community has recently 
been thoroughly revised and updated, going  by the name, \textit{\small{PARSEC}}(PAdova-TRieste Stellar Evolution Code). 
More details of this code can be found in \citet{Bressan2012,Bressan2013,Chen2014} and  will be briefly discussed later in the text.
In this paper, our main aim is to  use the  \textit{\small{PARSEC}} evolutionary  tracks of massive stars  to compute the quantities (the integrated light, 
ionizing photon budget and the supernova rates) of  young SSPs, thereby updating the SSPs used  by   \textit{\small{GRASIL}} in   predicting the panchromatic spectrum of star-forming galaxies. 
We also revisit  the thermal and non-thermal radio emissions.
We finally check  these new models  with  selected nearby well-observed  galaxies.
We perform an analysis of the resulting best-fit  SEDs, 
with the aim of obtaining a new set of SFR calibrations and investigating the 
dependence of the dust attenuation properties on galaxy types.

The structure of the rest of paper is as follows:
In Section~\ref{sec:ssp}, we use the new  \textit{\small{PARSEC}} 
code database of stellar evolutionary tracks of massive stars to compute the ionizing photon budget, the integrated light and the supernova rates predicted by young SSP models. Using the  integrated spectra of the SPPs in  \textit{\small{CLOUDY}}, we computed the nebular emissions.
In Section~\ref{sec:radio}, we revise (a) the prescription used in computing the thermal radio emission.
(b) the previous non-thermal radio emission model  originally described in \citet[][hereafter B02]{Bressan2002} while taking into account recent advances in CCSN explosion models.
In Section~\ref{sec:ssp_gra}, we check   our new radio emission models and SSPs  with  \textit{\small{GRASIL}} using selected well-observed nearby galaxies.
Finally, we  discuss the resulting  best fit SEDs of all galaxies studied in this paper in Section~\ref{sec:disc}. We draw our conclusions  in Section~\ref{sec:conc}.

Throughout this paper we adopt $\rm{12} + \rm{log}( \rm{O/H} ) = 8.69$
\citep{Asplund2009} as the solar oxygen abundance. Our models adopt a  \citet{Kennicutt1983} 
IMF  (slope of -1.4 for $0.1 - 1~\rm{M}_{\sun}$ and -2.5 for  $\geq 1~\rm{M}_{\sun}$).
The cosmological parameters we adopt assume
$H_{0} = 70~\rm{km} \; s^{-1} \; \rm{Mpc}^{-1}, \Omega_{\Lambda} = 0.7, \Omega_{M} = 0.3$

\section{Simple Stellar Populations with \textit{PARSEC} }
\label{sec:ssp}
\textit{\small{PARSEC}} is the latest version of the Padova-Trieste stellar evolution code with thorough update of
the major input physics including new and accurate homogenised opacity and equation of state tables
fully consistent with any adopted chemical composition.
More details may be found in \citet{Bressan2012, Bressan2013,Chen2014}.
The evolutionary tracks span a wide range in metallicities,
$0.0001 \leq Z \leq 0.04$, and initial masses, from very low $(M~=~0.1~\rm{M}_{\sun})$
 to very massive $(M~=~350~\rm{M}_{\sun})$ stars, 
starting from the pre-main sequence phase and ending at central carbon ignition.
 \citet{Tang2014} and \citet{Chen2015} computed new evolutionary tracks of massive stars and  tables of theoretical bolometric
 corrections that allow for the conversion from theoretical HR to the observed colour-magnitude diagrams.
A preliminary comparison of the new models with color-magnitude diagrams
of star-forming regions in nearby low metallicity dwarf irregular galaxies was performed by \citet{Tang2014}.
The full set of new evolutionary tracks and the corresponding isochrones may be found in
\url{http://people.sissa.it/~sbressan/parsec.html} and \url{http://stev.oapd.inaf.it/cgi-bin/cmd}, respectively.
\subsection{Integrated SSP spectra}
\label{sec:ssp_sed}
To build the integrated spectra of SSPs,
we adopt the spectral library compilation by \citet{Chen2015} who homogenized various  sets of  existing  stellar atmosphere  libraries encompassing a wide range of parameters
of both cool and hot stars (i.e. masses, evolutionary stages  and metallicities).
The atmosphere models  adopted by  \citet{Chen2015} are the  \textit{\small{ATLAS9}} models \citep{Kurucz1993,Castelli2004}, suitable for intermediate 
and low mass stars,  and the Phoenix models  \citep{Allard1997} 
for the coolest stars.  For hot massive stars \citet{Chen2015} generated new wind models with the   \textit{\small{WM-BASIC}}  code \citet{Pauldrach1986A} and adopted new models of WR stars from the Potsdam group \citep[see e.g.][]{PoWR2015}.
Since different sets of libraries have different metallicities, they were merged to obtain 
homogeneous sets of spectral libraries as described  in \citet{Girardi2002} and \citet{Chen2014}.
\begin{figure*}
\centering
\includegraphics[width=\linewidth]{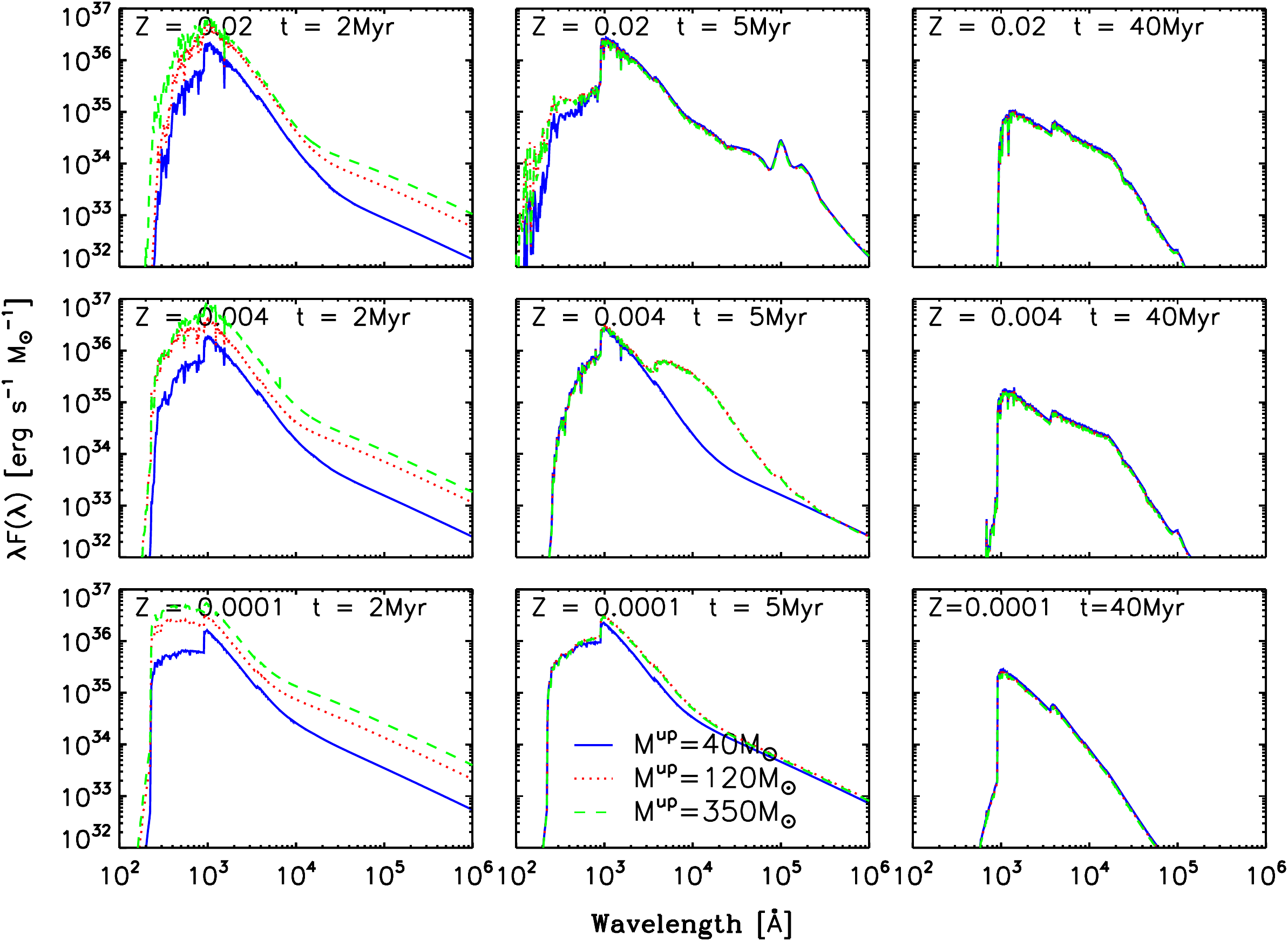}
\caption{SSP Integrated spectra per solar mass  at ages of $t~=~$2, 5 and 40~Myr, for   metallicities Z~=~0.02, 0.004 and 0.0001 and $\rm{M}_{up}$ of 40, 120 and 350~$\rm{M}_{\sun}$
indicated by the solid blue, dotted red  and dashed green lines respectively. 
At 2~Myr, the spectra for the various upper mass limits
at any of the three metallicities is quite distinguishable. At 5~Myr and Z = 0.02, it is  distinguishable
only at wavelengths below $912~\text{\AA}$, the maximum for  the lyman ionizing photons.
At this age, notice that the spectra for the $\rm{M}_{up}$ of 120 and 350$~\rm{M}_{\sun}$  below
$912~\mbox{\AA}$ are superimposed, owing to the fact they must have evolved off the main
sequence at this age. At 5~Myr and Z = 0.0001, the reverse is the case, the spectra is
distinguishable only at wavelengths above  $912~\text{\AA}$ because at this metallicity, stars
lifetimes are a bit longer than at Z = 0.02. At 40~Myr, the spectra for all upper mass limits 
and metalicities are all superimposed and the Lyman  ionizing photons are no longer produced.}
\label{fig:ssp_sed}
\end{figure*} 
The properties of SSP are obtained by integration along the corresponding isochrones
assuming a two-slopes power law Kennicutt initial mass function.  Results are presented for three different values of
the upper mass limit, $\rm{M}_{up}$ of 40,120, 350~$ \rm{M}_{\sun}$.
As an example, Figure~\ref{fig:ssp_sed} shows the synthesized spectra of star clusters at ages of $t~=~$2, 5 and 40~Myr and for three values of the 
initial metallicity, Z~=~0.02, 0.004 and 0.0001. 
The effect of assuming different  $\rm{M}_{up}$ values, at fixed total initial mass,
is more pronounced at an age of 2~Myr.
At 5~Myr, when stars with masses of 120~$\rm{M}_{\sun}$ and 350~$\rm{M}_{\sun}$ already left the main
sequence or already died out, the spectra of SSP with upper mass limits of 120~$\rm{M}_{\sun}$ and
350~$\rm{M}_{\sun}$ are superimposed, while those of 40~$\rm{M}_{\sun}$
remains distinct. At 40~Myr the spectra are almost independent of $\rm{M}_{up}$,
though it is evident that the spectrum of the SSP with larger $\rm{M}_{up}$ has a lower luminosity,
because of the higher mass budget stored in massive stars
We already note from this figure that the effects of $\rm{M}_{up}$ on the number of ionizing photons
disappears at ages larger than about 5~Myr, as discussed below.
\subsection{Ionizing Photon Budget}    
\label{sec:ssp_q}
\begin{figure}
\centering
\includegraphics[width=\linewidth]{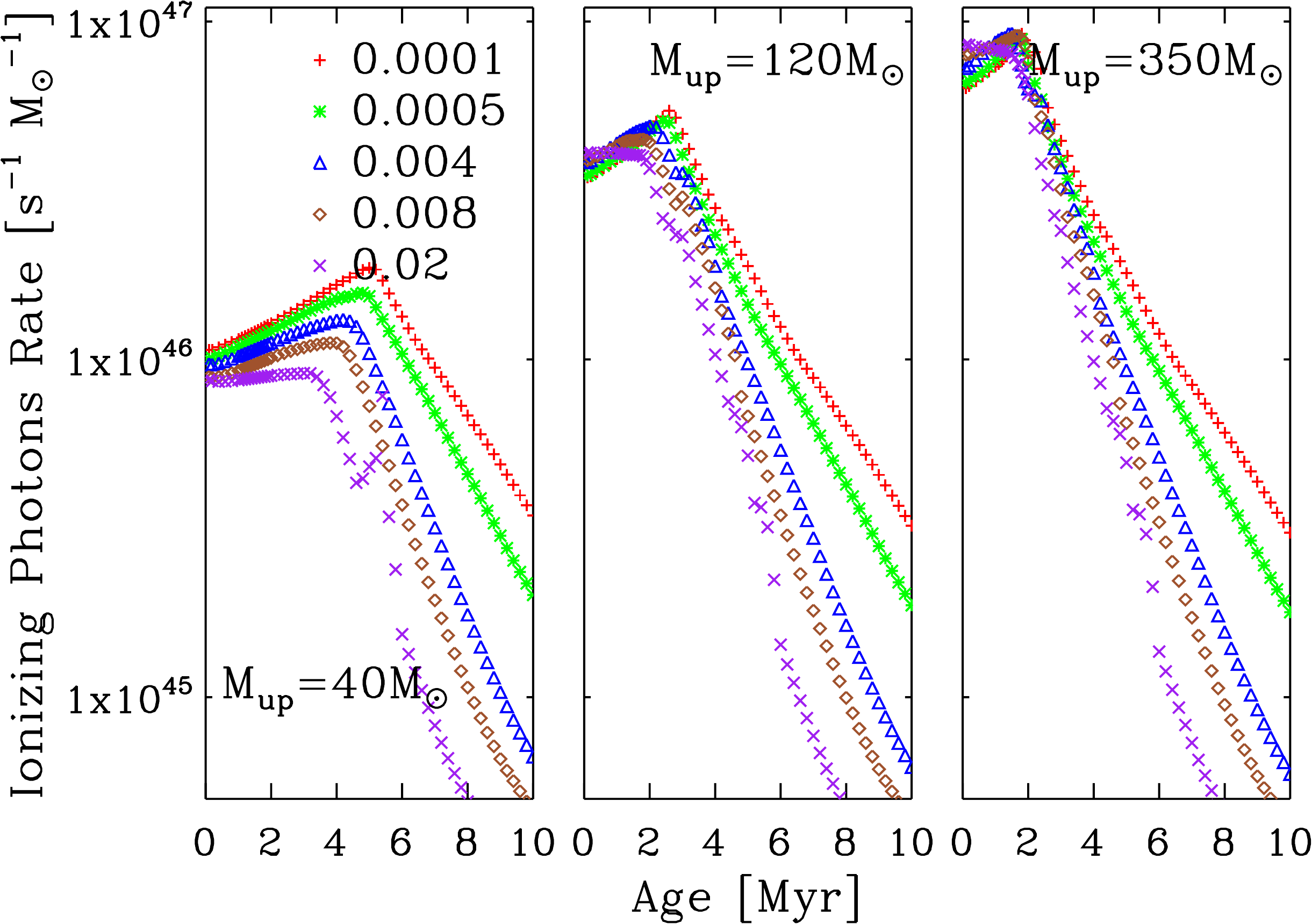}
\caption{Variation of the instantenous  number of ionizing photons per second per unit ionizing
solar mass of the cluster with age (Myr). Different symbols correspond to  different
metallicities as illustrated in the plot}.
\label{fig:ssp_q_inst}
\end{figure}

The number of Lyman ionizing photons per sec ($\rm{Q(H)}$)
and per unit mass emitted by  young stellar populations is
controlled by hot massive stars, i.e. O-B main sequence stars
and Wolf Rayet stars. This number is thus critically dependent on the shape of the
initial mass function in the domain of massive stars.
In Figure~\ref{fig:ssp_q_inst}, we show the time evolution of $\rm{Q(H)}$
for selected SSPs with different upper mass limit ($\rm{M}_{up}$) and metallicities.
As already said, the adopted IMF is a two-slope power law \citet{Kennicutt1983}.
The lower limit is $\rm{M}_{low}$~=~0.1 $\rm{M}_{\sun}$ and the upper limits are
$\rm{M}_{up}$~=~40 $\rm{M}_{\sun}$,
$\rm{M}_{up}$~=~120 $\rm{M}_{\sun}$ and
$\rm{M}_{up}$~=~350 $\rm{M}_{\sun}$.
The slope of the IMF is X~=~1.4 between
$\rm{M}_{low}$ and $\rm{M}$~=~1$\rm{M}_{\sun}$ and
X~=~2.5 between
$\rm{M}$~=~1$\rm{M}_{\sun}$ and
$\rm{M}_{up}$
From this figure we may appreciate the role of age, metallicity and IMF on the ionizing photon rate  $\rm{Q(H)}$.
For a given Z and $\rm{M}_{up}$, $\rm{Q(H)}$ generally increases to a maximum value and then, once a threshold age
is reached, it rapidly decreases to negligible values.
At fixed $\rm{M}_{up}$, both the maximum value of $\rm{Q(H)}$ and the threshold age decrease, at increasing metallicity.
In general this is also true at varying age, i.e. 
at given $\rm{M}_{up}$, $\rm{Q(H)}$ decreases at increasing metallicity.
However there are some cases where this is not true, in particular for the SSP of solar metallicity.
The effect of $\rm{M}_{up}$ is strong. A
star cluster with $\rm{M}_{up}$ of 350~$\rm{M}_{\sun}$ produces  about seven times more ionizing photons
than a cluster with a $\rm{M}_{up}$ of 40~$\rm{M}_{\sun}$, at constant total mass.
Moreover, the age to attain the maximum  $\rm{Q(H)}$ becomes lesser at increasing upper mass limit, reflecting the
larger relative weight of more massive stars in the ionizing photon budget. 
\subsection{Nebular Emission with \textit{PARSEC's} SSP}
\label{sec:nebular}
The integrated spectra of the SSPs are used to calculate the nebular emission from the surrounding
\ion{H}{II} regions which is then added to the original spectrum to obtain  the integrated spectra containing both the
stellar continuum (with absorption lines and eventually wind emission features) and the nebular features (continuum and lines).
For this purpose, star clusters are assumed to be  the central ionizing sources of the  \ion{H}{II} regions that are modelled using the photoionization code \textit{\small{CLOUDY}}  \citep{Ferland1996}.
As a further input to the \textit{\small{CLOUDY}} code, we  specify
that the \ion{H}{II} region is modelled as a thin shell of gas
with constant density, $n_H = 100~cm^{-3}$, placed at a distance $R_H$~=~15~pc from the central source.
The evolution of the ionizing star clusters is followed from 0.1~Myr to 20~Myr,
for  five different values of the metallicity (Z = 0.0001, 0.0005, 0.004, 0.008, 0.008 and 0.02)
and three values of $M_{up}$ of 40, 120 and 350 M$_{\sun}$.
We note that our goal is not that of providing a detailed dependence of a large ensemble of emission lines on the critical parameters of the \ion{H}{II} nebulae. Instead we aim at obtaining a reasonable estimate of the
intensities of the main recombination lines and of free-free emission
to increase the diagnostic capabilities of our population synthesis codes.
Line emission and the corresponding nebular continuum are much less dependent on the
characteristic properties of the \ion{H}{II} regions (e.g. ionization parameter, individual abundance of heavy elements etc.) than e.g. excitation lines, for which a more detailed set of initial parameters would be more appropriate
\citep[see e.g.][]{Panuzzo2003, Wofford2016}.
The main output of this process is a library of emission line intensities,
nebular continuum properties and electron temperatures (T$_e$) of the
\ion{H}{II} regions. Then, emission lines and nebular continuum are used to suitably complement the integrated SED of SSPs
from the far-UV to radio wavelengths.

\section {RADIO CONTINUUM EMISSION}
\label{sec:radio}
Radio emission associated with the presence of young massive stars
comprises essentially two components, the thermal radio emission
(also referred to as free-free emission)
and the non-thermal radio emission (also referred to as non-thermal radio emission).
The former emission comes from the nebular free electrons  originating  from the ionizing radiation of massive stars.
The non-thermal emission is instead believed to be 
synchrotron radiation that originated
from the interaction of relativistic electrons, produced in the
ejecta of core-collapsed supernovae (CSSN), with the ambient magnetic field.
The radio continuum is thus a tracer of the number of massive stars
formed (and exploded) and hence an optimal indicator of the very recent (if not the current) star formation rate.

The fact that the radio emission is a good SFR tracer is
supported by the remarkably tight correlation between FIR and non-thermal radio emission. 
At 1.49~GHz, this correlation is quantified by the q-parameter  \citep{Helou1985}:
\begin{equation}
	q = \rm{log} \frac{  F_{ \rm{FIR} } /(3.75 \times 10^{12} Hz)   }
	                 {  F_{\nu}(1.49 \rm{GHz}) / ( W \; m^{-2} \; Hz^{-1} )  }
	                   \approx 2.35 \pm 0.2
		\label{eq_q_ratio}
    \end{equation}
where $F_{ \rm{FIR} } = 1.26 \times 10^{-14} (2.58 S_{60\micron} + S_{100\micron}  ) \rm{W}
\rm{m}^{-2}$ \citep{Young1989},  $S_{60\micron}$ and $S_{100\micron}$ are IRAS
flux densities  in  $\rm{Jy}$.
\subsection{Thermal Radio Emission}
\label{sec:thermal}
\begin{figure}
\centering
 \includegraphics[width=\linewidth]{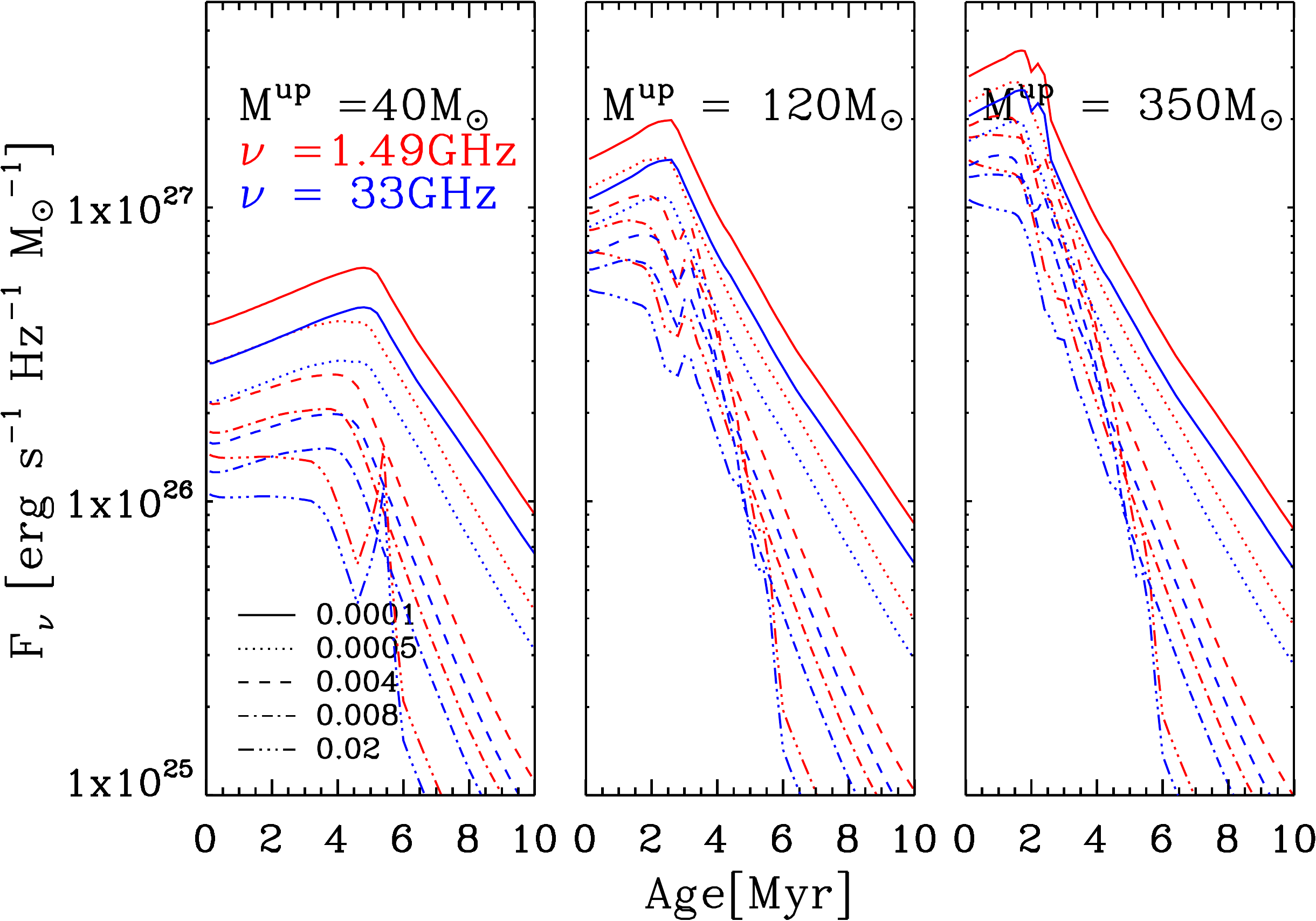}
 \caption{Variation of the 1.49GHz~(red) and 33GHz~(blue) thermal radio emission with age at different metallicities for different values of
$M_{up}$. The five different  line styles correspond to  the  five different values of metallicity.}
 \label{fig:therm_14_33}
\end{figure}

At sufficiently high frequency such that  free-free self-absorption is negligible,  
 the relation between 
 the specific luminosity  of
 free-free emission  $L_{\rm{ff}}(\nu)$
 and  the rate of ionizing photons $\rm{ Q(H) }$ can be written as
\citep{Rubin1968,Condon1992}
\begin{equation} 
 L_{ \rm{ff} }= \frac{ \rm{Q(H)} } {C_1} 
 \left( \frac{\rm{T}_e}{10^4\,\hbox{K}}\right)^{0.3}\,
 { \rm G_{dra} }(\nu,\hbox{T}_e)
\hspace{3pt} 
\label{eq_Lff_new_1}
\end{equation}
where $\rm{T}_{e}$ is the  electron
temperature, $C_1= \sqrt{3}/\pi\times6.86 \times 10^{26}$ and
${\rm G_{dra} }(\nu,\hbox{T}_e)$ is 
the  velocity averaged gaunt factor \citep{Draine2011}: 
\begin{eqnarray}  \nonumber
\hbox{G}_{ \text{dra} }(\nu, \rm{T}_e) = \\
\hbox{ln}  \left\{\rm{exp} \left[5.960 -{\frac{ \sqrt{3} } { \pi} }
\rm{ln} \left( Z_{i} {\frac{ \nu }{ \rm{GHz}} }
\left ( {\frac{ \rm{T_{e}} } { 10^4 \rm{K}} }\right) ^{-1.5}
\right)\right]+e  \right\}
\label{eq_gaunt_claudia}
\end{eqnarray}
where $\rm{ Z_{i} }$ is the charge of the ions in the \ion{H}{II} region.
An approximate velocity averaged gaunt factor were obtained earlier by  \citet{oster1961}:
\begin{eqnarray} \nonumber
\hbox{G}_{ \text{ost} }(\nu, \rm{T}_e) =  \\
\hbox{ln}  \left[ 4.955 \times 10^{-2}  \left( {\nu \over \hbox{GHz}} \right)^{-1}  \right]
+ 1.5 \  \hbox{ln} \left({ \hbox{T}_e \over \hbox{K} }\right)
\label{eq_gaunt_rubin}
\end{eqnarray}
Taking into account that
$\hbox{G}_{ \text{dra} }(\nu, \rm{T}_e)   \simeq  \sqrt{3}/\pi \times \hbox{G}_{ \text{ost} }(\nu, \rm{T}_e)$
and adopting a common approximation used in literature, we obtain for the free-free emission \citep{Condon1992}
\begin{equation}
\frac{L_{\rm ff}}{\hbox{erg}\,\hbox{s}^{-1}\,
\hbox{Hz}^{-1}}={\frac{Q(H)} {6.3\times~10^{25}}  }
\left( \frac{\rm{T}_e}{10^4\,\hbox{K}}\right)^{0.45}\,
\left( {\nu \over \hbox{GHz}} \right)^{-0.1}
\label{eq_Lff_new_2}
\end{equation}
The latter equation is often used by assuming
a fixed electronic temperature (10$^4$~K) to provide useful analytical approximations at radio frequencies especially in the lack of nebular emission (see discussion in B02).
Here we directly estimate the intensity of the thermal radio emission of our SSP from Cloudy with the procedure already described in Section~\ref{sec:nebular}.
In Figure~\ref{fig:therm_14_33}, we show the evolution of the 1.49~GHz and 33~GHz
thermal radio emission for different metallicities, Z = 0.0001, 0.0005, 0.004, 0.008 and 0.02,
and different $M_{up}$ of $40, 120 $  and  $ 350~\rm{M}_{\sun}$.
The effects of metallicity and
$M_{up}$ on the thermal emission are easily noted.
For example, at Z = 0.0001, the 1.4~GHz thermal emission for $M_{up} = 120~\rm{M}_{\sun}$ is about 3 - 7 times larger than that for $M_{up} = 40~\rm{M}_{\sun}$
In general by increasing the metallicity from Z = 0.0001 to Z = 0.02 the thermal radio emission decreases by a factor of about 3.
Instead by increasing  $M_{up}$  from 40 to 350~$\rm{M}_{\sun}$,
thermal emission increases by about one order of magnitude.
These important factors, that must be considered
in the calibration of the relation between star formation rate and thermal
radio emission, are discussed below.
\subsubsection{Metallicity effects}
Equations~\ref{eq_Lff_new_1}, \ref{eq_gaunt_claudia} and \ref{eq_Lff_new_2}
contain an explicit dependence on the electron temperature, that is generally neglected in analytical approximations
which assume a constant value of  $\rm{T_e}$~=~10$^4$~K. Since $\rm{T_e}$ is known to depend on the metallicity of the \ion{H}{II} regions, with a variation of more than a factor of two in the range of the observed values,
we provide in Appendix \ref{appendix_a}
some useful analytical relations.
We first list in Table~\ref{tab:C_2_C_3} 
the values of the electronic temperature derived using  \textit{\small{CLOUDY}}  for our SSP models at various  metallicities  and upper mass limits.
In Figure~\ref{fig:te_z_mup} of Appendix \ref{appendix_a}, we show the relations between oxygen abundance and the electronic temperature.
In this figure, crosses, 'X's and asterisks  indicate
$M_{up} = 40$, $120$ and $350~\rm{M}_{\sun}$ respectively.
The average of the empirical fits derived by \citet{Sanchez2012} for high-ionization  \ion{O}{III}  and for low-ionization \ion{O}{III}    zones is shown as the dashed black lines.
We  provide a multiple regression fitting relation (Equation~\ref{eq_te_z_mup}) between  $\rm{T_e}$, $\rm{M_{up}}$ and Z that could be easily included in analytical approximations.
\subsubsection{ $\rm{SFR}-\rm{Q(H)}$ calibration}
\begin{figure}
\centering
\includegraphics[width=\linewidth]{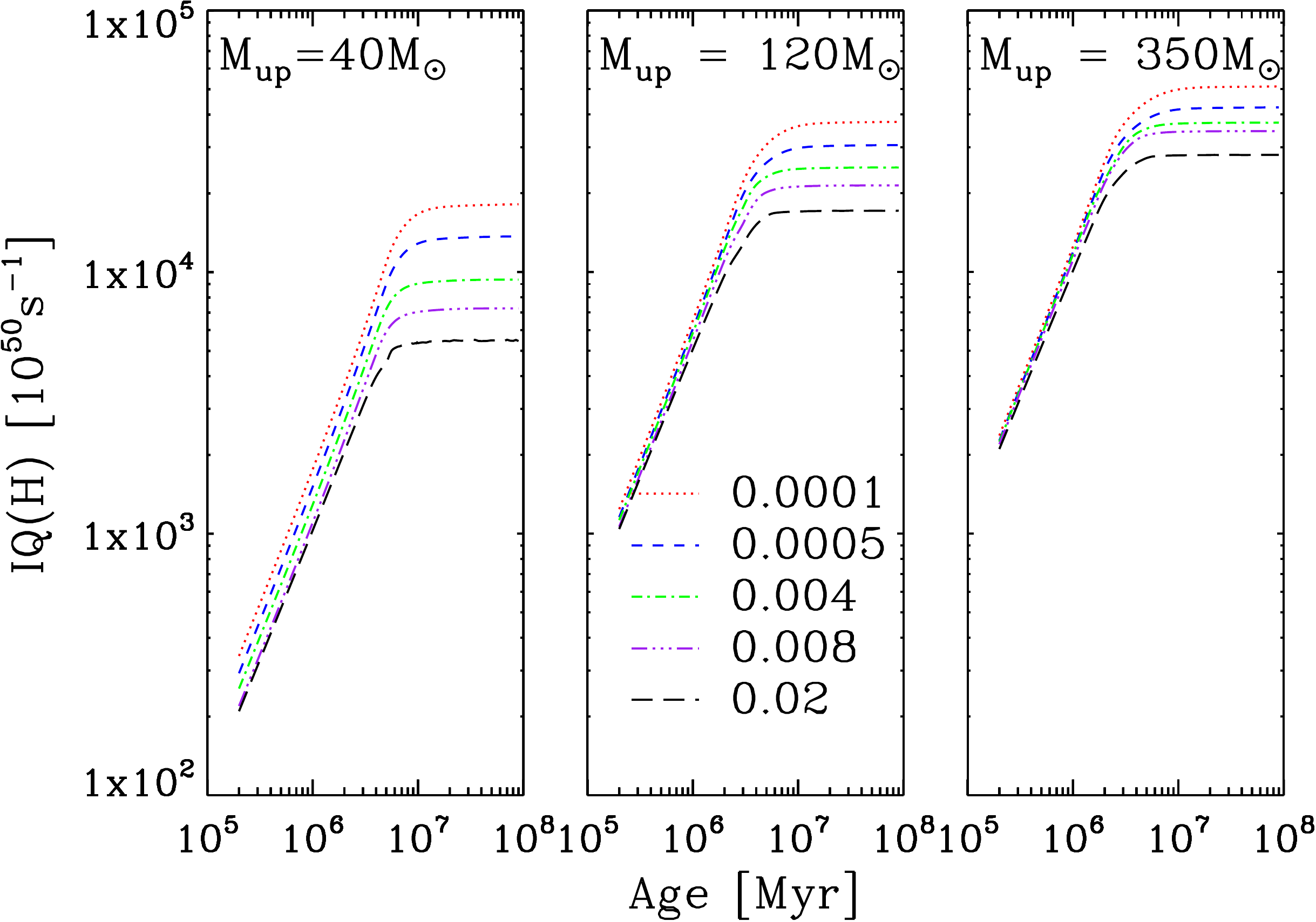}
\caption{ Evolution of the integrated  number of ionizing photons per second per unit solar  mass initially formed as a function of the age in yr. Different line colors represents different metallciities. The variation of Q(H) with metallicity decreasis at increasing  IMF upper mass limit.}
\label{fig:ssp_q_integ}
\end{figure}

In a young stellar system the ionizing photon budget is dominated by massive stars and thus there must be a tight
relation between the current star formation rate  and the  rate of ionizing photons, that ultimately produce
the thermal radio emission \citet{Condon1990}.
In this section, we provide a calibration of  the $\rm{SFR}-\rm{Q(H)}$ relation.
For this purpose we consider the integrated spectrum of a galaxy up to a time $t$:
\begin{equation}
f_{\nu}^{gal}(t,Z) = \int_{0}^{t} f_{\nu}^{SSP}(t-t',Z)SFR(t')dt'
\label{specgal}
\end{equation}
where SFR(t') is the star formation rate at time $t'$ and $f_{\nu}^{SSP}(t-t',Z)$ is the stellar spectrum of a SSP of age $t_{ssp}~=~t-t'$ and given metallicity $Z$
\begin{equation}
f_{\nu}^{SSP}(t_{ssp},Z) = \int_{M_{low}}^{M_{up}} \phi(m) f_{\nu}(m,t_{ssp},Z)dm
\label{specssp}
\end{equation}
In the latter equation, $ f_{\nu}(m,t_{ssp},Z)$ is the spectrum of an individual star in a SSP which depends on the fundamental parameters, its mass $m$, age $t_{ssp}$, metal abundance $Z$ and the IMF, $\phi(m)$.
Using the SSPs already described, a constant SFR of
10$~\rm{M_{\sun}}\;\rm{yr^{-1}}$ and adopting the  \citet{Kennicutt1983} IMF, 
we obtain the integrated number of ionizing photons $IQ(H)$, shown in figure~\ref{fig:ssp_q_integ}.
Since $IQ(H)$ is dominated by the most massive stars,
the number of ionizing photons emitted by a galaxy with constant SFR
will initially grow and soon saturate to a constant maximum value, when there is a balance between the newly formed ionizing massive stars and the ones that die.
Looking at Figure~\ref{fig:ssp_q_inst} we see that, almost independently from the metallicity, the characteristic time
of the saturation is set by the rapid drop of $IQ(H)$ above about 6~Myr.
The effect of $\rm{M}_{up}$ on  $IQ(H)$ is a direct consequence of the variation of the $Q(H)$ of SSPs shown in Figure \ref{fig:ssp_q_inst}.
The variation of $\rm{IQ(H)}$ with metallicity decreases with increasing upper mass limit.
After denoting by $C_2$  the calibration coefficient between SFR and Q(H) in the equation below
\begin{equation}
\left( {\frac{ \rm{SFR} } {   \rm{M}_{\sun} \; \rm{yr}^{-1}}    }\right) =  
C_2 \times   \left( \frac{ IQ(H) } { \rm{ s^{-1} } } \right)
\label{eq_sfr_q_obi_old}
\end{equation}
we collect in Table~\ref{tab:C_2_C_3} the values of $C_2$ obtained with our constant SFR models, for different SSPs parameters.
To illustrate the significant variations of  $C_2$ with metallicity at a given $M_{up}$,
we show in Figure~\ref{fig:C2_plot} the plot of $C_2$ against metallicity for 
$M_{up}$~=~40, 120 and 350 $M_{\sun}$
Using the values of $C_2$ given in Table~\ref{tab:C_2_C_3} for different $\rm{M}_{up}$ and $Z$ , we provide a multiple regression fitting relation
between $C_2$ , $\rm{M}_{up}$ and metallicity ($Z$), Equation~\ref{eq_c2_rel}.
As an example, using this fitting  relation,  for $M_{up} =  120~\rm{M}_{\sun} $ and Z = 0.02
Equation~\ref{eq_sfr_q_obi_old} is:
\begin{equation}
\left( {\frac{ \rm{SFR} } {   \rm{M}_{\sun} \; \rm{yr}^{-1}}    }\right) = 
\left( {\frac{ IQ(H) }{ s^{-1} } } \right)
\label{eq_sfr_q_obi}
\end{equation}
We may compare equation \ref{eq_sfr_q_obi} with the one obtained by
\citet{Murphy2012} using the Starburst99 stellar population model with a \citet{Kroupa2001} IMF,
a metallicity of Z = 0.02 and a constant star formation over 100~Myr.
\begin{equation}
\left( {\frac{ \rm{SFR} } {   \rm{M}_{\sun} \; \rm{yr}^{-1}}    }\right) = 
\left( {\frac{IQ(H) }{ s^{-1} }  } \right)
\label{eq_sfr_q_murphy}
\end{equation}
We see that the \citet{Murphy2012} calibration constant is fairly in agreement with  ours.
\subsubsection{SFR - $L_{ff}$ Calibration }
\label{sec:sfr_lff_cal}
By combining equation~\ref{eq_sfr_q_obi_old} and equation~\ref{eq_Lff_new_1}, we derive the
relation between the SFR and thermal radio emission:
\begin{equation}
 \frac{ \rm{SFR} }{ \rm{   M_{\sun} yr^{-1} }  }   = 
\frac{ L_{\rm{ff}}  }   { C_3 }
\left(    { \frac{ \rm{T}_e } {  10^4 \hbox{K} } }     \right)^{-0.3} \,
\left(    \frac{ 1}  { G_{dra} }   \right) 
\label{eq_lff_sfr_const_0}
\end{equation}
where $C_3 = 1/ (C_1 \times C_2)$.
The values of the coefficient $C_3$ for different SSP parameters are provided in Table~\ref{tab:C_2_C_3} and the variation of $C_3$ with metallicity and
$M_{up}$ is shown in Figure~\ref{fig:C3}.
Using these values of $C_3$ given in Table~\ref{tab:C_2_C_3} for different  $\rm{M}_{up}$ and $Z$, we also, as we did for the case $C_2$, provide a multiple regression fitting relation
between $C_3$, $M_{up}$ and metallicity ($Z$), Equation \ref{eq_c3_rel}.
As an example, using this fitting  relation  for $M_{up} =  120~\rm{M}_{\sun}$ and Z = 0.02  and assuming the already quoted approximation for the Gaunt factor,
Equation~\ref{eq_lff_sfr_const_0} becomes 
\begin{equation}
\frac{ \rm{SFR} }{ \rm{   M_{\sun} \; yr^{-1} }  }   = 
\left(  \frac{ L_{\rm{ff}}  }   { 2.77 \times 10^{27}  }   \right) 
\left(    { \frac{ \rm{T}_e } {  10^4 \hbox{K} } }     \right)^{-0.45} \,
\left( \frac{  \nu  }{  \rm{GHz} }\right)^{0.1}
\label{eq_lff_sfr_const_00}
\end{equation}
A similar equation is provided by \citet{Murphy2012}:
\begin{equation}
\frac{ \rm{SFR} }{ \rm{   M_{\sun} \;  yr^{-1} }  }   = 
\left(  \frac{ L_{\rm{ff}}  }   { 2.17 \times 10^{27}  }   \right) 
\left(\frac{ T_e } { 10^4 \rm{K} } \right)^{-0.45} \left( \frac{  \nu  }{  \rm{GHz}   }
\right)^{0.1}
\label{eq_lff_sfr_murphy}
\end{equation}
\subsubsection{SFR versus H$\alpha$  and H$\beta$ calibrations}
Using well-known relations between
$IQ(H)$ and the intensity of recombination  lines \citep{Osterbrock1989},
we may also obtain the corresponding calibrations for the SFR.
For H$\alpha$  and H$\beta$  we have
\begin{equation}
   \left( \frac{ \rm{IQ(H)} } { \rm{s}^{-1 } } \right)  =  7.31 \times 10^{11}
  \left( \frac{ L(H\alpha) } { \rm{erg} \; \rm{s}^{-1} } \right)
  \label{eq_q_ha}
\end{equation} 
\begin{equation}
  \left( \frac{ \rm{IQ(H)} } { \rm{s}^{-1 } } \right)  =  0.21 \times 10^{13} 
  \left( \frac{ L(H\beta) } { \rm{erg} \; \rm{s}^{-1} } \right)
\label{eq_q_hb}
\end{equation}
Using  the above relations in equation~\ref{eq_sfr_q_obi_old}, we write in  Appendix \ref{appendix_a} analytical equations for the  SFR-H$\alpha$ (equation~\ref{eq_sfr_ha_obi_fit}) and SFR-H$\beta$(equation~\ref{eq_sfr_hb_obi_fit}) calibrations, as a function of Z and M$_{up}$.
For the case with $M_{up}~=~120~\rm{M}_{\sun}$ and Z~=~0.02 we obtain  with these analytical equations
\begin{equation}
  \left( { \frac{ \rm{SFR} } { \rm{M}_{\sun} \; \rm{yr}^{-1} } }\right) =  4.79 \times 10^{-42}
   \left( {\frac{ L(H\alpha) }{ \rm{erg} \; \rm{s}^{-1} } } \right)
  \label{eq_sfr_ha_obi}
\end{equation}
\begin{equation}
  \left( { \frac{ \rm{SFR} } { \rm{M}_{\sun} \;   \rm{yr}^{-1} } }\right) =  0.14 \times 10^{-40}
  \left( {\frac{ L(H\beta) }{ \rm{erg} \; \rm{s}^{-1} } } \right)
\label{eq_sfr_hb_obi}
\end{equation}
Our  calibration coefficient in equation ~\ref{eq_sfr_ha_obi} is in good agreement with the value of $4.55\times{10^{-42}}$
obtained by \citet{Bicker2005} using \textit{GALEV} synthesis code.
\subsection{Non-Thermal Radio Emission}
\label{sec:nontherm}
\begin{figure*}
\centering
\includegraphics[width=\linewidth]{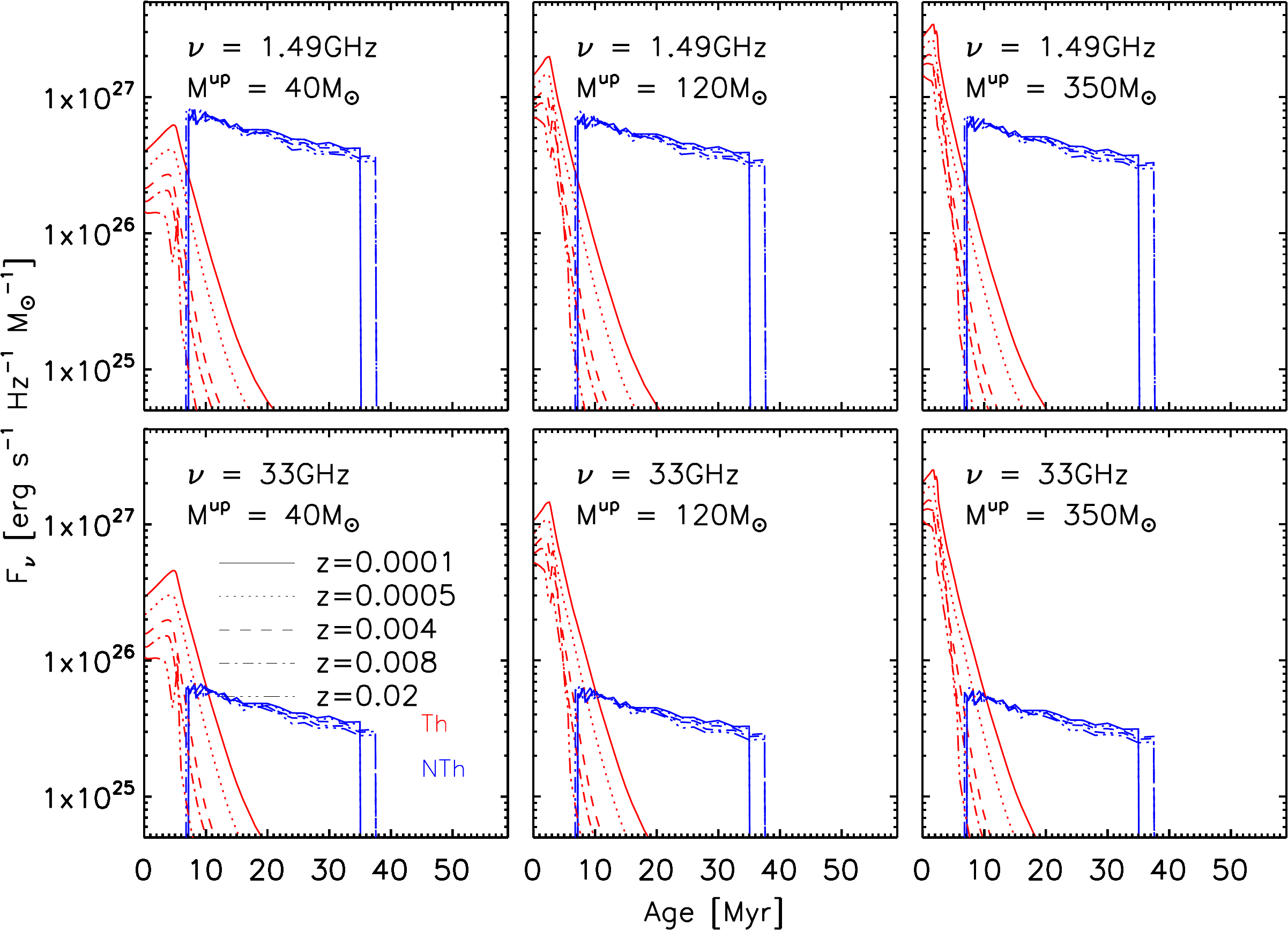}
 \caption{Variation of the 1.49~GHz and 33~GHz non-thermal (free-free) emission  with age at different  metallicities and different IMF upper mass limits. The red lines indicate thermal radio emission and the blue lines non-thermal radio emission. The plot for the thermal radio emission has already been shown in Figure~\ref{fig:therm_14_33} but added here for comparison with the non-thermal radio emission.
 As easily noticed in the figure, the non-thermal emission is almost unchanged for each of the different IMF upper mass limits, owing to the fact
that masses above 30~$\rm{M}_{\sun}$ collapse directly to BH and do not explode to produce non-thermal emission. For this same reason, the non thermal emission for all plots starts at about an age of 7~Myr.
For comparison purposes,
the plot of the thermal radio emission is included to show
it's pronounced variation with upper mass limits. At 1.4~GHz (upper panel), the non-thermal radio emission dominates  that at 33~GHz (lower panel) whereas at 33~GHz, the thermal emission dominates.
}
\label{fig:ntherm_14_33}
\end{figure*}

We know quite little about the source of  non-thermal (NT) radio emission in star-forming galaxies.  The mechanism is believed to be synchrotron emission from relativistic electrons that are accelerated by
ISM shocks in the outskirts of CCSN explosions \citep{Condon1992}.
B02 derived the following relation between NT radio emission $L^{NT}$  and the CCSN rate
\begin{eqnarray}\nonumber
\frac{ L_{NT}/\frac{ \nu_{CCSN} } { yr^{-1} }} {10^{30} ~\rm{erg} \; \rm{s}^{-1} \; \rm{Hz}^{-1}}  &=& E^{SNR}_{1.49} ( \frac{\nu}{1.49} )^{-0.5} + E^{NT}_{1.49} ( \frac{\nu}{1.49} )^{-\alpha} \\
	\label{eq_l_nt_bre}
\end{eqnarray}
In the above equation $\nu_{CCSN}$ is the CCSN rate, $E^{SNR}_{1.49}$ is the  average non thermal radio luminosity due to a young Supernova Remnant (SNR) and
$E^{NT}_{1.49}$ is the average injected energy in relativistic electrons per CCSN event.
Moreover, B02 estimated that, at 1.49~GHz,
\begin{equation}
	 E^{SNR} \approx 0.06E^{NT}.
		\label{eq_e_snr_bre}
\end{equation}
$E^{NT}_{1.49}$ can be estimated from observations. For our Galaxy B02 adopt $L^{NT} = 6.1 \times 10^{21} \;  \rm{W} \;  \rm{Hz}^{-1}$ \citep{Berkhuijsen1984}  at 0.4$ \rm{GHz}$ and  $ \nu_{CCSN} = 0.015$.  They obtain
for a radio slope $\alpha (  \frac{ -d \; \rm{log}S_{\nu} } { -d \; \rm{log} \nu  } ) = 0.8$, a value of $E^{NT}_{1.49}$ = 1.44.
We note that the final fate of massive stars is a critical assumption for estimating their contribution to thermal radio emission. The formation of NSs and BHs depends on the amount of mass lost by the massive star through stellar winds and on the hydrodynamics of the explosion. 
One of the the most troublesome facts is that the  variations shown by these pre-supernova stars in their structural properties, like for e.g. the Fe-core and O-core masses,
are non-monotonic and pronounced even within small differences in the ZAMS masses \citep{Sukhbold2014}.
Some  recent works in the  attempt to characterize the parameters of
successful and failed supernovae have  yielded some  structural parameters
that can be utilized in predicting the fate of SNe \citep[eg.][]{Ertl2015,Ugliano2012,Janka2012,Connor2011}. Using these parameters, 
\citet{Spera2015} and  Slemer et al. 2017 (to be submitted)
were able to characterize the final fate of  \textit{\small{PARSEC}} massive stars
for the different criteria adopted for successful CCSN explosion.
Following \citet{Spera2015} and {Slemer et al. 2017} 
we assume that stars with initial mass above about $M = 30~\rm{M}_{\sun}$ do not contribute to
non thermal radio emission, contrary to B02 where all stars of masses above
$M = 8~\rm{M}_{\sun}$ were thought to explode as CCSN.
The region between about  $24 - 30~\rm{M}_{\sun}$ is critical because the  explosion criterion by \citet{Connor2011} produces a much higher number of NSs than that produced by  \citet{Ertl2015} criterion. 
Finally we stress that we assume that  progenitors undergoing pair-instability SNe either collapse to a BH  or are  completely incinerated
by a thermonuclear explosion without producing relativistic electrons, similarly to what is assumed  for SNIa.
This assumption on the CCSN distribution with initial mass modifies the expected non thermal luminosity of star-forming galaxies and requires a re-calibration of the GRASIL model.
After accounting for the failed and successful SNe in the evaluation of non thermal emission from Equation \ref{eq_l_nt_bre}, we show in
Figure~\ref{fig:ntherm_14_33}  the variation of the 1.49~GHz non-thermal  radio emissions with age, for different upper mass limits  and metallicities.
We note that non-thermal radio emission is not sensitive to the  upper mass limit of the IMF, as long as it is larger than the assumed threshold for failed SN,
M~=~30$~\rm{M}_{\sun}$. In this case non thermal emission begins at an age of $\sim$7~Myr and not earlier ($\sim$3~Myr) as  is the case in  B02.
By coincidence, this critical time corresponds to about the epoch at which
thermal emission abruptly fades down, as depicted in Figure~\ref{fig:ntherm_14_33}.
Non thermal emission  remains effective up to about 35 - 38~Myr, depending on the metallicity.
In subsequent sections we will show how thermal and non-thermal radio emission are checked and eventually re-calibrated using the  \textit{GRASIL} code.

\section{Calibration of the new SSP suite with  \textit{\small{GRASIL}}}
\label{sec:ssp_gra}
In the previous sections, we described a new suite of SSPs that will supersede the ones used in the current version of  \textit{\small{GRASIL}}.
The new suite differs in many aspects from the previous one as we briefly list in the followings.
(1) The SSP are based on the most recent  \textit{\small{PARSEC}} stellar evolutionary tracks
with an updated physics, in particular new mass-loss recipes and finer and wider coverage in initial mass and metallicity;
(2) they include the most recent advances in our understanding of the CCSN explosion mechanism and account for the so called "\textit{failed SN}";
(3) a more accurate gaunt factor and metallicity-dependent electron temperature were incorporated in the
relation between the integrated ionization photon luminosity and thermal radio luminosity
(4) in the current suite, we may adopt several values of the IMF upper mass limit
Because of all these differences, we need to check and recalibrate some parameters of the
SED produced by  \textit{\small{GRASIL}}, in particular to check whether  we are able to reproduce
the canonical value of $q_{1.4GHz} = 2.35$, 
observed in a prototype normal star-forming galaxiy, see Eq.~\ref{eq_q_ratio}.
For this purpose, in the next section we will use  \textit{\small{GRASIL}} with the new SSPs to reproduce the
SEDs of a some  selected well studied galaxies.
\subsection {SED Modelling with  \textit{\small{GRASIL}}}
\label{sec:sed_gra}
 \textit{\small{GRASIL}} is a spectro-phometric code able to  predict the SED of galaxies from the FUV to the radio, including state-of-the-art
 treatment of dust reprocessing \citep{Silva1998, Silva2011, Granato2000},
production of radio photons by thermal and
non-thermal processes (BO2) and an updated treatment of PAH emission (\citet{Vega2005}). The reader is referred
to these original papers for a fully detailed description of the code.
It is also worth noting that nebular emission was already included in  \textit{\small{GRASIL}} by \citet{Panuzzo2003}
but with a completely different procedure which also accounted for different electron densities.
In this respect our procedure is more simple because nebular emission is added directly to the SSPs, but it allows a more versatile use of  \textit{\small{GRASIL}}.

For sake of convenience, we briefly summarise here the main  features of  \textit{\small{GRASIL}}.
The first step is to compute a chemical evolution model that provides the
star formation history (SFH) and the metallicity and mass of the gas.  
Other quantities are computed by the code \textit{CHE$-$EVO}, such as mass in stars, SN rates, detailed
elemental abundances, but they are not used for the spectro-photometric synthesis.
\textit{GRASIL's}  main task is to compute  the SED resulting from the  interaction between the stellar radiation from CHE$-$EVO  and dust, using a relatively flexible and realistic geometry where stars and dust are distributed in a spheroidal and/or a disk profiles. A spherical symmetric distribution with a King profile is adopted in the case of spheroidal systems while, for disk-like systems, a double exponential of the distance from the polar axis and from the equatorial plane is adopted.
The dusty environments are made up of dust (i) in interstellar HI clouds heated by the
general interstellar radiation of the galaxy (refered to as the  cirrus component),
(ii) associated with star-forming molecular clouds (MC)  and (iii) in the circumstellar shells of Asymptotic Giant Branch (AGB) stars.
 \textit{\small{GRASIL}} performs the radiative transfer of starlight through these environments.
We remind the reader that the reprocessing of starlight by dust in envelopes  of AGB stars is already taken into account   in our SSPs.
For the intrinsic dust properties,  \textit{\small{GRASIL}} adopts grains (graphite and silicate) in thermal equilibrium and a mixture of smaller grains  and PAHs fluctuating in temperature.
The dust parameters are tuned in order to match with the observed emissivity and extinction properties of the local ISM \citep{Vega2005}.
\subsubsection{M100: a close to solar metallicity galaxy}
\label{sec:m100}
\begin{figure*}
\begin{center}
\includegraphics[width=0.45\linewidth]{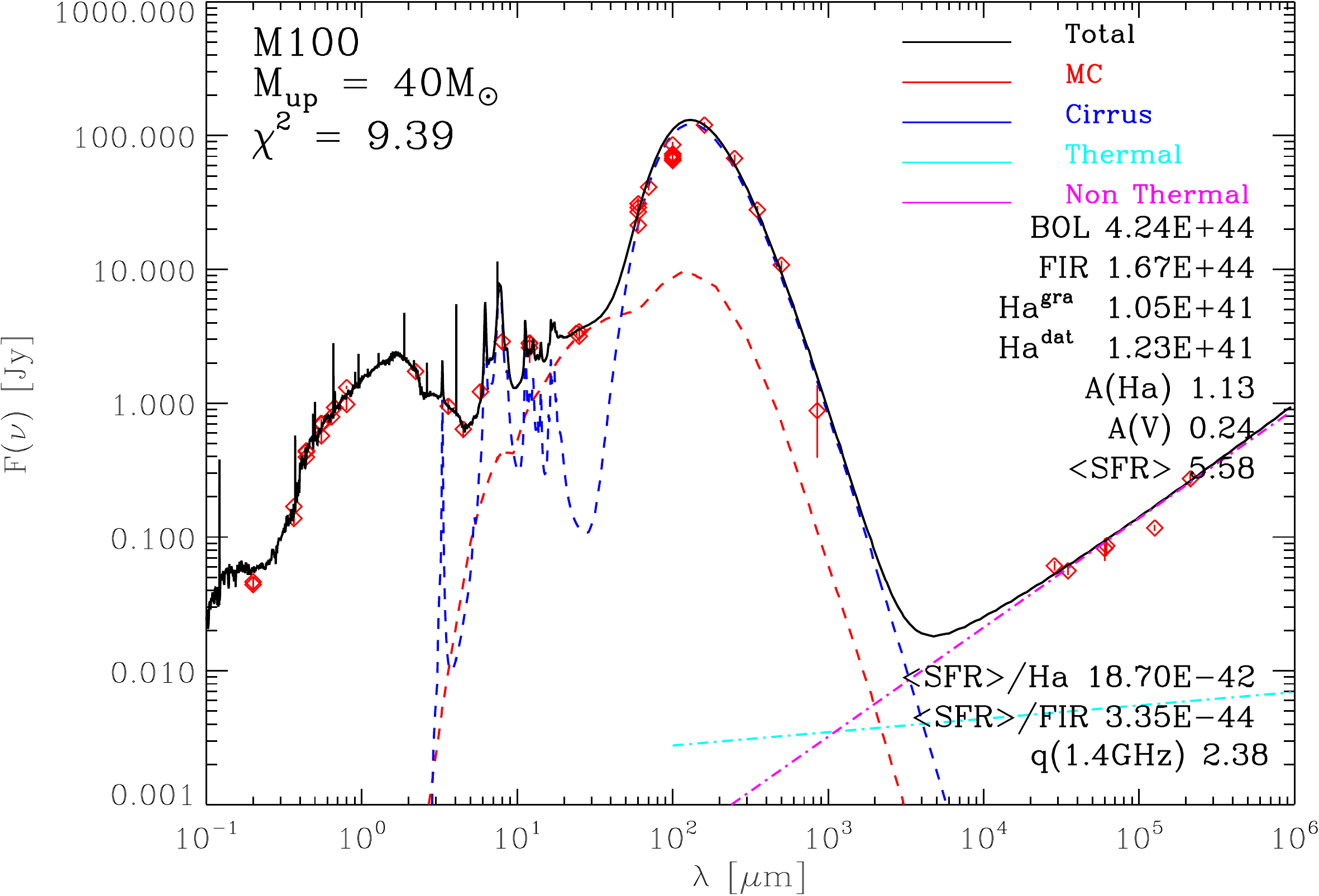}
\includegraphics[width= 0.45\linewidth]{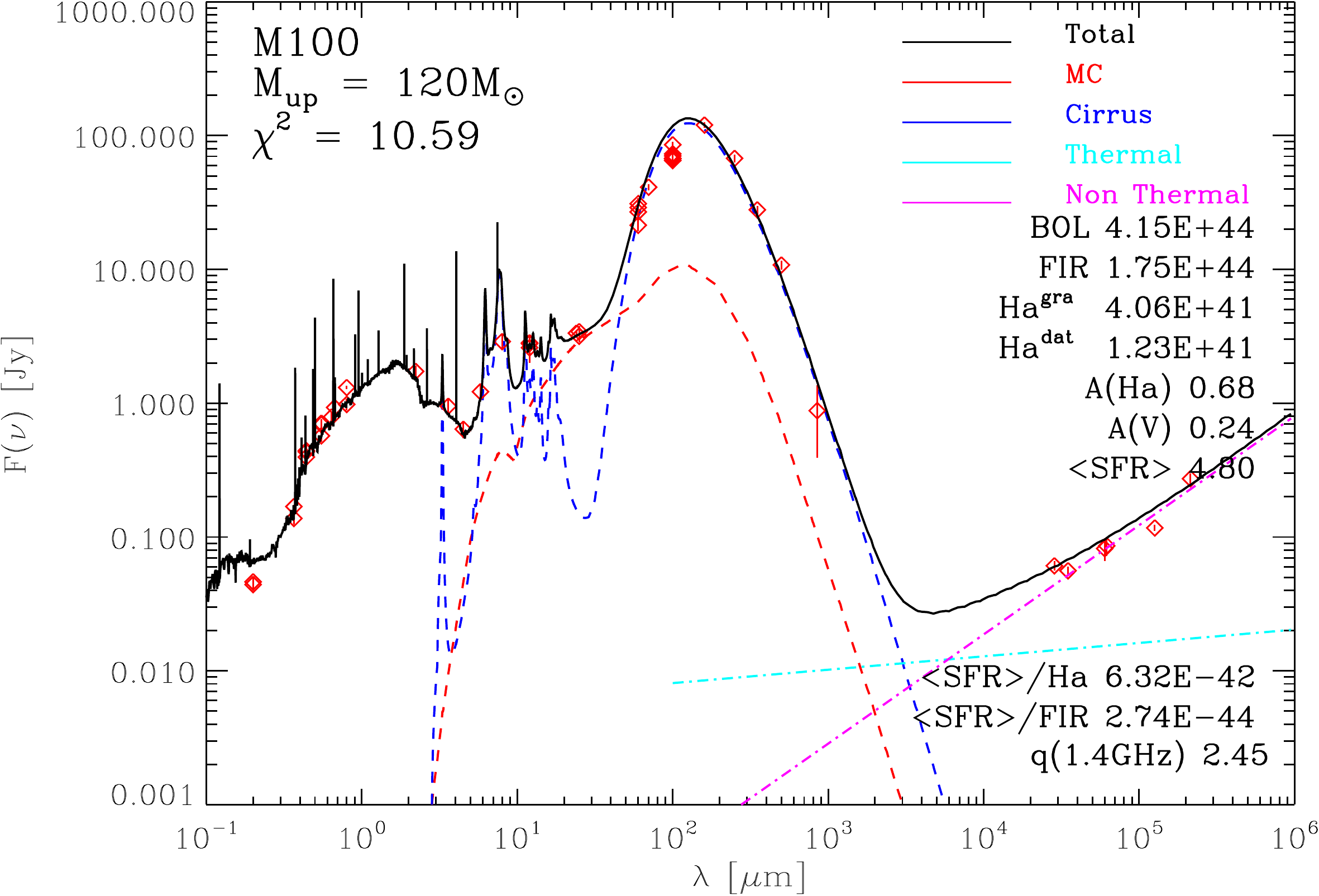}\\
\end{center}
 \caption{M100   \textit{\small{GRASIL}} best-fit SED for  $M_{up}$ = 40~$\rm{M}_{\sun}$ (right panel)
 and $M_{up} = 120~\rm{M}_{\sun}$(left panel).
 The different dust components, diffuse medium and molecular  clouds are represented by the dashed blue and red lines respectively. The thermal and non thermal radio components are represented by the cyan and purple dashed dotted lines. The thermal component can be seen to be negligible for the case of $M_{up} = 40~\rm{M}_{\sun}$. For the different $M_{up}$, note the differences in the predicted quantities  labelled in the plots (see text for more details), in particular the H$_{\alpha}$ luminosities, attenuations (A(H$_{\alpha}$)) and SFR(H$_{\alpha}$) calibration coefficient.
We estimated  $E^{NT}_{1.49} = 1.944$, a factor of 1.35 larger than that obtained in B02.
Our average value of q =  2.4  which is in line with the observed value of 2.35
(Equation~\ref{eq_q_ratio}) is an indication of our good calibration.
\label{fig:m100}
 }
\end{figure*}

\begin{figure}
\centering
\includegraphics[width=0.95\linewidth]{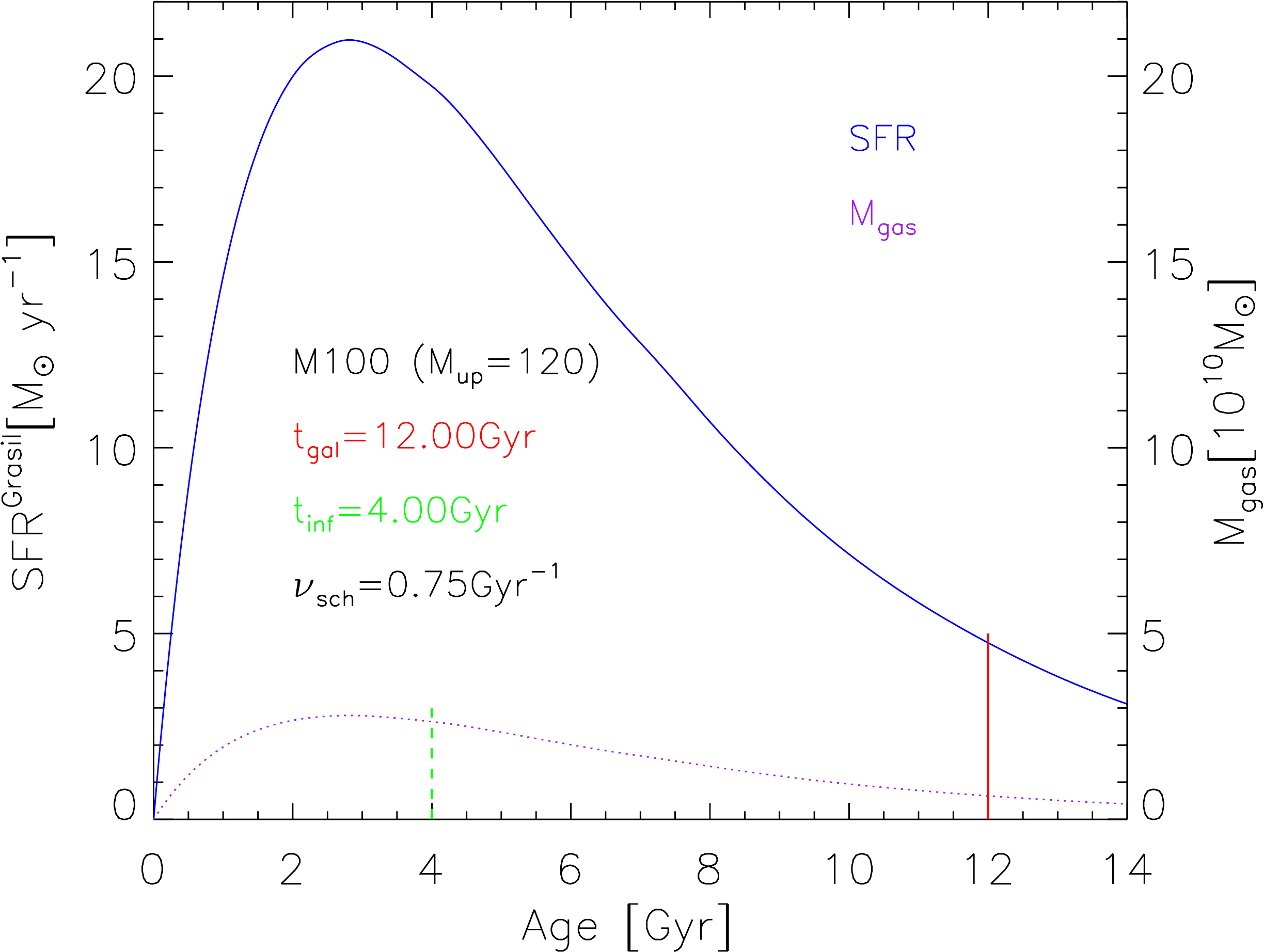}
\caption{ Star formation history of M100 adopted in our \textit{CHE-EVO} model.
 The model parametrs are labelled: galaxy's age (solid red line) , infall time (dashed green line) and schmidt-law efficiency.
 These parameters were adopted by \citet{Silva1998} based on constraints to reproduce the observed final mass of M100.
 The SFH is indicated by the solid blue lines while the gas mass history by the dotted purple line.
 } 
 \label{fig:M100_sfh}
\end{figure}

The galaxy M100 was selected from the Spitzer Nearby Galaxies Survey (SINGS) \citep{Kennicutt2003},  a collection of 75 nearby galaxies with a rich data coverage from the far-UV to the radio,
because it is one of the best sampled objects, including the intensities of main emission lines.
This will provide the opportunity to check, for the first time in  \textit{\small{GRASIL}}, the consistency of the SED continuum and emission lines fitting.
To obtain the chemical evolution model we adopt the parameters 
of the chemical evolution code  \textit{CHE-EVO} given
by \citet{Silva1998} who  was able to well-reproduce M100. They are infall time scale $t_{inf}~=~4.0$~Gyr, efficiency of the Schmidt-law star formation law $\nu_{sch}~=~0.75$
and the total infall mass  $m_{inf}$ = $2 \times 10^{11}~M_{\sun}$. 
Figure~\ref{fig:M100_sfh} shows the SFH resulting from the adopted chemical evolution parameters. The SFH is indicated by the solid blue line and the gas mass history by the dotted purple line.
 To perform the SED fit, we consider several
time steps along the chemical evolution of the galaxy
for which we built a library of SED models with  different
\textit{GRASIL} parameters. We note that since the observed SED refers to the whole galaxy,
the derived parameters correspond to a luminosity average of the physical properties,
as commonly obtained by this kind of fitting process.
As a particular check, we also investigate
the effect of changing the upper mass limit of the IMF, $\rm{M}_{up}~=~40$,120 and  $350~M_{\sun}$,
on some physical quantities derived from our best-fit SED, in particular the SFR.
The main  \textit{\small{GRASIL}} parameters varied  are the molecular gas fraction  $f_{mol}$ , the  escape time $t_{esc}$ and the  optical
depth of molecular clouds (MC) at 1$\micron$ $\tau_{1}$.
The best fits  were obtained by minimizing the merit function $\chi^2$ which is given by
\begin{equation}
\chi^2 = \frac{1}{N}  \sum^N_{i=1}
\left(   \frac{ F_{mod}(i) -   F_{obs}(i) } {Err(i)}     \right )^2
\label{eq;chisq}
\end{equation}
where $F_{mod}(i)$, $F_{obs}(i)$ and  $Err(i)$ are the model flux values, the observed fluxes and observational errors respectively. $N$ is the number of photometric bands used to obtain the best  fit.
Our  \textit{\small{GRASIL}} best-fit SEDs  of  M100 for
$M_{up} = 40~\rm{M}_{\sun}$ and $M_{up} = 120~\rm{M}_{\sun}$,
are shown in the left and right panels of Figure~\ref{fig:m100} respectively,
while the corresponding best
fit parameters are summarized in Table ~\ref{tab:bestfit_params}.
We do not show the case with $M_{up} = 350~\rm{M}_{\sun}$,
because
its thermal-radio emission was significantly exceeding the observed one.
This discrepancy could not be cured by
varying any other parameter in the fit.
This result is, by itself, quite interesting and shows that
for a normal star-forming galaxy it is difficult to have
an average IMF extending up to such high initial masses.
The different dust components in diffuse
medium (cirrus)  and molecular clouds are shown and indicated by the blue and red dashed lines, respectively. The thermal and non-thermal radio emission components are indicated by the cyan and purple dashed dotted  lines, respectively. The thermal component can be seen to be negligible for the case of $M_{up} = 40~\rm{M}_{\sun}$.
We include, for the first time in  \textit{\small{GRASIL}}, selected emission lines in the SED. 
The labels in the plots refer to some important quantities derived from our best fit:  the bolometric luminosity from UV to radio (BOL), the total IR luminosity from 3 - 1000$\micron$ (FIR), the predicted extinction uncorrected $H_{\alpha}$ luminosity
(Ha$^{gra}$), the observed $H_{\alpha}$ luminosity
(Ha$^{dat}$), $H_{\alpha}$ attenuation (A(H$\alpha$)), attenuation in the
V($0.55\micron$) band (A(V)), the SFR averaged over the  last 100 Myr
($<\rm{SFR}>$), the q-parameter as defined by equation~\ref{eq_q_ratio} (q(1.4GHz)), the coefficients of  the SFR(H$_{\alpha}$) and SFR(IR) calibrations ($<\rm{SFR}/Ha>$ and $<\rm{SFR}/ FIR>$ respectively).
A summary of other important quantities derived from our best fit is provided in
Tables~\ref{tab:dale_gra_derived_1_unsub}
and ~\ref{tab:dale_gra_derived_2_unsub}.
For both values of the adopted $M_{up}$, the best fits match very well the overall observed UV-radio SEDs.
However, for $M_{up}$ = 120$~\rm{M}_{\sun}$, the model over-predicts by factor of $\sim$3  the observed  (i.e. non corrected by extinction, hereafter 'transmitted') H${\alpha}$ luminosity, $1.23 \times 10^{41} \text{ergs}  \; \text{s}^{-1}$ (taken from \cite{Kennicutt2009}). This upper mass limit also produces a higher  UV emission and a slight excess of thermal emission. On the other hand, the observed  H${\alpha}$ luminosity is well reproduced in the case of $M_{up} = 40~\rm{M}_{\sun}$.
The over-predicted H${\alpha}$ luminosity resulting from adopting $M_{up}$ = 120~$\rm{M}_{\sun}$ could
be lowered by  increasing the escape time of young stars in their parent clouds, leading to a larger absorption in  the MCs. This will also
lower the predicted far-UV emission which, in the current best fit, is slightly larger than the observed value.
However, a larger absorption from MCs will increase the
predicted $24~\micron$ flux above the observed one. 
Note that at $24~\micron$ the cirrus component has a pronounced minimum and its contribution to the overall MIR emission is only a small fraction of the total.
Thus, the $24~\micron$ flux, being dominated by the MC emission, is indeed a strong diagnostic for the amount of light reprocessed by the MC component. 
From our best fit, we obtain a CCSN rate to NT radio luminosity ratio (Equation~\ref{eq_l_nt_bre}) that is about a factor of 1.35 larger than the value obtained by B02 using the previous radio model. That is,
\begin{equation}
E^{NT}_{1.49} (\rm{This work}) = 1.944 = E^{NT}_{1.49} (\rm{B02}) \times1.35
\end{equation}
We anticipate here that this value is confirmed by all subsequent SED fits.
We also note that this value is almost independent from the adoption of $M_{up}$
for the reasons already discussed previously.
The predicted average SFR resulting from the panchromatic fit
is only slightly affected by the adoption of a different value of  $M_{up}$.
By increasing $M_{up}$ from
40 to 120 $~\rm{M}_{\sun}$ the average SFR decreases by about 16~per~cent.
\subsubsection{NGC~6946: individual star-bursting regions}
In the previous section we were able to reproduce fairly well the observed SEDs  of the prototype  galaxy M100, in particular, we used  the best fit model  to calibrate the constant $E_{NT}$ of the  non-thermal radio emission model. 
In this section we wish to check the thermal component of our radio emission model. 
For this purpose, we compare our synthetic SEDs with those
of selected extranuclear regions  of NGC~6946, a well  studied starburst galaxy at a distance of 6.8~Mpc. This galaxy  is dominated by very young starburst regions and shows
relatively large thermal over non-thermal ratios \citep[eg][]{Israel1980,Heckman1983}, and thus is particularly well suited for a test of the free-free emission originating from star formation. \citet{Murphy2010} observed these regions at 1.4, 1.5, 1.7, 4.9, 8.5 and 33~GHz  and complemented their full SEDs with  existing data from the UV to the submm range.
For eight observed regions  \citet{Murphy2010} estimated an average 33~GHz thermal component of $\approx$ 85~per~cent of the total, while for one region (named extra-nuclear region 4),
this percentage was found to be significantly lower,  $\approx$ 42 ~per~cent, likely for the presence of a so-called anomalous dust emission component which suppresses the thermal component.
Particulary interesting for our purpose is that \citet{Murphy2010} also provide the intensity of the H$\alpha$ emission for each region, that can be directly compared to the predictions
of our model.

To model the SEDs of the extra-nuclear star-forming regions of NGC~6946
we adopt a spherical symmetric distribution with a King profile for the stars and dust.
Most important, we use the data not corrected for the local background emission of the galaxy,
though  \citep{Murphy2011} provide also data corrected for this emission. We thus adopt, for each region, a chemical evolution model composed by a star-burst superimposed to a quiescent star formating component. Our choice is dictated by the fact that the subtraction
by \citep{Murphy2011} may bias our results and we prefer to eventually split the contribution of the starburst and the old disk components directly from our fits.
\begin{figure}
\includegraphics[width=0.95\linewidth]{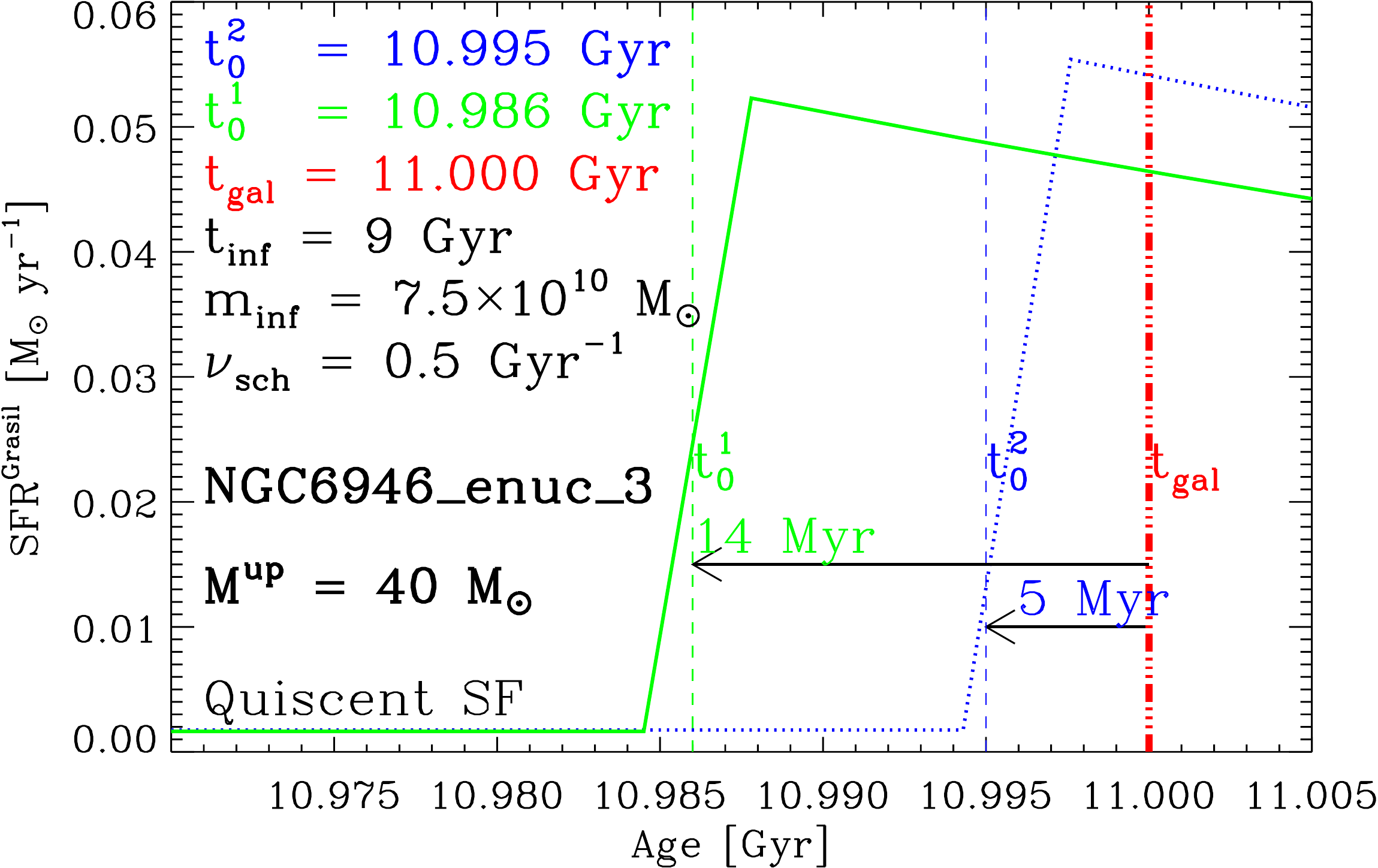}
\caption{SFH of one of the 8 extranuclear regions, NGC~6946\_enuc\_3 for $M_{up} = 40~M_{\sun}$ case.
A  starburst age of $t_{sb}^{1}$~=~14~Myr (green solid line) produced the best fit but we added $t_{sb}^{2}$=~5~Myr (blue dotted line) case only to illustrate  that we do not expect any contribution to the non-thermal emission from the starburst at this age according to our  radio model in  Figure~\ref{fig:ntherm_14_33}.
At burst ages  between 7 and 30~Myr, we expect a contribution to the non-thermal radio emission also from the starburst.
The  red  dotted-dashed line indicates the age (11.000~Gyr) at which the SED of this region was observed.
\label{fig:ngc6946_sfh}
}
\end{figure}
For purposes of clarity we show in Figure~\ref{fig:ngc6946_sfh}  a plot of the star formation history of the extra-nuclear region $\sharp$3 for $M_{up} = ~40~\rm{M}_{\sun}$.
The corresponding  chemical evolution model parameters are labelled in the left side of the figure.
For these models, the galaxy is observed at a given age $t_{gal}~=~11$~Gyr. The ongoing starburst has an exponentially declining SFR
that begins at  $t_0~=~t_{gal}-t_{sb}$, where t$_{sb}$ is the age of the starburst.
Two SFR examples are given for a starburst age of  t$_{sb}$~=~5~Myr and t$_{sb}$~=~14~Myr respectively. In the first case we are not expecting any contribution to the non-thermal emission from the starburst while, in the second case
since the age is larger than 7~Myr (see Figure~\ref{fig:ntherm_14_33}) we expect a contribution to the non-thermal radio emission also from the starburst. Thus any age between t$_{sb}$~=~ 7 and 30~Myr can provide SEDs with different percentages of thermal vs non-thermal radio emission.
We note that this can be used to mimic possible different contributions of non thermal emission from the underlying disk.
To obtain the  \textit{\small{GRASIL}} best fit models we mainly varied the following parameters:
the escape time of young stars from their birth clouds $t_{esc}$, the optical depth of molecular clouds at 1$\micron$
$\tau_1$, molecular gas fraction $f_{mol}$, the submm dust emissivity index $\beta$ and the start time of the burst $t_0$ (represented by the starburst age $t_{sb}$).
The parameters of the best fit SEDs are shown in  Table~\ref{tab:bestfit_params} while, the corresponding  plots,  are shown
in Figure~\ref{fig:ngc6946_sed_pg2_olga}. In this figure, left panels show the results obtained adopting $M_{up}~=~40M_{\sun}$ and  right panels show those for  $M_{up}~=~120~M_{\sun}$.
\begin{figure*}
\begin{center}
\includegraphics[width=0.9\linewidth]{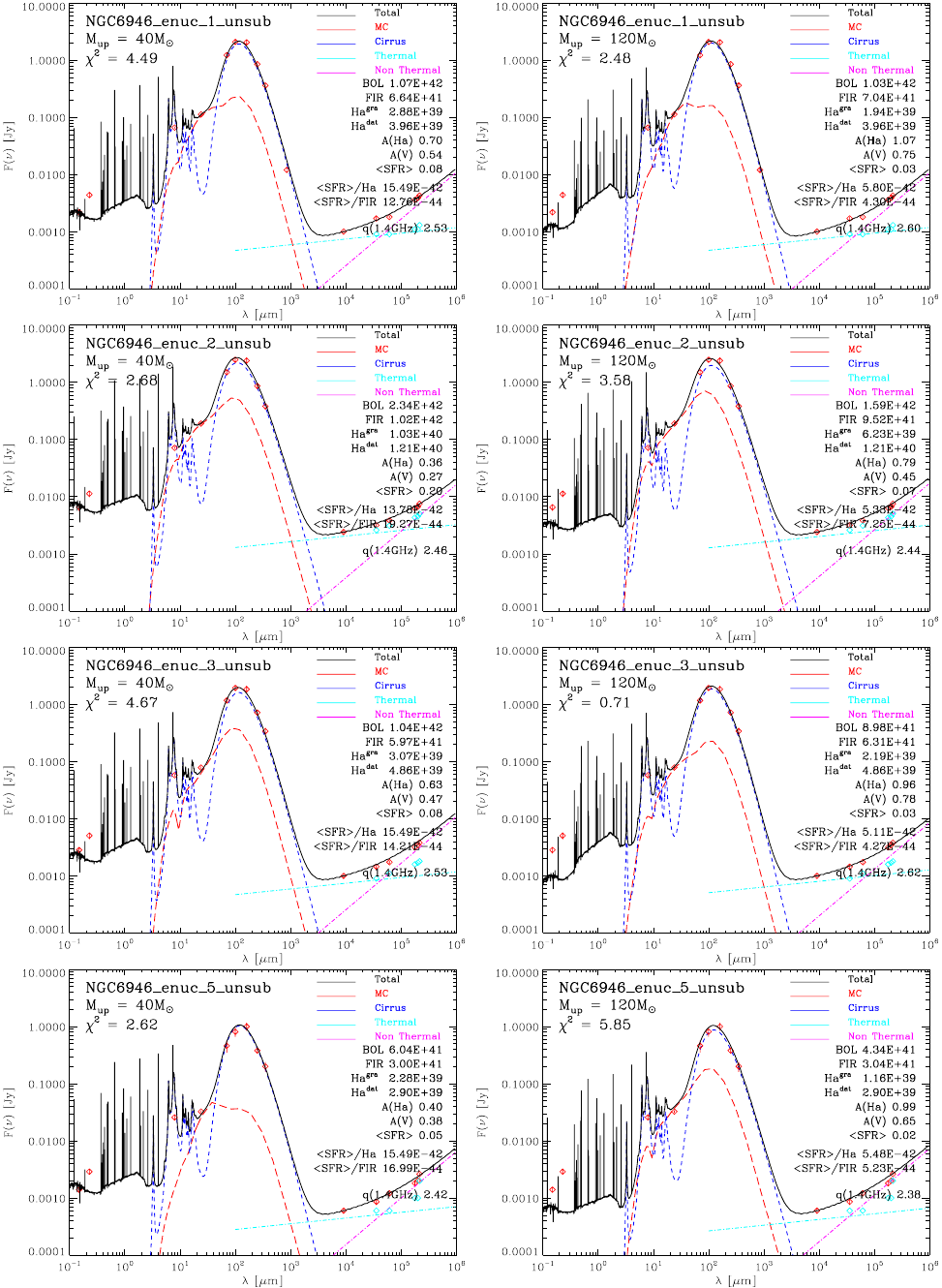}
\end{center}
 \caption{ \textit{\small{GRASIL}} best fit SEDs for the  extra-nuclear regions of NGC~6946 (red diamonds) for upper mass limit of  40 (left panels) and 120~$\rm{M}_{\sun}$(right panels).
To test our radio decomposition, we over-plotted on top our thermal radio component (dotted-dashed cyan lines) the local background subtracted radio data flux (cyan diamonds) by \citet{Murphy2011}.
 \label{fig:ngc6946_sed_pg2_olga}
}
\end{figure*}
\begin{figure*}
\begin{center}
\includegraphics[width= 0.9\linewidth]{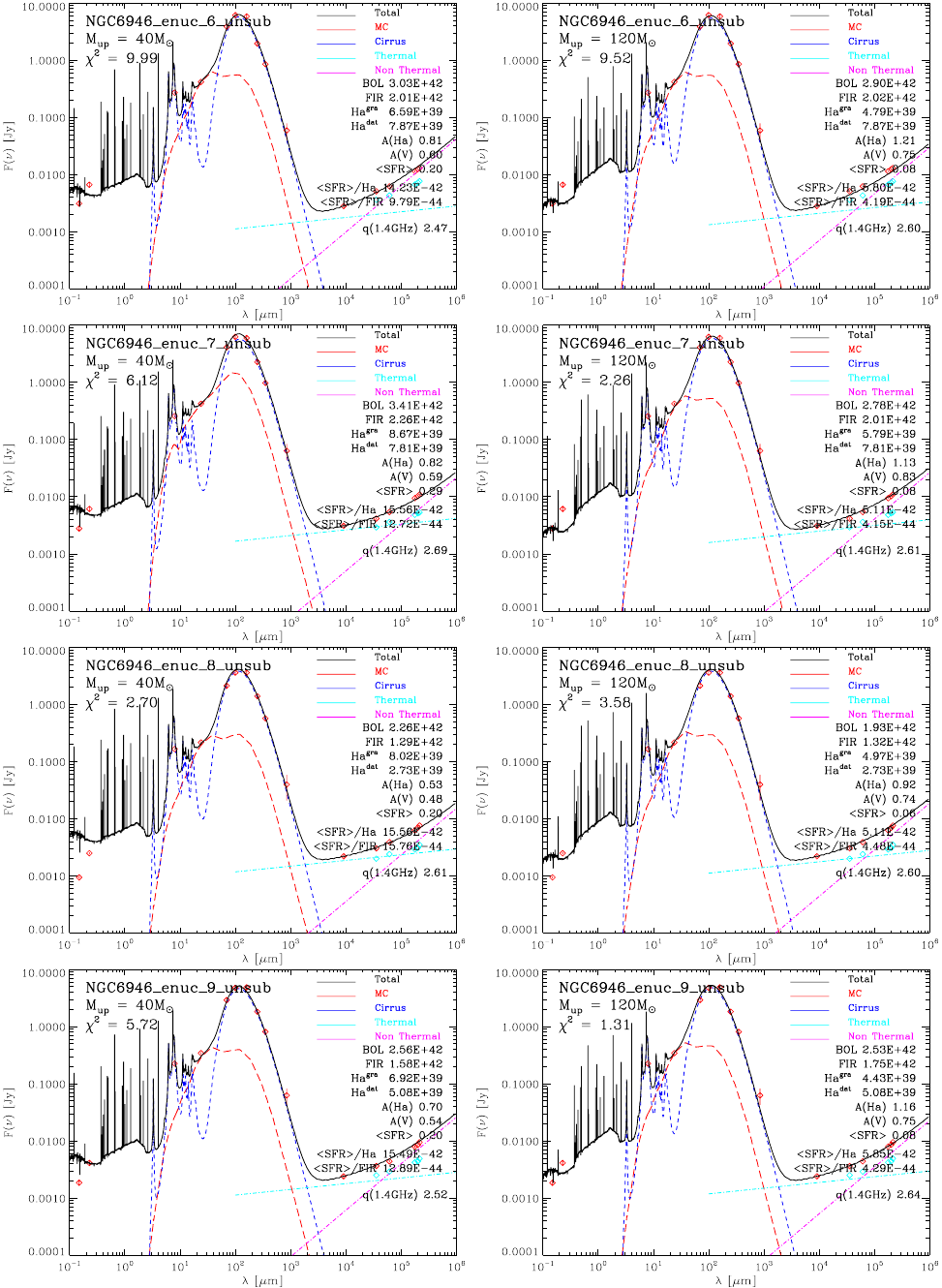}
\end{center}
 \contcaption{}
\end{figure*}
For all regions the observed IR SED can be fairly well reproduced by using either of the IMF upper mass limits.
In general, the  value of $q_{1.4}$  is found to lie between 2.5 and 2.6. This is about  0.2 dex larger than the  value observed in the  normal
star-forming galaxy M100 ($q_{1.4}$ ~=~ 2.35) implying that, in these star-forming regions, the ratio between radio and FIR luminosity
is about a factor 1.6 lower than in the normal star-forming galaxy M100. The estimated young starburst ages support the notion that
this is due to a lack of non-thermal radio emission as predicted by the original models by B02.
This is also evident from the radio slope which is found to be flatter than the value  observed in
the normal star-forming galaxy ($\alpha\simeq$- 0.8). It is worth reminding here that we adopted the calibration ($E^{NT}_{1.49}$) of non-thermal radio luminosity obtained from the fits to M100. We note also that since the radio emission is dominated by
the thermal component, which has no free parameters, the SFR in these regions is very well determined by the 33~GHz point, modulo the IMF.
Another characteristics of the fits is the difficulty of reproducing the UV data, in spite of a fairly well fit in all the other bands from mid-IR to radio bands. 
In particular the runs  with $M_{up} = 40~M\sun$ tend to show larger UV fluxes than the corresponding cases made with $M_{up} = 120~M\sun$,
which is against what could be expected from the dependence of the UV luminosity on the IMF.
This can be noted in Figure~\ref{fig:ngc6946_sed_pg2_olga} where we see that adopting
$M_{up}~=~120~M\sun$ we either reproduce or underpredic the far-UV  data while, adopting  $M_{up}~=~40~M\sun$, we  either reproduce or overpredict them.
At the same time, the average SFR in the models with $M_{up}~=~40~M\sun$ is about two to three times larger than that obtained with the
models that adopt $M_{up}~=~120~M\sun$.
This is in evident contrast with what we have found in the fits of normal galaxies where
we see that the SFR, obtained with different values of $M_{up}$, differ by 20~per~cent at maximum and it is a consequence of the young age of such star bursting regions, where the bolometric luminosity is dominated by the most massive stars.
Finally, the attenuation in the models with $M_{up}~=~40~M\sun$ is always lower than that obtained with the
models adopting $M_{up}~=~120~M\sun$.
As the non-thermal radio emission starts about 7~Myr after the beginning of the burst of star formation (see Figure~\ref{fig:ntherm_14_33}), we have the opportunity to perform
an accurate decomposition of the radio flux into thermal and non thermal components.
We present in Table~\ref{tab:dale_gra_derived_2_unsub} the thermal fraction resulting from this decomposition and we show  the two radio components in Figure~\ref{fig:ngc6946_sed_pg2_olga}.
To show the quality of  our decomposition,  we also show  in Figure~\ref{fig:ngc6946_sed_pg2_olga} and only for the case with $M_{up}~=~120~M\sun$, the background subtracted radio data fluxes (cyan diamonds) by \citet{Murphy2011}.
In the next section, we will discuss the  SFR calibrations  derived  using  our best fit  model  of the galaxies studied in this work. 
\begin{table}
\caption {
 \textit{\small{GRASIL}} best fit parameters (for upper mass limits of 40 and 120$~\rm{M}\sun$) for the
modelled SEDs of M100 and NGC~6956 SF regions.
\label{tab:bestfit_params}
}
\small{ 
\centering
\begin{tabular}{lccccc}
\hline
ID   &    $ t _{gal}/t _{sb}$     &   $t_{esc}$    &   $\beta$  &    $\tau_{1}$   &   $f_{mol}$\\
      &            (Gyr/Myr)         &    (Myr)          &                 &                        &                  \\
 (1)  &            (2)         &    (3)          &     (4)              &     (5)           &     (6)         \\
\hline
\hline
     &   &  &     $M_{up}$   &  = 40~\rm{M}\sun  &  \\
 \hline
M100 &12.0 &2.5 &2.0 &12.0 &0.10 \\
 \hline
  NGC~6946\_1 & 8.0 & 1.0 & 2.1 & 9.07 & 0.30\\
  NGC~6946\_2 & 9.0 & 0.5 & 1.7 & 12.00 & 0.05\\
  NGC~6946\_3 & 8.0 & 1.0 & 2.0 & 24.48 & 0.20\\
  NGC~6946\_5 & 8.0 & 0.3 & 2.0 & 4.15 & 0.25\\
  NGC~6946\_6 & 12.0 & 1.2 & 2.2 & 5.33 & 0.40\\
  NGC~6946\_7 & 7.0 & 1.0 & 2.1 & 14.81 & 0.30\\
  NGC~6946\_8 & 7.0 & 0.6 & 2.1 & 6.12 & 0.22\\
  NGC~6946\_9 & 8.0 & 1.0 & 2.1 & 5.33 & 0.30\\
 \hline
 \hline
     &   &  &     $M_{up}$   &  = 120~\rm{M}\sun  &  \\
  \hline
M100 &12.0 &0.9 &2.0 &14.81 & 0.10 \\
   \hline
  NGC~6946\_1 & 11.0 & 0.6 & 2.2 & 5.33 & 0.12\\
  NGC~6946\_2 & 11.0 & 1.0 & 1.7 & 16.60 & 0.15\\
  NGC~6946\_3 & 8.0 & 0.9 & 2.2 & 12.00 & 0.14\\
  NGC~6946\_5 & 18.0 & 1.0 & 1.8 & 18.75 & 0.50\\
  NGC~6946\_6 & 11.0 & 0.7 & 2.2 & 5.33 & 0.12\\
  NGC~6946\_7 & 8.0 & 1.0 & 2.2 & 5.33 & 0.16\\
  NGC~6946\_8 & 8.0 & 0.9 & 2.2 & 5.33 & 0.12\\
  NGC~6946\_9 & 9.0 & 0.6 & 2.2 & 5.33 & 0.12\\
\hline
\end{tabular}
}
Column (2) gives  the age of the galaxy ($t _{gal}$) in Gyr for the normal star-forming galaxy M100  and the age of the burst ($t _{sb}$) in Myr for the NGC~6946 star-bursting regions.
Parameters in other columns are as described in text.
\end{table}

\section{Discussion}
\label{sec:disc}
In this section we discuss the  results obtained from the  best fits of the SEDs of
M100 and the extra-nuclear regions of NGC~6946.
We begin with the SFR calibrations obtained with the new SSPs and then discuss
the impact of the newly added emission line prediction on the resulting attenuation.
\subsection{Star formation rate calibrations from  \textit{\small{GRASIL}} best fits}
\label{sec:sfr_cals}
We first show in Table~\ref{tab:dale_gra_derived_1_unsub} the luminosities of the best fit models in the selected photometric bands, from the UV to radio wavelengths. The upper panel refers to the results obtained adopting
$M_{up}~=~40~M_{\sun}$ while, the lower panel refers to those obtained
with $M_{up}~=~120~M_{\sun}$.
The FIR luminosity in column~11 was obtained by integrating the IR specific luminosity from 3 to 1000~$\micron$.
We also show the predicted intrinsic (suffix $^{int}$) and transmitted (suffix  $^{tra}$) intensities of the H$\beta$ and H$\alpha$ emission lines,
the value of $q_{1.4}$ (Equation~\ref{eq_q_ratio}), the relative contribution to the
3 - 1000~$\micron$ FIR luminosity by the molecular clouds component and by the cirrus component, and the ratio between the  3 - 1000~$\micron$ FIR  and the bolometric luminosity.
In the columns of radio luminosities, we  enclose in parenthesis the fractional contribution 
of the thermal radio component to the total radio emission, derived from the models.

\begin{figure}
\includegraphics[width= 1\linewidth]{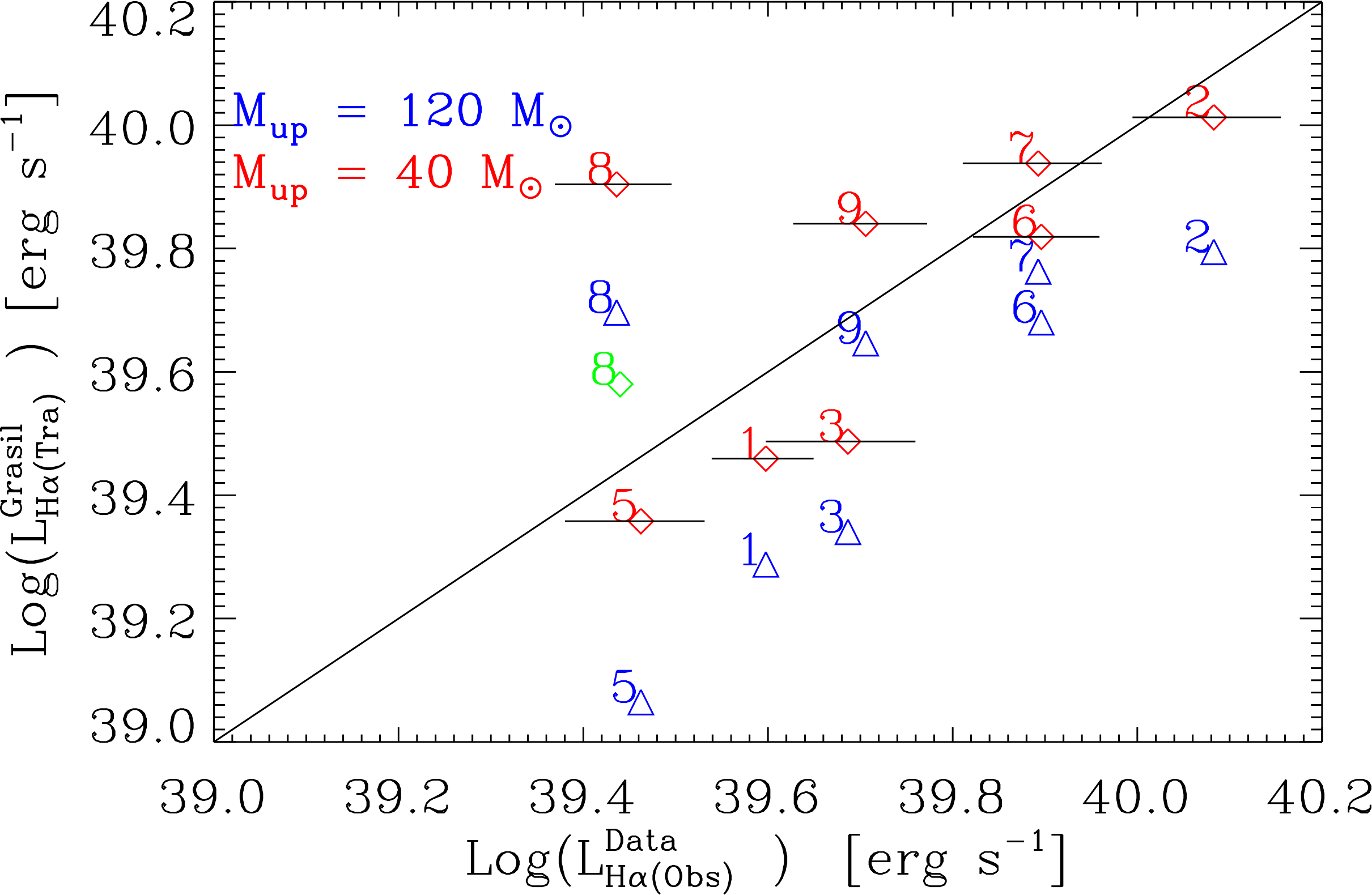}
 \caption{Comparisons between our model's transmitted intensities of the H$\alpha$ emission lines  with the observed ones, for $M_{up}~=~40 M_{\sun}$ (red daimonds)  and $M_{up}~=~120$$M_{\sun}$  (blue triangles) cases.
The solid line indicates the one-to-one correlation whereas the horizontal lines indicate the errors in the observed H$\alpha$ line.
The green daimond indicates the case with $M_{up}~=~350~\rm{M}_{\sun}$ for the extra-nuclear region 8.
\label{fig:ha_rat}
 }
\end{figure}

\smallskip
\noindent
{\sl Effects of Mup\\}
By construction, i.e. since we are discussing best fits SEDs, the predicted luminosities obtained with the two different
$M_{up}$ are very similar, but for those in the emission lines and in the UV bands. These are the most difficult to
model because they are the most sensitive to the attenuation. This is not the only reason however, because even the intrinsic
H$\alpha$ and H$\beta$ luminosity differs by a significant factor in the two $M_{up}$ cases. Indeed, in the two cases the  ionizing photon flux (and the far and near-UV) differ much more than the corresponding bolometric luminosities.
Said in another way, one may be able to obtain the same bolometric flux with two different IMF upper limits but the amount of
ionizing photons may be significantly different, and this mainly affect the intensity of the recombination lines  and  the free-free radio emission.  Note however that the 33~GHz luminosity does not show the same strong dependence  on the adopted Mup as that shown by the recombination lines, and this is likely due to the fact that a non negligible contribution of the non-thermal luminosity is present even at this high radio frequency, which should be larger in the case of lower $M_{up}$.
The best fit transmitted intensities of the H$\alpha$ emission lines are compared with the observed ones in
Figure~\ref{fig:ha_rat}, for both cases of $M_{up}$.
We do not include M100  in this plot to better show the case of the star-forming regions.
The solid line indicates the one-to-one correlation.
We see that, in the case of extra-nuclear regions,  the predicted values of the case with $M_{up}~=~40~\rm{M}_{\sun}$ are  larger than those of the case with $M_{up}~=~120~\rm{M}_{\sun}$. 
In general our best fit models are able to reproduce the observed values with an accuracy of about 50~per~cent.
Looking to individual regions, we see that regions   $\sharp$1 ,$\sharp$2, $\sharp$3, $\sharp$5, $\sharp$6  are compatible with $M_{up}=40~\rm{M}_{\sun}$ while, for regions $\sharp$9 and $\sharp$7,
a larger value of $M_{up}$ seems suggested, up to $M_{up}~=~120~\rm{M}_{\sun}$. From the figure it also appears that, in order to fit region $\sharp$8, an even  higher IMF upper mass limit
(see the green daimond in Figure~\ref{fig:ha_rat} for the case of $M_{up}~=~350~\rm{M}_{\sun}$) should be used.
Furhermore, we note that the in all cases where we are able to reproduce the observed H$\alpha$ emission  we are also able to reproduce fairly well the observed FUV luminosity, which is also strictly related to the intrinsic ionizing photon flux. On the contrary the NUV flux is not well reproduced by our models, likely indicating that its origin is not strictly related to the star formation process.
Since all model reproduce the observed bolometric luminosity and, in particular, the 24~$\micron$  and the 33~GHz data, we conclude that, by
combining in a consistent way this relevant information, it is possible to obtain a fairly robust estimate of the upper mass limit of the IMF . 
Indeed we stress that the FIR luminosity and those at H$\alpha$, 33~GHz and 24~$\micron$  bracket, on one  side, the intrinsic ionizing photon flux and, on the other, the attenuation and re-emission by molecular clouds where the stars reside during their initial phase, which are the most sensitive to the fractional contribution of the most massive stars.

\begin{figure*}
\begin{center}
\includegraphics[width= 0.4\linewidth]{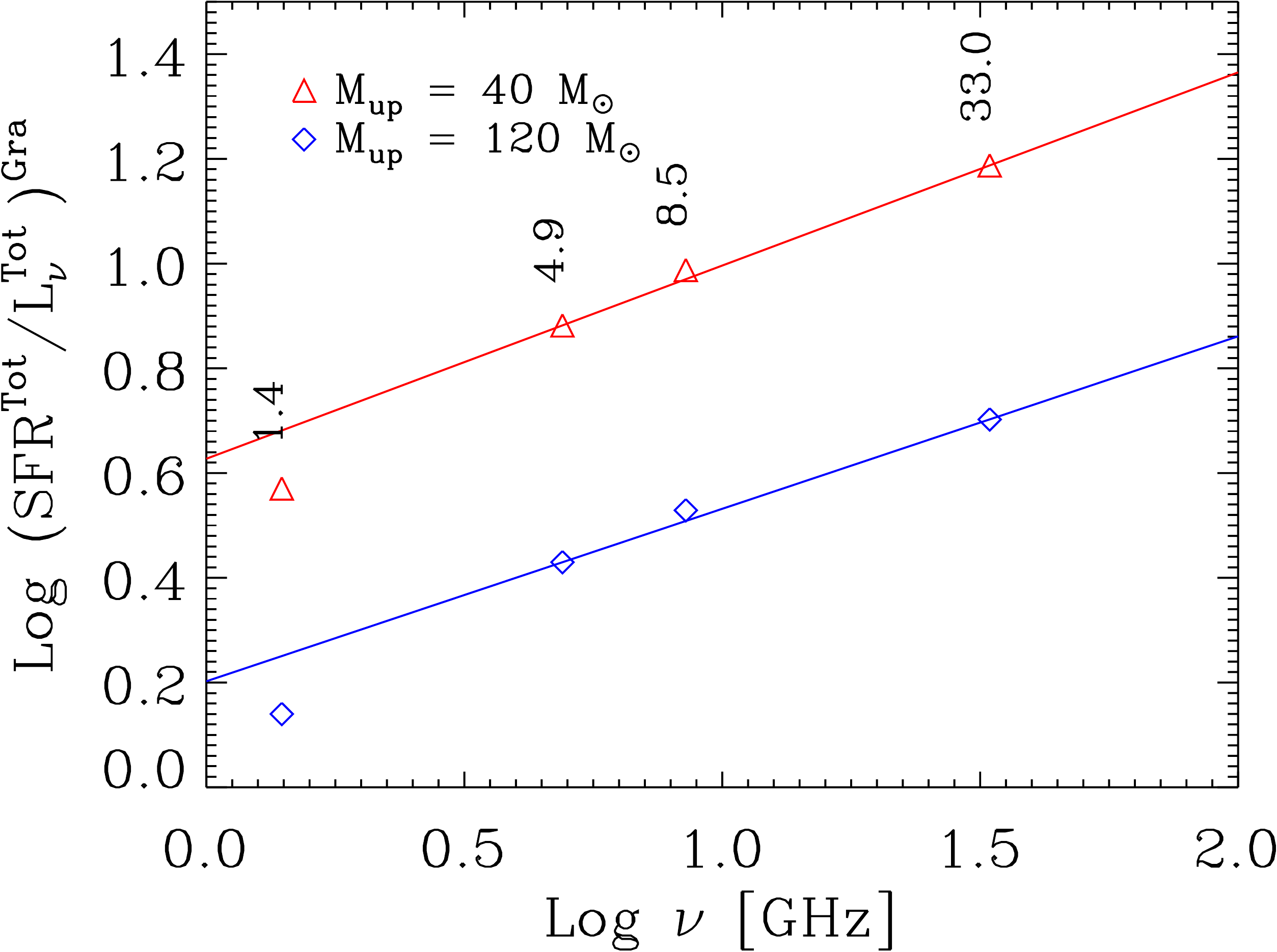}
\includegraphics[width= 0.4\linewidth]{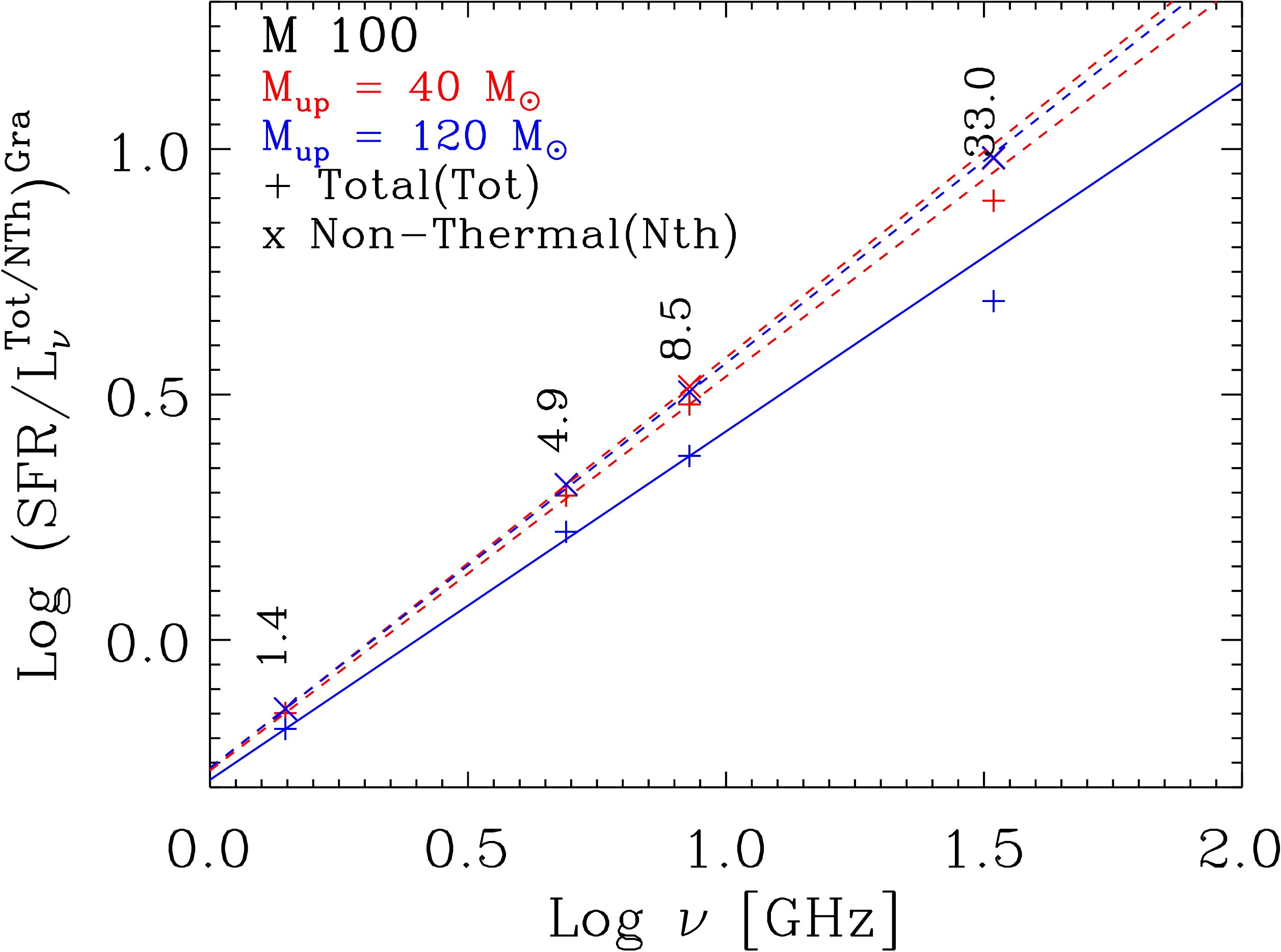}
\end{center}
\caption{
\textit{Left panel}: SFR calibration constants C($\nu$), in units of 
$\rm{M}_{\sun} \rm{yr}^{-1} 10^{28}\rm{erg}^{-1} \; \rm{s} \;  \rm{Hz}$,
 at $\nu$~=~1.4, 4.9, 8.5 and 33.0$\rm{GHz}$ (Table~\ref{tab:dale_gra_derived_2_unsub}), 
for the star-bursting regions and using  the total radio emission. Triangles  and  daimonds indicate  cases of $M_{up}~=~40~\rm{M}_{\sun}$  and $M_{up}~=~120~\rm{M}_{\sun}$ respectively.
\textit{Right panel}: The same as in the left panel but for M 100 and using  both the total radio (crosses)  and 
non-thermal radio emission (xs). 
In both panels, all red symbols and lines indicate  cases of $M_{up}~=~40~\rm{M}_{\sun}$  
and  blue symbols and lines cases of  $M_{up}~=~120~\rm{M}_{\sun}$.
Also all lines represent least-square fits.
\label{fig:ngc6946_sfr_cal_olga_pg1}
 }
\end{figure*}

As far as the prototype  normal galaxy M100  is concerned, we remind that the  observed H$\alpha$ luminosity
was matched by the model  with $M_{up}~=~40~M_{\sun}$.
It is important, at this stage, to remind that these conclusions should be taken with care because the current SSP adopted do not account for binary evolution. Indeed it is known that binary evolution can produce more ionizing photons at later stages than single star evolution,
because of the loss of the envelope by binary interaction \citep{Wofford2016}.
In particular we may say that regions $\sharp$1 ,$\sharp$2, $\sharp$3, $\sharp$5, $\sharp$6  may be reproduced fairly well without invoking a high $M_{up}$ and/or significant effects by binary evolution. The latter effect could instead explain the location of region number $\sharp$8 that cannot be reproduced even with $M_{up}~=~350~M_{\sun}$, as indicated by the green diamond in Figure~\ref{fig:ngc6946_sfr_cal_olga_pg1}).
This effect, which add a new dimension to the problem, is under study in  \textit{\small{PARSEC}} and it will be included in the nearest  future.

\smallskip
\noindent
{\sl The q ratio.\\}
The second thing that we note by inspecting Table~\ref{tab:dale_gra_derived_1_unsub} is that the value of the
q$_{1.4}$ does not significantly depend on the upper limit of the IMF. This is because a significant fraction of both the bolometric luminosity and the radio luminosity are contributed by stars less massive than M~=~$40~M_{\sun}$. Indeed after a few Myr,  the upper mass of the SSP is already significantly lower than M~=~$120~M_{\sun}$.
Furthermore, the q$_{1.4}$ ratio is generally higher in the extranuclear regions of NGC~6946 than in the normal SF galaxy M100, by  a factor of $\simeq$1.4.
We remind that in our model, CCSNs, responsible for the non thermal emission, are produced only by the mass range $10~M_{\sun}$  $\leq$  M$\leq$ ~$30~M_{\sun}$ and  the individual star-bursting regions have an age which is younger than the threshold for  CCSN production.
Thus they did not yet reach a stationary equilibrium for non-thermal radio emission and not only their radio slopes appear flatter but also their
q$_{1.4}$ are higher
than that of normal star-forming galaxies B02.

\smallskip
{\noindent \sl The SFR calibration. \\}
Using the luminosity in the different bands, as presented in
Table~\ref{tab:dale_gra_derived_1_unsub}, and the average SFR obtained from the  \textit{\small{GRASIL}} fits, we show in
Table~\ref{tab:dale_gra_derived_2_unsub}, the corresponding SFR calibrations C(b)~=~SFR/L(b), where {\sl{b}} indicates the band.
The average SFR is derived  by considering the last 100 Myr time interval for the normal star-forming galaxy  M100,
and the age of the burst  in the star-forming regions of NGC~6946, which is shown
in column 16 of Table~\ref{tab:dale_gra_derived_2_unsub} for the NGC~6946 extranuclear regions.
This choice is suggested by the fact that within an entire galaxy, the star-forming regions distribute continuously with time while, this is obviously not the case for individual
regions.
The average SFR ($\rm{M}_{\sun} \; \rm{yr}^{-1}$) so computed is shown in the second column of
Table~\ref{tab:dale_gra_derived_2_unsub}. These calibrations are directly obtained by  \textit{\small{GRASIL}}.
For sake of comparison, we also add, in Table~\ref{tab:dale_gra_derived_2_unsub}, some calibrations that
can be obtained by using the un-attenuated fluxes derived from  \textit{\small{GRASIL}} best fit models, either directly or by using the analytical relations obtained by means of the SSPs. These are useful to compare the differences between the SFR derived from a panchromatic fit that realistically account for ISM processes to analytical fits made with SSP fluxes.
For example, the FUV, H$\beta $ and  H$\alpha$ values shown in column 3, 4 and 6,  refer to the calibrations obtained by dividing the average SFR by the corresponding un-attenuated  intrinsic fluxes of the models, listed in Table~\ref{tab:dale_gra_derived_1_unsub}.
For H$\beta $ and  H$\alpha$  we also show, in columns 5 and 7, the values obtained by inserting their corresponding intrinsic fluxes
in the analytical calibrations obtained from simple SSPs models, Equations~\ref{eq_sfr_hb_obi_fit} and  \ref{eq_sfr_ha_obi_fit} respectively.
For the 33~GHz radio emission,  we  add in column~15, the value of the calibration C(33)$^{SSP}$~=~SFR/L(33~GHz), directly derived from the SSP models (Equation~\ref{eq_sfr_lff_const_fit_b}).
The last rows in the first and second sections, indicated by $< NGC~6946 >$, show the median values of all SFR calibrations derived from the star-bursting regions of NGC~6946.
In the third section we collect some common  SFR calibrations from the literature, with the corresponding bibliographic sources given in the caption.
In Figure~\ref{fig:ngc6946_sfr_cal_olga_pg1} we plot
the calibration constants C($\nu$) at $\nu$~=~1.4, 4.9, 8.5 and 33.0~$\rm{GHz}$, in units of 
$\rm{M}_{\sun} \rm{Yr}^{-1} 10^{28}\rm{erg}^{-1} \; \rm{s} \rm{Hz}$,
against the corresponding radio frequencies, for both the star-bursting regions (left panel) and M100 (right panel) and for both cases of  $M_{up}$.
The relations (in logarithmic units) are almost linear above 1.4~GHz, and we show the best fit regressions as solid lines.
Concerning the dependence on M$_{up}$, there is a clear difference between the starburst and the normal star formation regime in the sense that, in the former case
the constant is about a factor of three larger independently from the frequency.
Again this effect is due to the young age of the starburst regions coupled with the
dependence of the thermal radio emission on M$_{up}$. Indeed for the normal SFR regime,
the dependence on M$_{up}$ becomes significant only at high frequencies, where the thermal emission start dominating.
From the fits we derive the following general relations between SFR (in $\rm{M}_{\sun} \rm{yr}^{-1}$) and radio luminosity (in  $\rm{erg} \; \rm{s}^{-1} \rm{Hz}^{-1}$).
For Star-bursting regions, with average metallicity $\sim$ 0.006, we have at any frequency between 1.4~$\rm{GHz}$ and 33~$\rm{GHz}$, to better than 10~per~cent,
\begin{eqnarray}
\rm{log}(\frac{SFR}{L\nu})^{tot} ~=~ 0.43 \ \rm{log}(\nu)- 0.96 \ \rm{log}   
(\frac{  M_{up} }  {120}   )        - 27.91
\label{eq_sfr_tot_rad_fit}
\end{eqnarray}
while, for  the normal star-forming galaxy, M100, with slightly more than solar metallicity, we have for the total radio emission
\begin{eqnarray}
\rm{log}(\frac{SFR}{L\nu})^{tot}~=~0.73 \ \rm{log}(\nu)-0.16 \ \rm{log}
(\frac{  M_{up} }  {120}   )  	- 28.31
\label{eq_sfr_tot_rad_fit_m100}
\end{eqnarray}
and for  the non-thermal radio emission, we have 
\begin{eqnarray}
\rm{log}(\frac{SFR}{L\nu})^{nth} ~=~ 0.58 \ \rm{log}(\nu)-0.43 \ \rm{log}
(\frac{  M_{up} }  {120}   ) 	- 28.33
\label{eq_sfr_nth_rad_fit_m100}
\end{eqnarray}
Relations  similar to Equations~\ref{eq_sfr_tot_rad_fit}, ~\ref{eq_sfr_tot_rad_fit_m100}
and  ~\ref{eq_sfr_nth_rad_fit_m100} above were provided by \citet{Murphy2011}, assuming Kroupa IMF (with $M_{up}=100~ \rm{M}_{\sun}$ ), solar metallicity and a constant SFR over a timescale of 100~Myr. For total radio emission
\begin{equation}
\left( \frac{ SFR} {  L\nu } \right)^{tot}_{Mur}  =   10^{-27}
\left(  2.18\nu^{-0.1} 
(\frac{T_e}{10^4})^{0.45} + 15.1\nu^{-0.85}   \right)^{-1}
\label{eq_sfr_tot_rad_murphy}
\end{equation}
and for non-thermal  radio emission
\begin{equation}
\left( \frac{ SFR } {  L\nu } \right)^{nth}_{Mur} =    10^{-27}( 6.64 \times 10^{-29} \nu^{0.85}  )
\label{eq_sfr_nth_rad_murphy}
\end{equation}
These relations are in between our relations in  \ref{eq_sfr_tot_rad_fit}  and
\ref{eq_sfr_tot_rad_fit_m100}, running only $\lesssim$~20~per~cent lower than our one for normal star-forming galaxies, below 33~GHz.
However at frequencies higher than 33~GHz the difference grows
rapidly and at 100~GHz our values are $\sim$40 per~cent and $\sim$200 per~cent higher than that of \citet{Murphy2011} for starburst and normal galaxies, respectively.
Interestingly, the relation derived by  \cite{Schmitt2006} with Starburst99,
 assuming  a Salpeter IMF (with $M_{up}~=~100~\rm{M}_{\sun}$), solar metallicity and a continuous SFR of 1 $\rm{M}_{\sun} \;  yr^{-1}$ 
 \begin{equation}
\left( \frac{ SFR} {  L\nu }  \right)^{tot}_{Sch}~=~ 10^{-27}( 8.55\nu^{-0.8}  + 1.6\nu^{-0.1} )
\label{eq_sfr_tot_rad_schmitt}
\end{equation}
agrees  well with our relations in \ref{eq_sfr_tot_rad_fit}  and
\ref{eq_sfr_tot_rad_fit_m100}  at frequencies from about  33~GHz to 100~GHz.
Above 100~GHz the dust emission contribution becomes dominant.

It is interesting to compare the  \textit{\small{GRASIL}} calibration at  33~GHz  with that derived from the SSP models (Equation~\ref{eq_sfr_lff_const_fit_b}), i.e.  columns~14 and column~15 
in Table~\ref{tab:dale_gra_derived_2_unsub}.
The values show some discrepancies, especially for normal galaxies.
However we remind that the values in column~15 refer to free-free emission alone while the 33~GHz flux in column~14 may include a non negligible contribution by synchrotron emission.
If we take this contamination into account, using the percentage contribution of the thermal radio component enclosed in parenthesis,
the corrected values  of column 14 are in fair agreement with those predicted by
Equation~\ref{eq_sfr_lff_const_fit_b}. This is expected because, at these wavelengths, attenuation, including also the free-free one \citep{Vega2008}, should not be important.
The above overall agreement indicate a surprising robustness of the
radio calibration, irrespective of the differences in the underlying  models.
This will be particularly important for the high redshift galaxies especially in their early
evolutionary phases.

As a further comparison, we provide  at  the bottom of Table~\ref{tab:dale_gra_derived_2_unsub} other SFR calibrations taken from the literature.
We remind the reader that, in this work, we adopt a \citet{Kennicutt1983}  IMF  
and a typical time-scale for averaging the SFR rate   
of 100~Myr for normal galaxies and equal to the age of the starburst in the extra-nuclear star bursting regions. The details on the parameters adopted by different authors
may be found in the quoted papers.
\subsection{Dust attenuation properties}
\label{sec:dust_attens}
It has been already  shown in \citep{Silva1998, Granato2000} that the attenuation
is the result of the interplay between the extinction properties of dust grains and the
distribution  between dust and stars in space and in time.
Further  discussion can also be found in
\citep{Charlot_Fall2000,Panuzzo2003}.
We now investigate the properties of the attenuation curves
of the galaxies studied here, with particular interest to understand
its dependence on the galaxy type, i.e. starburst vs. normal regime, and also
the effects of using different parameters in the analysis, such as
M$_{up}$ in the IMF.

\begin{figure}
\includegraphics[width= 1\linewidth]{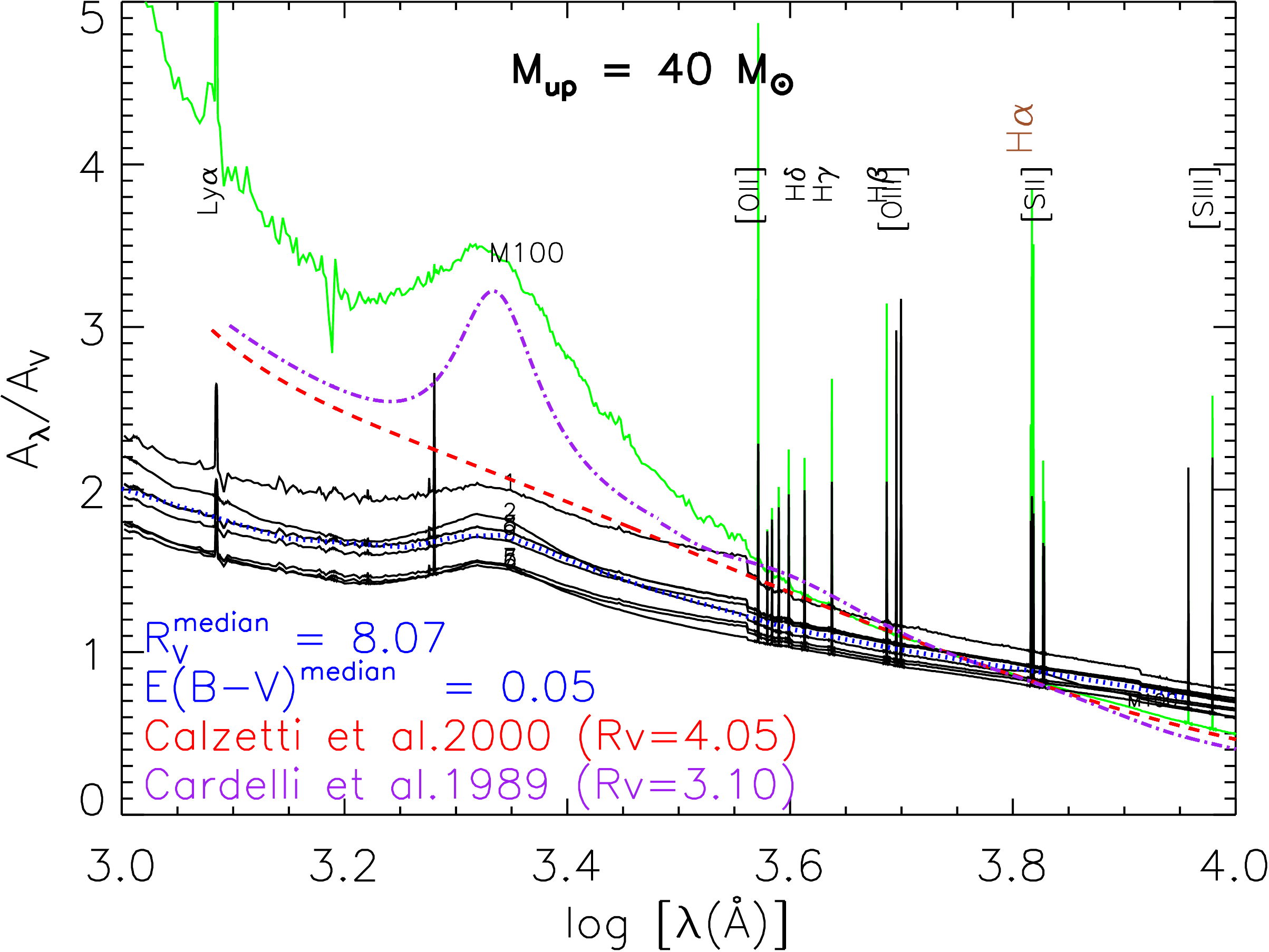}\\
\includegraphics[width= 1\linewidth]{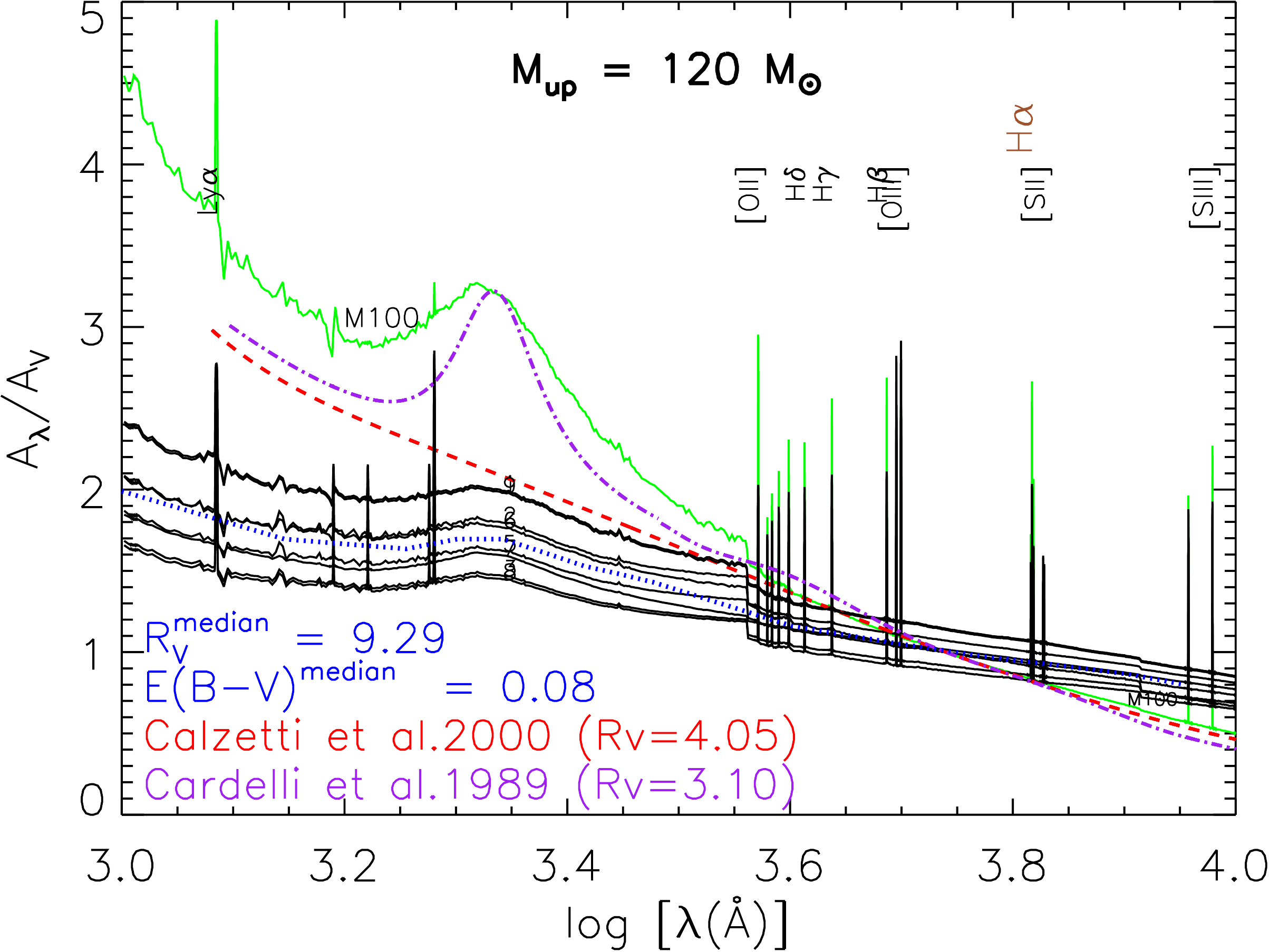}
\caption{Attenuation curves,  A$_\lambda$/A$_V$, for M100  and NGC~6946 extra-nuclear  regions, for the cases of $M_{up}~=~40$ (top panel) and $M_{up}~=~120~\rm{M}_{\sun}$ (bottom panel). 
The black solid lines are the attenuations curve of the eight regions of NGC~6946 while the
dotted blue line represents the median of these curves.
The median values of  $R_{V}$ and ${E(B-V)}$ for the star-forming regions are also given in the labels. The green line is the attenuations curve  of M100.
For comparison purposes, we added the attenuation curves of  \citet{Calzetti2000} (red dashed line)
and \citet{Cardelli1989} (purple dotted-dashed line)
The later attenuation seems to agree very  well  with that of M100 at wavelengths above  $\sim$ 4000~\AA.
\label{fig:attens_1}
}
\end{figure}

We first show in Figure~\ref{fig:attens_1}, the attenuation curves, A$_\lambda$/A$_V$, of M100 (green solid line) and of the NGC~6946 extra-nuclear  regions studied in this work (black solid lines).
They are simply derived by comparing the transmitted galaxy flux to the intrinsic one, as a function of the wavelength.
The top panel refers to the case of $M_{up}~=~40~\rm{M}_{\sun}$ while the case with $M_{up}~=~120~\rm{M}_{\sun}$ is shown in the bottom panel.
The dotted blue lines represent the median curve of the NGC~6946 extra-nuclear regions  obtained by sampling the curves of individual regions at selected wavelengths.
The median values of the quantities $E(B-V)$ and R$_{V}$~=~A$_V$/E(B-V) for the star-bursting regions are  also given in the
labels. We have also  labelled  some prominent emission lines ($H\alpha$ line  is in brown).
For comparison purposes, we added the attenuation curves of  \citet{Calzetti2000} (red dashed lines)
and \citet{Cardelli1989} (purple dotted-dashed lines).
We also collect in Columns (2 - 9)  of  Table~\ref{tab:dale_olga_attens}
the attenuation in H$\alpha$,  H$\beta$, in the FUV (0.16~$\micron$) and NUV (0.20~$\micron$), in the \textit{B}-band(0.45~$\micron$),
and in the interpolated continuum at H$\beta$ (named  A48), in the 
\textit{V}-band(0.55~$\micron$) and in the interpolated continuum at H$\alpha$ (named A65).
Columns 10 and 11 give the optical depth of the molecular cloud at 1~$\micron$ derived from  \textit{\small{GRASIL}} and the $R_{V}$ ratio, respectively. In
Column 12 we show the difference A$_{48}$-A$_{65}$ which corresponds to $E(48-65)$.
Using the Balmer decrement method for the  $H\beta$  and  $H\alpha$ lines, we also obtained the same quantity, $E(H\beta - H\alpha)$:
\begin{equation}
E(H\beta - H\alpha) = 1.086 \left( {\rm ln}   (\frac{  H\beta^{int} }  {  H\alpha^{int}  } )    
  -  {\rm ln}  (\frac{  H\beta^{tra}  }   {  H\alpha^{tra}  }) \right)
\label{eq:ebv_lin}
\end{equation}
and show its value in Column 13.
We assume an intrinsic line intensity ratio, $(H\alpha^{int}/H\beta^{int})$~=~2.86, which is valid for  $T_e$ =10,000~K \citep{Hummer1987}.
Column 14 gives the escape time of young stars from their birth cloud in Myr.
Column  15  gives the FIR flux in $\rm{ergs}  \, \rm{s}^{-1} \, \rm{cm}^{-2}$ in the 40 - 120~$\micron$ interval derived using the 60 and 120~$\micron$ fluxes \citep{Helou1988}.

\begin{figure}
\includegraphics[width= 0.95\linewidth]{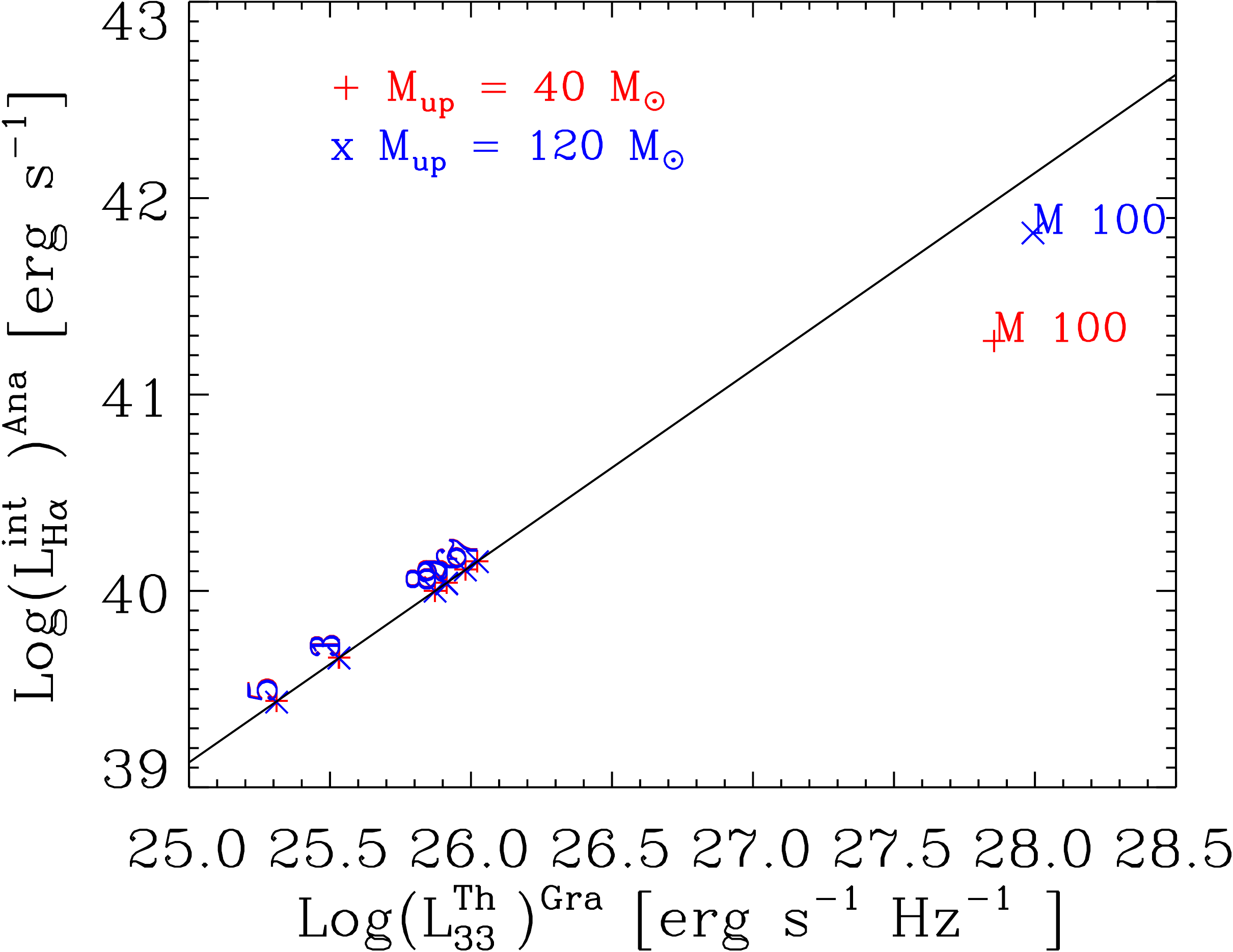}\\
\caption{Plot of our model's 33~GHz thermal radio luminosity  against the  intrinsic H$\alpha$ luminosity.
Solid line represents linear fit for the star-bursting regions.
The H$\alpha$ luminosity here is derived from the thermal  radio luminosity using
Equations~\ref{eq_Lff_new_1} and \ref{eq_c2_rel}.
We adopted a median  thermal fraction of 78.5~per~cent  in estimating the thermal radio component  of  the 33~GHz total radio luminosity.
We also added in this plot the M100 which has  an average  thermal fraction of 16.0~per~cent(42.0~per~cent)
for   M$_{up}$ ~=~ 40~$\rm{M}_{\sun}$(M$_{up}$ ~=~ 120~$\rm{M}_{\sun}$).
 \label{fig:Ha_int_L33}
} luminosity (L33 in erg s?1Hz?1) is presented in Figure 12. 
\end{figure}

\begin{figure}
\includegraphics[width= 0.95\linewidth]{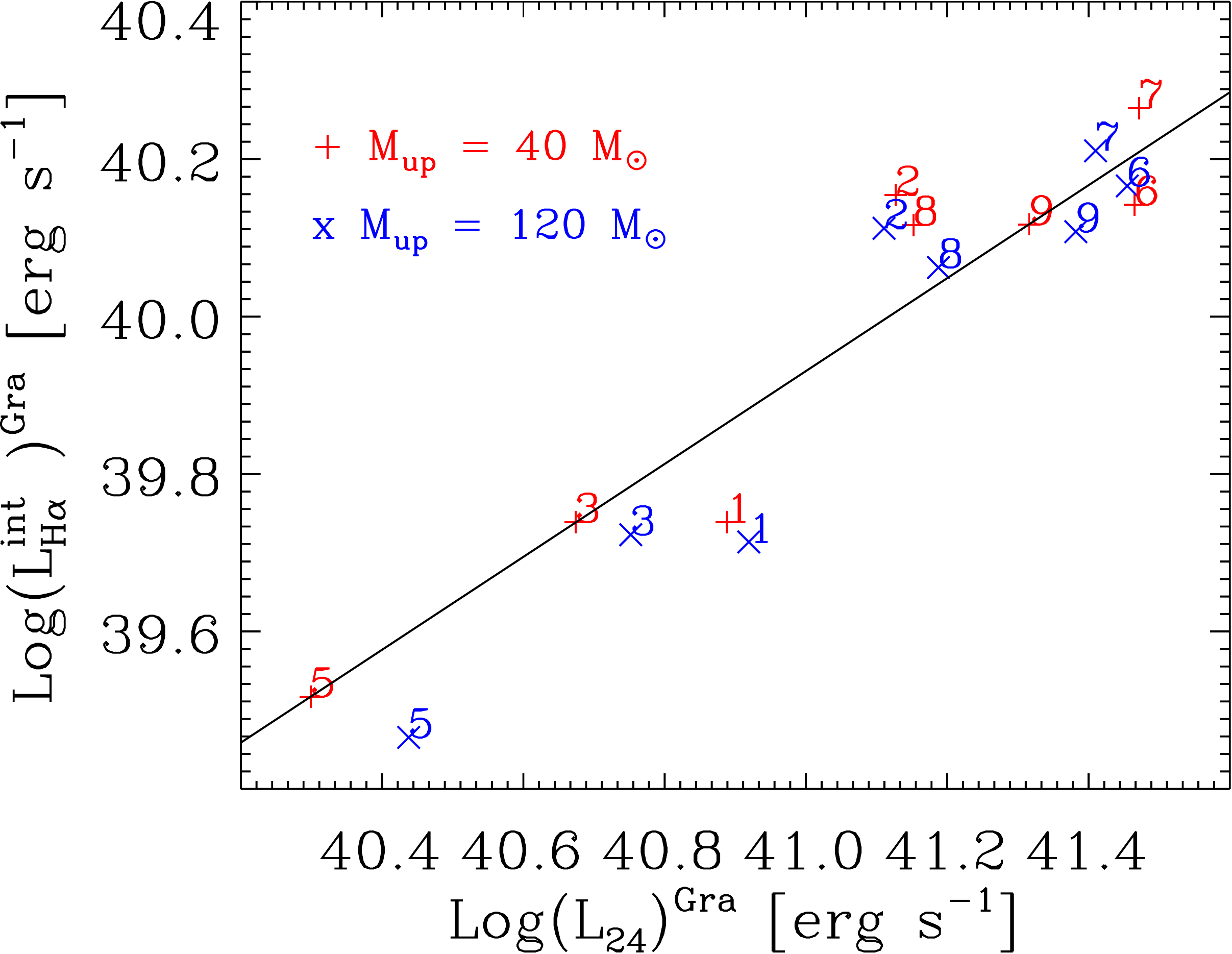}
\caption{Plot of the 24~$\micron$ luminosity against the  intrinsic H$\alpha$ luminosity, both
 extracted directly  from our best-fit models.
 Solid line represents linear fit. We excluded M100 in this plot so as to better represent the star-bursting regions.
 \label{fig:Ha_int_L24}
}
\end{figure}

\begin{figure}
\includegraphics[width= 1\linewidth]{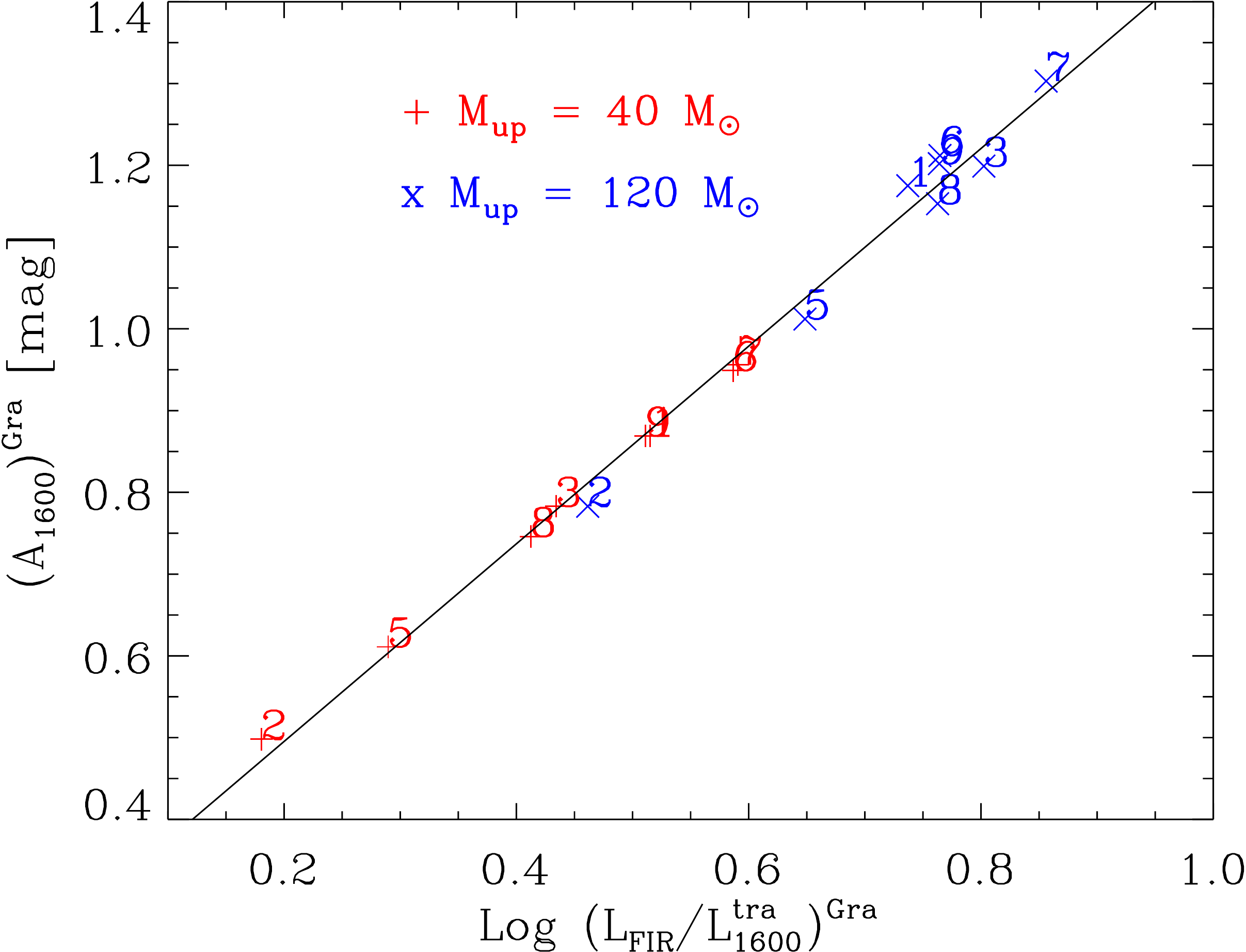}
\caption{
Plot of attenuation at 1600~\AA ~against the ratio,  $FIR/F_{1600}$, both  extracted from  our best-fit models.
The solid black line represents  our linear fit as given in Equation~\ref{eqn:A1600_Fir}. The red crosses  and blue  'X's  indicate M$_{up} ~=~ 40$ and 120~$\rm{M}_{\sun}$ cases respectively.
\label{fig:A1600_IRX}
}
\end{figure}

Several points can be noted by inspection of  Figure~\ref{fig:attens_1} and Table~\ref{tab:dale_olga_attens}.
First of all, we see that, while in normal galaxies the attenuation in the lines is significantly higher than that in the surrounding continuum,
(a factor of 4 on average  for M100), this is not true for the star bursting regions where this factor is not more than 1.3.
The fact that the  attenuation in the lines is significantly higher than that in the surrounding continuum is well known and it is a manifestation that
young stars are more dust enshrouded than older stars. However, in the case of the star-bursting regions the underlying continuum is essentially produced
by the same stellar populations, with a small contribution from the older populations, so that the attenuation in the lines is not much different from that in the continuum.
To get more insights from the models we compare the characteristic R values in the lines and in the nearby continuum.
We see that for star-forming galaxies,  R$_{H\alpha}$~=~A$_{H\alpha}/($A$_{H\beta}$-A$_{H\alpha}$) is much larger than
the value obtained in the corresponding interpolated underlying continuum regions (R$_{65}=A_{65}/(A_{48}-A_{65}$).
For example, for M100,  R$_{H\alpha}$~=~16(10) while R$_{65}$~=~4(4), for M$_{up}$~=~40(120)~M$_{\sun}$.
In the starburst regions,  R$_{H\alpha}$ and  R$_{65}$ have similar values between  6 and 10.
The high values so determined for R$_{H\alpha}$ would suggest a high neutral absorption in these objects but this is not the case because we have not modified the grain size distribution which, by default in  \textit{\small{GRASIL}} is the one of the Galactic ISM.
We also see that in M100, R$_{65}$, in the continuum, has a value  of $\sim$3, compatible with the common extinction laws provided in literature \citep{Calzetti2000,Cardelli1989}.
Instead this effect is due to the fact that the attenuation within molecular clouds is so high (see the values of $\tau_{1\micron}$ in  Table~\ref{tab:dale_olga_attens}) that the
emission of young stars is almost completely absorbed, until they escape from the clouds.
This effect mimics the neutral absorption.
This effect is enhanced for the emission lines because they are generated only by stars younger than about 6~Myr, so that the fraction of line flux absorbed
by molecular clouds with respect to the total one, is even higher. Adopting a $t_{esc}$ larger than 7~Myr will erase line emission in our model, irrespective of the galaxy types considered here.
A major consequence of the age selective attenuation  described above is that it leads to wrong estimates of the attenuation in the line emission, if one uses methods involving
attenuation-affected observables, like the Balmer decrement method.
A more accurate method to estimate the  intrinsic H$\alpha$ luminosity (hence the
H$\alpha$ attenuation) is by using the
33~GHz and  24~$\micron$ luminosities, since they are optimal tracers of the radiation emitted by the most massive.
Adopting a median value of 78.5~per~cent as the percentage contribution of the thermal radio component in young star-bursts, we first obtain the thermal radio emission component  of our model's 33~GHz total radio emission luminosity. Using the analytical relation given in Equation~\ref{eq_Lff_new_1}
(T$_{e}$ and gaunt factors used here are  as  given in Equations~\ref{eq_te_z_mup}
and ~\ref{eq_gaunt_claudia} respectively),
we derive the number of ionizing photons corresponding to this luminosity.
With  Equation~\ref{eq_c2_rel}, we then estimate  the corresponding intrinsic H$\alpha$ luminosity.
A plot of the  estimated  intrinsic H$\alpha$ luminosity (in ${\rm erg} \; {\rm s}^{-1}$) against our model's
33~GHz thermal  radio luminosity (L$_{33}$ in ${\rm erg} \; {\rm s}^{-1} {\rm Hz}^{-1}$)  is presented in Figure~\ref{fig:Ha_int_L33}.
We added in this figure also the normal SF galaxy M100, with  an average  thermal fraction of 16.0~per~cent(42.0~per~cent) for   M$_{up}$~=~40~$\rm{M}_{\sun}$(M$_{up}$~=~120~$\rm{M}_{\sun}$), respectively.
The fitting relation (solid line)  considering only the star bursting regions for both cases of M$_{up}$, is given by
\begin{equation}
{\rm log}( {\rm L_{H\alpha}^{int} } ) =   {\rm log}( 0.785{\rm L(33) } )  + 14.10
\label{eqn:ha_33}
\end{equation}
The corresponding attenuation at H$\alpha$ can be obtained by using the intrinsic value provided by equation \ref{eqn:ha_33}
\begin{equation}
 {\rm A_{H\alpha}} =   -2.5{\rm log}\left(\frac{H\alpha^{obs}}{0.785{\rm L_{33} }} \right)  + 35.25
\label{eqn:ah_33}
\end{equation}
As mentioned earlier, this attenuation is larger than the one that can be obtained with the Balmer decrement method because the latter
might not accounts for the strong obscuration in the early life of massive stars
(see Table~\ref{tab:dale_olga_attens}). Thus, in presence of an estimate of the radio flux at 33~GHz, Equation \ref{eqn:ah_33} should be preferred for young starbursts.
A similar relation, though with somewhat larger scatter, can be obtained for the
24~${\micron}$ flux,
which is shown in Figure~\ref{fig:Ha_int_L24}. 
The relation between  the model's  intrinsic H$\alpha$ (in ${\rm erg} \; {\rm s}^{-1}$) and   24~$\micron$  luminositiy,  L$_{24}$~=~$\lambda$L$_\lambda$ in ${\rm erg} \; {\rm s}^{-1}$, is
\begin{equation}
{\rm log}( {\rm L_{H\alpha}^{int} } ) =  0. 59{\rm log}( {\rm L_{24} } )  + 15.73
\label{eqn:ha_24}
\end{equation}
almost independently from M$_{up}$.
Correspondingly, the derived attenuation in H$\alpha$ is
\begin{equation}
{\rm A_{H\alpha}} =   -2.5{\rm log}\left( \frac{H\alpha^{obs}}{1.48\rm L_{24}} \right)  + 39.33
\label{eqn:ah_24}
\end{equation}
It is also interesting to provide a relation  between the attenuation at 1600~\AA~(Column~4
of Table~\ref{tab:dale_olga_attens}) and the ratio of 
FIR (Column~11 of Table~\ref{tab:dale_gra_derived_1_unsub})
to F$_{1600}^{tra}$ (Column~3 of Table~\ref{tab:dale_gra_derived_1_unsub}).
The plot is shown in Figure~\ref{fig:A1600_IRX}, along with the linear fitting relation 
obtained for the star-bursting regions (solid line), for both M$_{up}$ cases.
\begin{equation}
A_{1600} ~=~ 1.209 {\rm log}\left (\frac{ FIR } { F_{1600}^{obs} }\ \right)  + 0.254
\label{eqn:A1600_Fir}
\end{equation}
We note that the normal galaxy M100 is  out from the relation. This may likely be due to the non negligible contribution to the FIR intermediate age stellar populations that are not contributing to the far UV.

\section{Conclusions}
\label{sec:conc}
In this paper we have  used the recent  \textit{\small{PARSEC}} tracks to  compute  
the integrated stellar light, the ionizing photon budget and the supernova rates predicted by young SSP models for different IMF upper mass limits.
The SSPs spans a wide range in metallicities, from  0.0001
to  0.04 and initial masses ranging from 0.1 to 350M$\sun$.
In building the integrated spectra, we have adopted the spectral library compiled by \citet{Chen2015}.
This library is a result of homogenising different sets of libraries through a process fully described in \citet{Girardi2002} and \citet{Chen2014}.
Using  this integrated spectra in the photoionization code  \textit{\small{CLOUDY}}, we have also calculated
and included in the integrated spectra the nebular continum and the intensities  of some selected emission lines.
With the new SSPs we can then be able to  predict the panchromatic spectrum and the main recombination lines of star-forming galaxies of any type, with  \textit{\small{GRASIL}}.

We made also two major revisions in the radio emission model.
First, we revisited the relation between free-free radio emission, number of ionizing photons and radio frequency and  came up with a relation that  (a) takes into account  the full dependence of the electron temperature  on the metallicity and the effect of considering different IMF upper mass limits  and (b) incorporated a more accurate gaunt factor term. 
The  free-free  radio emission shows a significant variation with metallicity and IMF upper mass limit, decreasing by a factor of about 3 when Z increases from 0.0001 to 0.02 and inceasing by  more than an order of of magnitude when the mass limit is increased from 40 to 350~$M\sun$, at constant total mass.
These differences cannot be negelected, especially when calibrating SFRs using tracers that depend  strongly on the ionizing photon budget.
Hence, the  corresponding SFR calibrations (H$\alpha$, H$\beta$ and thermal radio emission)  take into account this  dependence on metallicity and IMF upper limits.

Second, we  revised  the non-thermal radio emission model originally described by B02, taking into account  recent advances in the CCSN explosion mechanisms  which indicates a range of stellar masses where the stars fail to explode, the so called {\sl Failed SNe}.
We adopt a threshold value  of 30~M$\sun$, above which stars do not produce non-thermal radio emission. This has two immediate effects: (a)~the beginning of the non-thermal radio emission is delayed by about 7~Myr, $\sim$ the lifetime of a 30~M$\sun$ star;
(b)~the non-thermal radio emission is independent from the IMF upper mass limit as long as it is above 30~M$\sun$. It is also assumed that  pair-instability SNe, discussed by Slemer et al.~(2017) do not produce synchrotron radiation.

With these revisions in the new SSPs and in the radio emission model,
there is the need to check some parameters of the  \textit{\small{GRASIL}} code.
We first re-calibrate the  proportionality constant between the supernova rate and the corresponding non-thermal radio luminosity, $E_{1.49}^{NT}$), using one of the best-sampled  nearby normal star-forming galaxy, M100. 
We are able to  reproduce very well  the far-UV to radio SED of this galaxy, for IMF upper mass limits of 40~$M_{\sun}$ and 120~$M_{\sun}$,  with an average  $q_{1.4}$ of 2.42.
We obtain a value of  $E_{1.49}^{NT}$ = 1.94 which is larger than that obtained previously by B02 by a factor of 1.35, but it is similar to that obtained by \citet{Vega2005}.
We find that using M$_{up}~=~350~\rm{M}_{\sun}  $ produces  a thermal  radio emission
that significantly exceeds the observed one. This excess thermal radio emission cannot be cured by varying any other parameter in the fit. This result suggests that,  for a normal star-forming galaxy, it is difficult to have an average IMF extending up to such high initial masses. 
What depends strongly on the upper mass limit of the IMF
is the number of ionizing photons, which beqars on the intensity of the recombination lines and on the thermal radio emission. Indeed, in spite of
being able to reproduce the  FIR and 1.4~GHz radio emission,
the model that adopts M$_{up}$~=~120~$M_{\sun}$ over predicts the H$\alpha$ emission by a factor of about three while, in the case of M$_{up}$~=~40~$M_{\sun}$, the discrepancy is only of a few percent. 

We then check the new thermal radio emission model against the  well studied thermal radio dominated star-forming regions in NGC~6946.  Even in this case we are able to reproduce very well the observed SEDs from NIR to radio wavelengths for both cases of the IMF upper mass limits.
In fitting the SEDs of these thermal radio-dominanted star-bursting  regions, we adopt  the value of  $E_{1.49}^{NT}$ resulting from the non-thermal radio calibration with the  normal star-forming galaxy M100. The estimated values of the  $q_{1.4}$ ratio in these regions lie between 2.5 and 2.6, implying a relatively lower non-thermal emission than in normal star-forming galaxies. The additional evidence of flatter radio slopes supports the notion that there is  lack of the non-thermal emission  as predicted by the original non-thermal radio emission models by B02. The resulting ages of the bursts, range from 7 to 12 Myr, confirming that these observations can be used to determine the star-burst ages.
The fit obtained with the two IMF upper mass limits exhibit  interesting differences, in particular in the  predicted H$\alpha$ and UV luminosities which are the most sensitive to the IMF and to dust attenuations.  We show that, by combining  information from the FIR,  24~$\micron$,   33~GHz and H$\alpha$, we can determine the preferred value for M$_{up}$.
This  is the first time, to the best of our knowledge, that the  IMF upper mass limit can be consistently determined.
However, for region $\sharp$8 we cannot reproduce the observed  H$\alpha$  flux even with
M$_{up}~=~350~\rm{M}_{\sun}$. This could be a case where 
binary evolution,  which is known to produce an excess of ionizing photons due to  the loss of the envelope of massive stars  by binary interaction, may be required.

With luminosities and averaged SFR extracted from  our  best fit models
(Table~\ref{tab:dale_gra_derived_1_unsub}), we derive  
SFR calibrations in different bands
for both normal galaxies and star-burst regions
(Table~\ref{tab:dale_gra_derived_2_unsub}). 
Since thermal radio emission and that in the recombination lines depend
strongly on the upper mass limit of the IMF,  we provide multiple regression fitting relations with M$_{up}$ and metallicity (e.g. Equations \ref{eq_sfr_lff_const_fit_b}, \ref{eq_sfr_ha_obi_fit} \ref{eq_sfr_hb_obi_fit}).

Finally, exploiting the  realistic treatment of dust performed by  \textit{\small{GRASIL}}, we
investigate the properties of the attenuation curves of the galaxies studied in this work with the aim of understanding  their dependence on the galaxy type.
We find that, while in the normal SF galaxy M100  the attenuation  in the lines is significantly higher than that in the surrounding continuum, for the star bursting regions of NGC~6946 the  two attenuations are similar.
A common property is that the predicted R values obtained for the
${H\alpha}$ line is large, mimicking a significant neutral absorption.
For star bursting regions, large values are found also for the continuum while,
in the case of M100, the R value of the continuum is normal.
Large R values have been found in high redshift star-forming galaxies by
\citet{Fan_2014}, using a completely different population synthesis tool.
A major disturbing consequence of the age selective attenuation, which could not be revealed in a foreground screen model, is that it leads to  wrong  estimates of  attenuation when using methods involving observed line fluxes, like  the Balmer decrement.
Of course, more accurate methods are those that combine UV or line fluxes
with fluxes that are not affected by attenuation, such as
the well known  relation between the 
FUV (at 1600~\AA) attenuation and  the flux ratio ratio, FIR/F$_{1600}$.
For the star-forming regions we revisit the former relation (Equation \ref{eqn:A1600_Fir}) 
and we provide new relations between the  H$\alpha$ attenuation  and the observed  H$\alpha$and  24~$\micron$  (Equation \ref{eqn:ah_33}) or 33~GHz~(Equation \ref{eqn:ah_24}) fluxes.
The above mentioned  relations, which we show to be almost independent from M$_{up}$, can be extremely  useful in estimating the attenuations in young high redshift  galaxies.
For this purpose we are working on a larger galaxy sample to increase
the statistical significance of our results.

\section*{Acknowledgments}
We thank   ?? for helpful discussions.
A. Bressan and L. Girardi acknowledge financial support
from INAF through grants PRIN-INAF-2014-14.
P. Marigo and Y. Chen acknowledge support from ERC Consolidator Grant "STARKEY", G.A. n. 615604.
F. Perrotta was supported by the RADIOFOREGROUNDS grant of the European Union's Horizon
2020 research and innovation programme  COMPET-05-2015, grant agreement number 687312.
\bibliography{radio_paper_final_bibdatabase}

\begin{thebibliography}{66}
\expandafter\ifx\csname natexlab\endcsname\relax\def\natexlab#1{#1}\fi

\bibitem[{{Allard} {et~al}\mbox{.}(1997){Allard}, {Hauschildt}, {Alexander}, \&
  {Starrfield}}]{Allard1997}
{Allard} F., {Hauschildt} P.~H., {Alexander} D.~R., {Starrfield} S., 1997,
  \araa, 35, 137

\bibitem[{{Asplund} {et~al}\mbox{.}(2009){Asplund}, {Grevesse}, {Sauval}, \&
  {Scott}}]{Asplund2009}
{Asplund} M., {Grevesse} N., {Sauval} A.~J., {Scott} P., 2009, \araa, 47, 481

\bibitem[{{Berkhuijsen}(1984)}]{Berkhuijsen1984}
{Berkhuijsen} E.~M., 1984, \aap, 140, 431

\bibitem[{{Bicker} \& {Fritze-v.~Alvensleben}(2005)}]{Bicker2005}
{Bicker} J., {Fritze-v.~Alvensleben} U., 2005, \aap, 443, L19

\bibitem[{{Bressan} {et~al}\mbox{.}(2013){Bressan}, {Marigo}, {Girardi},
  {Nanni}, \& {Rubele}}]{Bressan2013}
{Bressan} A., {Marigo} P., {Girardi} L., {Nanni} A., {Rubele} S., 2013, in
  European Physical Journal Web of Conferences, Vol.~43, European Physical
  Journal Web of Conferences, p. 3001

\bibitem[{{Bressan} {et~al}\mbox{.}(2012){Bressan}, {Marigo}, {Girardi},
  {Salasnich}, {Dal Cero}, {Rubele}, \& {Nanni}}]{Bressan2012}
{Bressan} A., {Marigo} P., {Girardi} L., {Salasnich} B., {Dal Cero} C.,
  {Rubele} S., {Nanni} A., 2012, \mnras, 427, 127

\bibitem[{{Bressan}, {Silva} \& {Granato}(2002){Bressan}, {Silva}, \&
  {Granato}}]{Bressan2002}
{Bressan} A., {Silva} L., {Granato} G.~L., 2002, \aap, 392, 377

\bibitem[{{Calzetti} {et~al}\mbox{.}(2000){Calzetti}, {Armus}, {Bohlin},
  {Kinney}, {Koornneef}, \& {Storchi-Bergmann}}]{Calzetti2000}
{Calzetti} D., {Armus} L., {Bohlin} R.~C., {Kinney} A.~L., {Koornneef} J.,
  {Storchi-Bergmann} T., 2000, \apj, 533, 682

\bibitem[{{Calzetti} {et~al}\mbox{.}(2007){Calzetti}, {Kennicutt},
  {Engelbracht}, {Leitherer}, {Draine}, {Kewley}, {Moustakas}, {Sosey}, {Dale},
  {Gordon}, {Helou}, {Hollenbach}, {Armus}, {Bendo}, {Bot}, {Buckalew},
  {Jarrett}, {Li}, {Meyer}, {Murphy}, {Prescott}, {Regan}, {Rieke}, {Roussel},
  {Sheth}, {Smith}, {Thornley}, \& {Walter}}]{Calzetti2007}
{Calzetti} D. {et~al.}, 2007, \apj, 666, 870

\bibitem[{{Cardelli}, {Clayton} \& {Mathis}(1989){Cardelli}, {Clayton}, \&
  {Mathis}}]{Cardelli1989}
{Cardelli} J.~A., {Clayton} G.~C., {Mathis} J.~S., 1989, \apj, 345, 245

\bibitem[{{Castelli} \& {Kurucz}(2004)}]{Castelli2004}
{Castelli} F., {Kurucz} R.~L., 2004, ArXiv Astrophysics e-prints

\bibitem[{{Charlot} \& {Bruzual}(1991)}]{Charlot1991}
{Charlot} S., {Bruzual} A.~G., 1991, \apj, 367, 126

\bibitem[{{Charlot} \& {Fall}(2000)}]{Charlot_Fall2000}
{Charlot} S., {Fall} S.~M., 2000, \apj, 539, 718

\bibitem[{{Chen} {et~al}\mbox{.}(2015){Chen}, {Bressan}, {Girardi}, {Marigo},
  {Kong}, \& {Lanza}}]{Chen2015}
{Chen} Y., {Bressan} A., {Girardi} L., {Marigo} P., {Kong} X., {Lanza} A.,
  2015, \mnras, 452, 1068

\bibitem[{{Chen} {et~al}\mbox{.}(2014){Chen}, {Girardi}, {Bressan}, {Marigo},
  {Barbieri}, \& {Kong}}]{Chen2014}
{Chen} Y., {Girardi} L., {Bressan} A., {Marigo} P., {Barbieri} M., {Kong} X.,
  2014, \mnras, 444, 2525

\bibitem[{{Chiosi}, {Bertelli} \& {Bressan}(1988){Chiosi}, {Bertelli}, \&
  {Bressan}}]{Chiosi1988}
{Chiosi} C., {Bertelli} G., {Bressan} A., 1988, \aap, 196, 84

\bibitem[{{Condon}(1992)}]{Condon1992}
{Condon} J.~J., 1992, \araa, 30, 575

\bibitem[{{Condon} \& {Yin}(1990)}]{Condon1990}
{Condon} J.~J., {Yin} Q.~F., 1990, \apj, 357, 97

\bibitem[{{Draine}(2011)}]{Draine2011}
{Draine} B.~T., 2011, {Physics of the Interstellar and Intergalactic Medium}

\bibitem[{{Ertl} {et~al}\mbox{.}(2015){Ertl}, {Janka}, {Woosley}, {Sukhbold},
  \& {Ugliano}}]{Ertl2015}
{Ertl} T., {Janka} H.-T., {Woosley} S.~E., {Sukhbold} T., {Ugliano} M., 2015,
  ArXiv e-prints

\bibitem[{{Fan} {et~al}\mbox{.}(2014){Fan}, {Lapi}, {Bressan}, {Nonino}, {De
  Zotti}, \& {Danese}}]{Fan_2014}
{Fan} L.-L., {Lapi} A., {Bressan} A., {Nonino} M., {De Zotti} G., {Danese} L.,
  2014, Research in Astronomy and Astrophysics, 14, 15

\bibitem[{{Ferland}(1996)}]{Ferland1996}
{Ferland} G., 1996, {Hazy: A Brief Introduction to CLOUDY}. Univ. of Kentucky
  Physics Department Internal Report

\bibitem[{{Girardi} {et~al}\mbox{.}(2002){Girardi}, {Bertelli}, {Bressan},
  {Chiosi}, {Groenewegen}, {Marigo}, {Salasnich}, \& {Weiss}}]{Girardi2002}
{Girardi} L., {Bertelli} G., {Bressan} A., {Chiosi} C., {Groenewegen} M.~A.~T.,
  {Marigo} P., {Salasnich} B., {Weiss} A., 2002, \aap, 391, 195

\bibitem[{{Granato} {et~al}\mbox{.}(2000){Granato}, {Lacey}, {Silva},
  {Bressan}, {Baugh}, {Cole}, \& {Frenk}}]{Granato2000}
{Granato} G.~L., {Lacey} C.~G., {Silva} L., {Bressan} A., {Baugh} C.~M., {Cole}
  S., {Frenk} C.~S., 2000, \apj, 542, 710

\bibitem[{{Heckman} {et~al}\mbox{.}(1983){Heckman}, {van Breugel}, {Miley}, \&
  {Butcher}}]{Heckman1983}
{Heckman} T.~M., {van Breugel} W., {Miley} G.~K., {Butcher} H.~R., 1983, \aj,
  88, 1077

\bibitem[{{Helou} {et~al}\mbox{.}(1988){Helou}, {Khan}, {Malek}, \&
  {Boehmer}}]{Helou1988}
{Helou} G., {Khan} I.~R., {Malek} L., {Boehmer} L., 1988, \apjs, 68, 151

\bibitem[{{Helou}, {Soifer} \& {Rowan-Robinson}(1985){Helou}, {Soifer}, \&
  {Rowan-Robinson}}]{Helou1985}
{Helou} G., {Soifer} B.~T., {Rowan-Robinson} M., 1985, \apjl, 298, L7

\bibitem[{{Hummer} \& {Storey}(1987)}]{Hummer1987}
{Hummer} D.~G., {Storey} P.~J., 1987, \mnras, 224, 801

\bibitem[{{Israel} \& {Kennicutt}(1980)}]{Israel1980}
{Israel} F.~P., {Kennicutt} R.~C., 1980, \aplett, 21, 1

\bibitem[{{Janka}(2012)}]{Janka2012}
{Janka} H.-T., 2012, Annual Review of Nuclear and Particle Science, 62, 407

\bibitem[{{Kennicutt}(1983)}]{Kennicutt1983}
{Kennicutt}, Jr. R.~C., 1983, \apj, 272, 54

\bibitem[{{Kennicutt}(1998)}]{Kennicutt1998}
{Kennicutt}, Jr. R.~C., 1998, \araa, 36, 189

\bibitem[{{Kennicutt} {et~al}\mbox{.}(2003){Kennicutt}, {Armus}, {Bendo},
  {Calzetti}, {Dale}, {Draine}, {Engelbracht}, {Gordon}, {Grauer}, {Helou},
  {Hollenbach}, {Jarrett}, {Kewley}, {Leitherer}, {Li}, {Malhotra}, {Regan},
  {Rieke}, {Rieke}, {Roussel}, {Smith}, {Thornley}, \&
  {Walter}}]{Kennicutt2003}
{Kennicutt}, Jr. R.~C. {et~al.}, 2003, \pasp, 115, 928

\bibitem[{{Kennicutt} {et~al}\mbox{.}(2009){Kennicutt}, {Hao}, {Calzetti},
  {Moustakas}, {Dale}, {Bendo}, {Engelbracht}, {Johnson}, \&
  {Lee}}]{Kennicutt2009}
{Kennicutt}, Jr. R.~C. {et~al.}, 2009, \apj, 703, 1672

\bibitem[{{Kroupa}(2001)}]{Kroupa2001}
{Kroupa} P., 2001, \mnras, 322, 231

\bibitem[{{Kurucz}(1993)}]{Kurucz1993}
{Kurucz} R.~L., 1993, in Astronomical Society of the Pacific Conference Series,
  Vol.~44, IAU Colloq. 138: Peculiar versus Normal Phenomena in A-type and
  Related Stars, {Dworetsky} M.~M., {Castelli} F., {Faraggiana} R., eds., p.~87

\bibitem[{{Lawton} {et~al}\mbox{.}(2010){Lawton}, {Gordon}, {Babler}, {Block},
  {Bolatto}, {Bracker}, {Carlson}, {Engelbracht}, {Hora}, {Indebetouw},
  {Madden}, {Meade}, {Meixner}, {Misselt}, {Oey}, {Oliveira}, {Robitaille},
  {Sewilo}, {Shiao}, {Vijh}, \& {Whitney}}]{Lawton2010}
{Lawton} B. {et~al.}, 2010, \apj, 716, 453

\bibitem[{{Li} {et~al}\mbox{.}(2010){Li}, {Calzetti}, {Kennicutt}, {Hong},
  {Engelbracht}, {Dale}, \& {Moustakas}}]{Li2010}
{Li} Y., {Calzetti} D., {Kennicutt} R.~C., {Hong} S., {Engelbracht} C.~W.,
  {Dale} D.~A., {Moustakas} J., 2010, \apj, 725, 677

\bibitem[{{L{\'o}pez-S{\'a}nchez} {et~al}\mbox{.}(2012){L{\'o}pez-S{\'a}nchez},
  {Dopita}, {Kewley}, {Zahid}, {Nicholls}, \& {Scharw{\"a}chter}}]{Sanchez2012}
{L{\'o}pez-S{\'a}nchez} {\'A}.~R., {Dopita} M.~A., {Kewley} L.~J., {Zahid}
  H.~J., {Nicholls} D.~C., {Scharw{\"a}chter} J., 2012, \mnras, 426, 2630

\bibitem[{{Maeder} \& {Meynet}(1988)}]{Maeder1988}
{Maeder} A., {Meynet} G., 1988, \aaps, 76, 411

\bibitem[{{Maraston}(2005)}]{Maraston2005}
{Maraston} C., 2005, \mnras, 362, 799

\bibitem[{{Marigo} \& {Girardi}(2007)}]{Marigo2007}
{Marigo} P., {Girardi} L., 2007, \aap, 469, 239

\bibitem[{{Marigo} {et~al}\mbox{.}(2017){Marigo}, {Girardi}, {Bressan},
  {Rosenfield}, {Aringer}, {Chen}, {Dussin}, {Nanni}, {Pastorelli},
  {Rodrigues}, {Trabucchi}, {Bladh}, {Dalcanton}, {Groenewegen},
  {Montalb{\'a}n}, \& {Wood}}]{Marigo2017}
{Marigo} P. {et~al.}, 2017, \apj, 835, 77

\bibitem[{{Murphy} {et~al}\mbox{.}(2012){Murphy}, {Bremseth}, {Mason},
  {Condon}, {Schinnerer}, {Aniano}, {Armus}, {Helou}, {Turner}, \&
  {Jarrett}}]{Murphy2012}
{Murphy} E.~J. {et~al.}, 2012, \apj, 761, 97

\bibitem[{{Murphy} {et~al}\mbox{.}(2011){Murphy}, {Condon}, {Schinnerer},
  {Kennicutt}, {Calzetti}, {Armus}, {Helou}, {Turner}, {Aniano}, {Beir{\~a}o},
  {Bolatto}, {Brandl}, {Croxall}, {Dale}, {Donovan Meyer}, {Draine},
  {Engelbracht}, {Hunt}, {Hao}, {Koda}, {Roussel}, {Skibba}, \&
  {Smith}}]{Murphy2011}
{Murphy} E.~J. {et~al.}, 2011, \apj, 737, 67

\bibitem[{{Murphy} {et~al}\mbox{.}(2010){Murphy}, {Helou}, {Condon},
  {Schinnerer}, {Turner}, {Beck}, {Mason}, {Chary}, \& {Armus}}]{Murphy2010}
{Murphy} E.~J. {et~al.}, 2010, \apjl, 709, L108

\bibitem[{{O'Connor} \& {Ott}(2011)}]{Connor2011}
{O'Connor} E., {Ott} C.~D., 2011, \apj, 730, 70

\bibitem[{{Oster}(1961)}]{oster1961}
{Oster} L., 1961, Reviews of Modern Physics, 33, 525

\bibitem[{{Osterbrock}(1989)}]{Osterbrock1989}
{Osterbrock} D.~E., 1989, {Astrophysics of gaseous nebulae and active galactic
  nuclei}. University Science Books, Mill Valley, CA

\bibitem[{{Panuzzo} {et~al}\mbox{.}(2003){Panuzzo}, {Bressan}, {Granato},
  {Silva}, \& {Danese}}]{Panuzzo2003}
{Panuzzo} P., {Bressan} A., {Granato} G.~L., {Silva} L., {Danese} L., 2003,
  \aap, 409, 99

\bibitem[{{Pauldrach}, {Puls} \& {Kudritzki}(1986){Pauldrach}, {Puls}, \&
  {Kudritzki}}]{Pauldrach1986A}
{Pauldrach} A., {Puls} J., {Kudritzki} R.~P., 1986, \aap, 164, 86

\bibitem[{{Rubin}(1968)}]{Rubin1968}
{Rubin} R.~H., 1968, \apj, 154, 391

\bibitem[{{Sander} {et~al}\mbox{.}(2015){Sander}, {Shenar}, {Hainich},
  {G{\'{\i}}menez-Garc{\'{\i}}a}, {Todt}, \& {Hamann}}]{PoWR2015}
{Sander} A., {Shenar} T., {Hainich} R., {G{\'{\i}}menez-Garc{\'{\i}}a} A.,
  {Todt} H., {Hamann} W.-R., 2015, \aap, 577, A13

\bibitem[{{Schmitt} {et~al}\mbox{.}(2006){Schmitt}, {Calzetti}, {Armus},
  {Giavalisco}, {Heckman}, {Kennicutt}, {Leitherer}, \& {Meurer}}]{Schmitt2006}
{Schmitt} H.~R., {Calzetti} D., {Armus} L., {Giavalisco} M., {Heckman} T.~M.,
  {Kennicutt}, Jr. R.~C., {Leitherer} C., {Meurer} G.~R., 2006, \apj, 643, 173

\bibitem[{{Silva} {et~al}\mbox{.}(1998){Silva}, {Granato}, {Bressan}, \&
  {Danese}}]{Silva1998}
{Silva} L., {Granato} G.~L., {Bressan} A., {Danese} L., 1998, \apj, 509, 103

\bibitem[{{Silva} {et~al}\mbox{.}(2011){Silva}, {Schurer}, {Granato},
  {Almeida}, {Baugh}, {Frenk}, {Lacey}, {Paoletti}, {Petrella}, \&
  {Selvestrel}}]{Silva2011}
{Silva} L. {et~al.}, 2011, \mnras, 410, 2043

\bibitem[{{Spera}, {Mapelli} \& {Bressan}(2015){Spera}, {Mapelli}, \&
  {Bressan}}]{Spera2015}
{Spera} M., {Mapelli} M., {Bressan} A., 2015, \mnras, 451, 4086

\bibitem[{{Sukhbold} \& {Woosley}(2014)}]{Sukhbold2014}
{Sukhbold} T., {Woosley} S.~E., 2014, \apj, 783, 10

\bibitem[{{Sutherland} \& {Dopita}(1993)}]{Sutherland1993}
{Sutherland} R.~S., {Dopita} M.~A., 1993, \apjs, 88, 253

\bibitem[{{Tang} {et~al}\mbox{.}(2014){Tang}, {Bressan}, {Rosenfield},
  {Slemer}, {Marigo}, {Girardi}, \& {Bianchi}}]{Tang2014}
{Tang} J., {Bressan} A., {Rosenfield} P., {Slemer} A., {Marigo} P., {Girardi}
  L., {Bianchi} L., 2014, \mnras, 445, 4287

\bibitem[{{Ugliano} {et~al}\mbox{.}(2012){Ugliano}, {Janka}, {Marek}, \&
  {Arcones}}]{Ugliano2012}
{Ugliano} M., {Janka} H.-T., {Marek} A., {Arcones} A., 2012, \apj, 757, 69

\bibitem[{{Vega} {et~al}\mbox{.}(2008){Vega}, {Clemens}, {Bressan}, {Granato},
  {Silva}, \& {Panuzzo}}]{Vega2008}
{Vega} O., {Clemens} M.~S., {Bressan} A., {Granato} G.~L., {Silva} L.,
  {Panuzzo} P., 2008, \aap, 484, 631

\bibitem[{{Vega} {et~al}\mbox{.}(2005){Vega}, {Silva}, {Panuzzo}, {Bressan},
  {Granato}, \& {Chavez}}]{Vega2005}
{Vega} O., {Silva} L., {Panuzzo} P., {Bressan} A., {Granato} G.~L., {Chavez}
  M., 2005, \mnras, 364, 1286

\bibitem[{{Wofford} {et~al}\mbox{.}(2016){Wofford}, {Charlot}, {Bruzual},
  {Eldridge}, {Calzetti}, {Adamo}, {Cignoni}, {de Mink}, {Gouliermis},
  {Grasha}, {Grebel}, {Lee}, {{\"O}stlin}, {Smith}, {Ubeda}, \&
  {Zackrisson}}]{Wofford2016}
{Wofford} A. {et~al.}, 2016, \mnras, 457, 4296

\bibitem[{{Young} {et~al}\mbox{.}(1989){Young}, {Xie}, {Kenney}, \&
  {Rice}}]{Young1989}
{Young} J.~S., {Xie} S., {Kenney} J.~D.~P., {Rice} W.~L., 1989, \apjs, 70, 699

\bibitem[{{Zhu} {et~al}\mbox{.}(2008){Zhu}, {Wu}, {Cao}, \& {Li}}]{Zhu2008}
{Zhu} Y.-N., {Wu} H., {Cao} C., {Li} H.-N., 2008, \apj, 686, 155

\end{thebibliography}
\bibliographystyle{mn2e}
\bsp	

\appendix
\section{Variation of electron temperature and the constants, $C_2$ and $C_3$,  with IMF Upper mass limits and metallicity}
\label{appendix_a}
\begin{table}
\caption{Calibration constants, $C_2$ and $C_3$, for different metallicities and upper mass limits.
\label{tab:C_2_C_3}
}
\small{ 
\centering
\begin{tabular}{lcccc}
\hline
$\rm{Z}$    &        $IQ(H)$    &       $C_2$     &    $T_e$    &     $C_3$ \\ 
 &       $(10^{54}$  &      $ (10^{-54} )$     &  $(10^{4} )$       &    $ (10^{26} )$ \\
(1)&(2) &(3)&(4)&(5) \\
\hline
 $M_{up} ~=~ 40~\rm{M}_{\sun}$   & &\\
\hline
0.0200 & 0.554 & 18.03 & 0.64 & 1.47\\
0.0080 & 0.738 & 13.56 & 0.89 & 1.96\\
0.0040 & 0.943 & 10.61 & 0.11  & 2.51\\
0.0005 & 1.376 & 7.27 & 0.16 & 3.66\\
0.0001 & 1.823 & 5.48  & 0.17 & 4.85\\
\hline
 $M_{up} ~=~ 120~\rm{M}_{\sun}$    &&    \\
\hline
0.0200 & 1.730  & 5.78 & 0.65 & 4.60\\
0.0080 & 2.153 & 4.64  & 0.94 & 5.73\\
0.0040 & 2.520 & 3.97 & 0.12  & 6.70\\
0.0005 & 3.065  & 3.26  & 0.17  & 8.15\\
0.0001& 3.761  & 2.66 & 0.19 & 10.00\\
\hline
  $M_{up} ~=~ 350~\rm{M}_{\sun}$     &&    \\
\hline
0.0200 & 2.819  & 3.54  & 0.65 & 7.50\\
0.0080 & 3.469 & 2.88  & 0.95 & 9.22\\
0.0040 & 3.739 & 2.67  & 0.12 & 9.94\\
0.0005 & 4.269  & 2.34 & 0.17  & 11.35\\
0.0001 & 5.144 & 1.94 & 0.20 & 13.68\\
\hline
\end{tabular}

Column(1): metallicty.
Column(2): production rate of ionizing photons per unit solar mass formed.
Column(3): calibration  constant ($C_2$) of the SFR-IQ(H) relation as given in
equation~\ref{eq_sfr_q_obi_old}.
Column(4): electron temperature in K computed using  \textit{\small{CLOUDY}}.
Column(5): calibration  constant ($C_3$) of the $SFR$-$L_{ff}$ relation as given 
in equation~\ref{eq_lff_sfr_const_0}.
} 
\end{table}
\begin{figure}
\includegraphics[width=\linewidth]{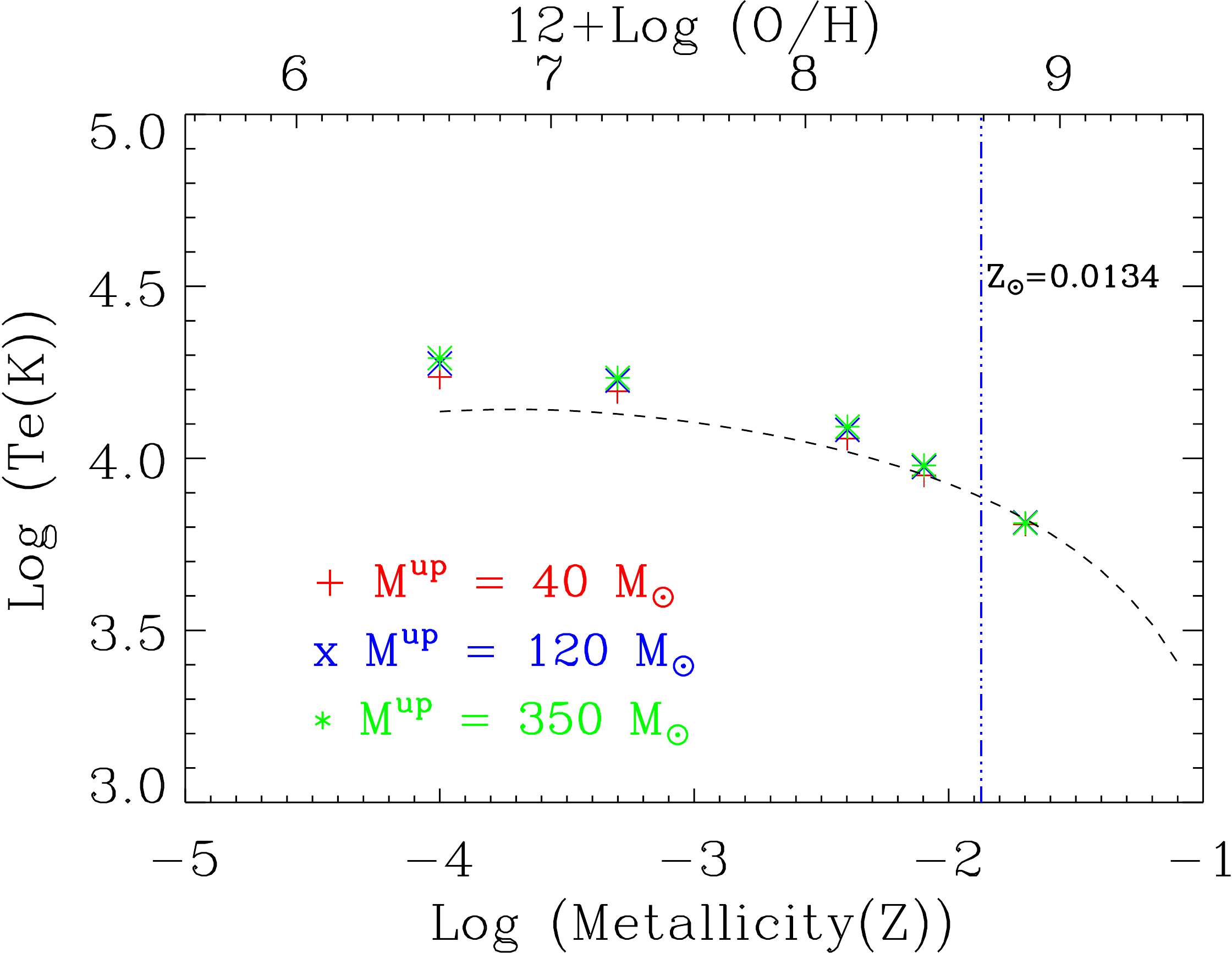}
\caption{Plot of electron temperature  T$_e$ (obtained from  \textit{\small{CLOUDY}}) against  metallicity Z  for  $M_{up}$~=~40, 120 and 350$~\rm{M}_{\sun}$ indicated
by the red 'X's, blue crosses and green xterisks respectively.
The lower  axis is the metallicity while the upper one is the corresponding oxygen abundance, $\rm{x} ~=~ 12 + \rm{log(O/H)}$. The blue dashed-dotted line indicates the metallicity at Z~=~0.0134 which we adopted in this work as the solar metallicity.
The average of the empirical fits derived by \citet{Sanchez2012} for high-ionization  \ion{O}{III}  and for low-ionization \ion{O}{II} zones is shown by the black  dashed line.
\label{fig:te_z_mup}
} 
\end{figure}
\begin{figure}
\centering
\includegraphics[width=1\linewidth]{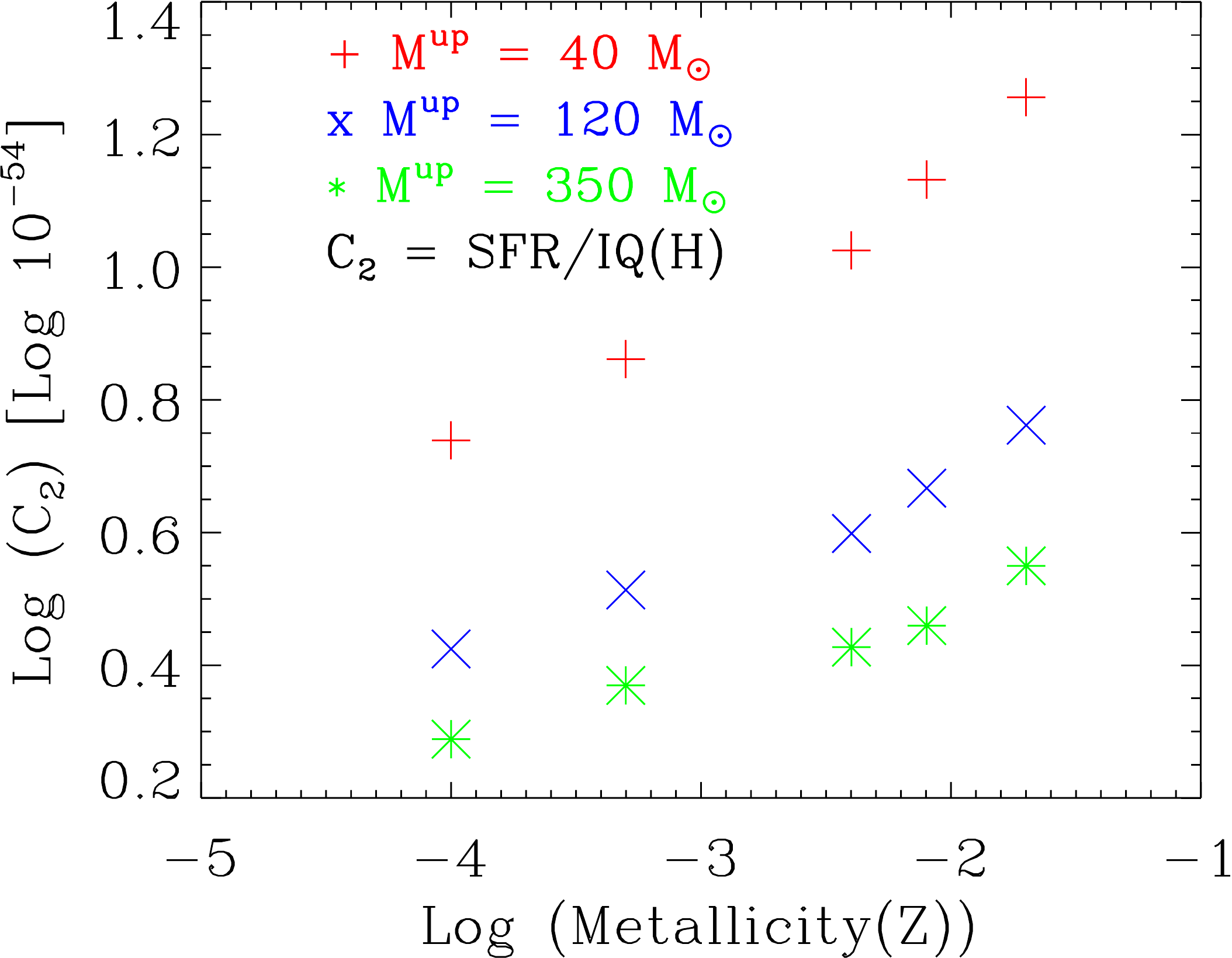}
\caption{Variation of the constant ${C_2}$ with age at different
 metallicities (Z = 0.0001, 0.0005, 0.004, 0.008 and 0.02) and different upper mass limits.
The symbols used to indicate the different  $M_{up}$ are as in Figure~\ref{fig:te_z_mup}.
These variations of ${C_2}$, depending on the set of upper limits used, can severely underestimate or overestimate the estimated SFR using the thermal radio luminosity as a tracer.
As clearly noticed, there is a large difference between $C_2$ obtained with $M_{up} = 40~\rm{M}_{\sun}$ and $M_{up} = 120~\rm{M}_{\sun}$ and that obtained with $M_{up} ~=~ 120~\rm{M}_{\sun}$ and $M_{up} = 350~\rm{M}_{\sun}$.  
$C_2$ is as defined  in equation~\ref{eq_sfr_q_obi_old}.
\label{fig:C2_plot}
} 
\end{figure}
\begin{figure}
\centering
\includegraphics[width=1\linewidth]{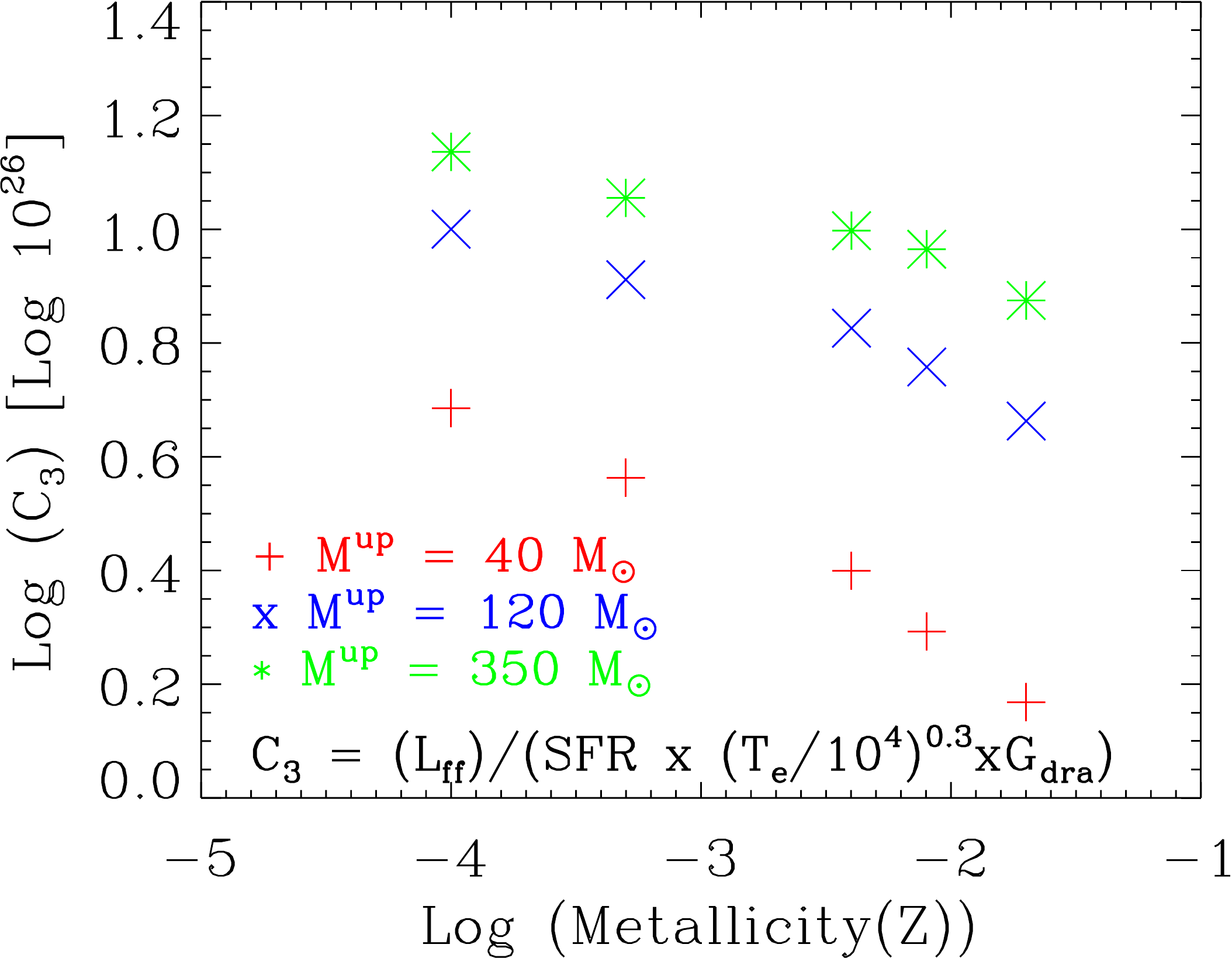}
\caption{Same as in Figure~\ref{fig:C2_plot} but for the constant $C_3$.
$C_3$ is as defined  in equation~\ref{eq_lff_sfr_const_0}.
 \label{fig:C3}
  }  
\end{figure}
In Figure~\ref{fig:te_z_mup},
we show the plot of electron temperature  T$_e$ (obtained from  \textit{\small{CLOUDY}}) against  metallicity Z  for  $M_{up}$~=~40, 120 and 350. The oxygen abundance corresponding to the metallicity Z, written as $\rm{x} = 12 +\ log (\rm{O/H})$
where $\rm{O/H}~=~\rm{O/H}_{\sun} + \log (Z/Z_{\sun}$, 
is shown on the upper axis.
To  convert from Z to (O/H), we adopt the solar oxygen abundance given by \citet{Asplund2009},
$\rm{x}_{\sun} = 12+ \log (\rm{O/H})_{\sun} = 8.69 \pm  0.05$ and  $Z_{\sun} = 0.0134$
and thus x ~=~ 8.69 + log(Z/0.0134).
The blue vertical dotted line indicates this  solar metallicity value of 
0.0134.
The average of the empirical fits derived by \citet{Sanchez2012} for high-ionization  \ion{O}{III}  and for low-ionization \ion{O}{II} zones, using the \textit{MAPPINGS III} code \citep{Sutherland1993}, is shown as the dashed black lines.
In this figure, our  $T_e$ values ('X's, crosses and xterisks)  computed with   \textit{\small{CLOUDY}} for different metallicities are compared with the above-mentioned empirical fit obatined by \citet{Sanchez2012}. It can be seen that our $T_e$ values start deviating from those of \citet{Sanchez2012} at low metallicities.
A multiple regression fitting relation between  $T_e$, $M_{up}$ and Z that can be easily included in analytical approximations, is provided by  equation \ref{eq_te_z_mup}.
\begin{equation}
\rm{log} \left(   \frac{\rm{T}_e(\rm{Z},\rm{M}_{up})}{10^4~K}   \right)
=  -0.13    - 0.19\rm{log}(Z/0.02) + 0.03\rm{log}(\rm{M}_{up}/120) 
\label{eq_te_z_mup}
\end{equation}
Table~\ref{tab:C_2_C_3} lists the values of the constants $C_2$ and $C_3$
in equations \ref{eq_sfr_q_obi_old}  and \ref{eq_lff_sfr_const_0},
obtained with our constant SFR models
for different SSPs parameters. These values are shown in Figure \ref{fig:C2_plot}.
The multiple regression fitting relation is given by Equation 
\ref{eq_c2_rel}.
\begin{equation}
\rm{log}(C_{2}) =   0.16\rm{log}(Z/0.02) - 0.61\rm{log}(M_{up}/120)  - 53.18
\label{eq_c2_rel}
\end{equation}
Table~\ref{tab:C_2_C_3} also lists the values of the constants $C_3$
in equation \ref{eq_lff_sfr_const_0}.
These values are shown in Figure \ref{fig:C3}.
The corresponding multiple regression fitting relation is given by Equation 
\ref{eq_c3_rel}.
\begin{equation}
\rm{log}(C_{3}) =   0.61\rm{log}(M_{up}/120)  - 0.16\rm{log}(Z/0.02)  + 26.61
\label{eq_c3_rel}
\end{equation}
Using the relation ~\ref{eq_c2_rel} in Equation~\ref{eq_sfr_q_obi_old} we obtain the integrated ionizing photon flux of a
star-forming region of arbitrary constant metallicity and IMF upper mass limit
\begin{eqnarray}
\rm{log}( IQ(H)) =  \nonumber  \\
 -0.16\rm{log}(Z/0.02) + 0.61\rm{log}(M_{up}/120)  + 53.18]+ \rm{log}(SFR)
\label{eq_sfr_q_obi_fit}
\end{eqnarray}
while, using the relation \ref{eq_c3_rel} in Equation~\ref{eq_lff_sfr_const_0}
 ($\rm{T}_{e}$ in this Equation is  in $10^4~K$), we may get the corresponding SFR-thermal radio  calibration:
\begin{eqnarray}
\rm{log} ( SFR/ L(\nu)_{\rm{ff}}  ) =  0.60\rm{log}(M_{up}/120) -  0.10\rm{log}(Z/0.02) \nonumber  \\
 - \rm{log} ( G_{dra}  )   - 26.57
\label{eq_sfr_lff_const_fit_b}
\end{eqnarray}
where  SFR and  $\nu$ are in  units of $\rm{M}_{\sun} \rm{yr}^{-1}$  and GHz respectively.
Similarly for SFR vs. H$\alpha$ calibration we may write:
\begin{equation}
\rm{log} (SFR/H\alpha)  =  0.16\rm{log}(Z/0.02) - 0.61\rm{log}(M_{up}/120)-41.32
\label{eq_sfr_ha_obi_fit}
\end{equation}
and  for SFR vs. H$\beta$ calibration:
\begin{equation}
\rm{log} (SFR/H\beta)  =  0.16\rm{log}(Z/0.02) - 0.61\rm{log}(M_{up}/120)- 40.87
\label{eq_sfr_hb_obi_fit}
\end{equation}
Comparisons between SFR calibrations  derived using Equations
~\ref{eq_sfr_lff_const_fit_b} , \ref{eq_sfr_ha_obi_fit} and \ref{eq_sfr_hb_obi_fit} above
and those obtained directly from our model  are given in Table~\ref{tab:dale_gra_derived_2_unsub} and discussed in Section~\ref{sec:sfr_cals}.
The former case is indicated by the superscript $^{ssp}$ in this table.

\begin{landscape}
\section{Best-fits derived quantities for M100 and  NGC~6946 Star-forming regions}
\label{appendix_b}
\begin{table}
\caption{  \textit{\small{GRASIL}} Best-fit  model  luminosities.
 \label{tab:dale_gra_derived_1_unsub}
 }
\small{
\centering
\begin{tabular}{lcccccccccccccccccc}
\hline
\multicolumn{1}{|c|}{ID} &
\multicolumn{1}{c|}{${FUV_{i}}$} &
\multicolumn{1}{c|}{${FUV_{t}}$} &
\multicolumn{1}{c|}{$H\beta_{i}$} &
\multicolumn{1}{c|}{$H\beta_{t}$} &
\multicolumn{1}{c|}{$H\alpha_{i}$} &
\multicolumn{1}{c|}{$H\alpha^{ \rm{data} }$} &
\multicolumn{1}{c|}{$H\alpha_{t}$} &
\multicolumn{1}{c|}{$24$} &
\multicolumn{1}{c|}{$70$} &
\multicolumn{1}{c|}{$FIR$} &
\multicolumn{1}{c|}{$1.4$} &
\multicolumn{1}{c|}{$4.9$} &
\multicolumn{1}{c|}{$8.5$} &
\multicolumn{1}{c|}{$33$} &
\multicolumn{1}{c|}{$q_{1.4}$} &
\multicolumn{1}{c|}{$MC$} &
\multicolumn{1}{c|}{$CIR$} &
\multicolumn{1}{c|}{$FIR/BOL$} \\
(1) & (2)&(3) &(4) &(5) &(6) &(7) &(8) &(9) &(10) &(11) &(12)  &(13)&(14)&(15)&(16)&(17)&(18)&(19) \\
&     $10^{41}$&        $10^{41}$&         $10^{39}$&        $10^{39}$&       $10^{39}$&        $10^{39}$&      $10^{39}$&       $10^{41}$&        $10^{41}$&        $10^{42}$&            $10^{26}$&        $10^{26}$&        $10^{26}$&         $10^{26}$&         &&& \\
\hline
\hline
   $M_{up}~=~40~\rm{M}_{\sun}  $ & & & & & & & & &     \\
\hline
  M100 & 653.10 & 319.0 & 102.15 & 33.70 & 298.61 & 123.0 & 105.0 & 130.0 & 616.0 & 167.0  & 793.0 (2) & 285.0 (5) & 186.0 (8) & 71.6 (18) & 2.38 & 19 & 81 & 0.39\\
  \hline
   NGC~6946\_1 & 4.52 & 2.03 & 1.92 & 0.95 & 5.48 & 3.96 & 2.88 & 0.77 & 2.81 & 0.66  & 2.28 (24) & 1.1 (44) & 0.87 (54) & 0.55 (75) & 2.53 & 25 & 75 & 0.62\\
  NGC~6946\_2 & 10.65 & 6.73 & 5.01 & 3.47 & 14.28 & 12.10 & 10.30 & 1.34 & 4.50 & 1.02  & 4.11 (37) & 2.3 (59) & 1.89 (68) & 1.33 (85) & 2.46 & 35 & 65 & 0.44\\
  NGC~6946\_3 & 4.52 & 2.20 & 1.92 & 1.02 & 5.48 & 4.86 & 3.07 & 0.47 & 2.85 & 0.60  & 2.28 (24) & 1.1 (44) & 0.87 (54) & 0.55 (75) & 2.53 & 28 & 72 & 0.58\\
  NGC~6946\_5 & 2.71 & 1.54 & 1.15 & 0.76 & 3.29 & 2.90 & 2.28 & 0.20 & 1.28 & 0.30  & 1.37 (24) & 0.7 (44) & 0.52 (54) & 0.33 (75) & 2.42 & 12 & 88 & 0.50\\
  NGC~6946\_6 & 12.47 & 5.22 & 4.86 & 2.17 & 13.87 & 7.87 & 6.59 & 2.92 & 8.41 & 2.01  & 7.99 (16) & 3.6 (32) & 2.68 (42) & 1.55 (65) & 2.47 & 27 & 73 & 0.66\\
  NGC~6946\_7 & 13.93 & 5.79 & 6.47 & 2.87 & 18.40 & 7.81 & 8.67 & 2.96 & 9.42 & 2.26  & 5.36 (36) & 3.0 (58) & 2.45 (67) & 1.71 (84) & 2.69 & 37 & 63 & 0.66\\
  NGC~6946\_8 & 9.90 & 4.99 & 4.59 & 2.66 & 13.08 & 2.73 & 8.02 & 1.42 & 5.75 & 1.29  & 3.81 (36) & 2.1 (58) & 1.74 (67) & 1.21 (84) & 2.61 & 21 & 80 & 0.57\\
  NGC~6946\_9 & 10.80 & 4.86 & 4.61 & 2.29 & 13.10 & 5.08 & 6.92 & 2.07 & 6.58 & 1.58  & 5.47 (24) & 2.7 (44) & 2.09 (54) & 1.33 (75) & 2.52 & 24 & 76 & 0.62\\
 \hline
\hline
   $M_{up}~=~120~\rm{M}_{\sun}  $ & & & & & & & & &     \\
\hline
  M100 & 729.94 & 379.0 & 260.66 & 131.0 & 759.97 & 123.0 & 406.0 & 120.0 & 687.0 & 175.0  & 734.0 (7) & 291.0 (15) & 204.0 (21) & 98.4 (42) & 2.45 & 18 & 82 & 0.42\\
  \hline
  NGC~6946\_1 & 3.77 & 1.29 & 1.82 & 0.63 & 5.17 & 3.96 & 1.94 & 0.83 & 3.33 & 0.70  & 2.26 (24) & 1.1 (44) & 0.87 (54) & 0.55 (76) & 2.60 & 212& 78 & 0.69\\
  NGC~6946\_2 & 6.76 & 3.29 & 4.53 & 2.07 & 12.95 & 12.10 & 6.23 & 1.29 & 4.29 & 0.95  & 4.23 (35) & 2.3 (57) & 1.91 (66) & 1.33 (84) & 2.44 & 42 & 58 & 0.60\\
  NGC~6946\_3 & 3.01 & 0.99 & 1.85 & 0.71 & 5.28 & 4.86 & 2.19 & 0.57 & 3.02 & 0.63  & 1.96 (30) & 1.0 (52) & 0.82 (61) & 0.55 (81) & 2.62 & 21& 79 & 0.70\\
  NGC~6946\_5 & 1.74 & 0.68 & 1.02 & 0.38 & 2.92 & 2.90 & 1.16 & 0.27 & 1.25 & 0.30  & 1.48 (21) & 0.7 (40) & 0.54 (49) & 0.33 (72) & 2.38 & 29 & 71 & 0.70\\
  NGC~6946\_6 & 10.69 & 3.49 & 5.12 & 1.55 & 14.66 & 7.87 & 4.79 & 2.85 & 9.27 & 2.02  & 6.32 (24) & 3.1 (44) & 2.43 (54) & 1.55 (76) & 2.60 & 26 & 74 & 0.70\\
  NGC~6946\_7 & 9.24 & 2.80 & 5.73 & 1.87 & 16.24 & 7.81 & 5.79 & 2.57 & 8.99 & 2.01  & 6.06 (30) & 3.2 (52) & 2.55 (61) & 1.71 (81) & 2.61 & 24 & 76 & 0.72\\
  NGC~6946\_8 & 6.57 & 2.28 & 4.07 & 1.61 & 11.55 & 2.73 & 4.97 & 1.54 & 6.26 & 1.32  & 4.30 (30) & 2.2 (52) & 1.81 (61) & 1.21 (81) & 2.60 & 21  & 79 & 0.68\\
  NGC~6946\_9 & 9.15 & 3.03 & 4.51 & 1.43 & 12.82 & 5.08 & 4.43 & 2.41 & 8.06 & 1.75  & 5.00 (28) & 2.5 (49) & 2.02 (58) & 1.33 (79) & 2.64 & 26& 74 & 0.69\\
\hline
\hline
\end{tabular}
 Col.(1): ID.
 Col.(2-7): luminosities at FUV  (0.16~$\micron$), $H\beta$ and $H\alpha$.
 $^{i}$ and  $^{t}$ indicates intrinsic and attenuated transmitted luminosities respectively.
 Col.(8): observed attenuation uncorrected $H\alpha$ luminosity.
 Cols.(9 and 10): luminosities at 24~$\micron$ and 70~$\micron$.
 Col.(11): total (3-1000~$\micron$)IR  luminosity.
 Cols.(12-15): radio luminosities at 1.4, 4.9, 8.5 and 33~GHz.
 Enclosed in parenthesis is the fraction in ~per~cent of the thermal radio component
 to the total radio emission.
Col.(16): q-parameter as defined by equation~\ref{eq_q_ratio}.
 Cols.(17 and 18) are the MC and cirrus contribution (in per~cent) to the
 total IR  luminosity respectively.
Col.(19): ratio of the total IR and  the bolometric luminosities
 All UV, optical and infrared luminosities are in  $ \rm{erg} \; \rm{s}^{-1}$ while all radio
 luminosities  are in  $ \rm{erg} \; \rm{s}^{-1}\; \rm{Hz}^{-1}$.
 The corresponding SFR calibrations at these  luminosities are given in Table~\ref{tab:dale_gra_derived_2_unsub}.
 }  
\end{table}
\end{landscape}
\begin{landscape}
\begin{table}
\caption{ \textit{\small{GRASIL}} Best-fit derived SFRs and their  calibrations at various bands.
 \label{tab:dale_gra_derived_2_unsub}
 } 
\small{ 
\centering
\begin{tabular}{lccccccccccccccc}
\hline
 \hline
\multicolumn{1}{|c|}{$ID$}   &
\multicolumn{1}{c|}{$<SFR>$} &
\multicolumn{1}{c|}{$C(FUV_{i} )$} &
\multicolumn{1}{c|}{ $C(H\beta_{i}) $ } &
\multicolumn{1}{c|}{ $C(H\beta_{i})^{ssp}$ } &
\multicolumn{1}{c|}{$C(H\alpha_{i}) $} &
\multicolumn{1}{c|}{$ C(H\alpha_{i})^{ssp}$ } &
\multicolumn{1}{c|}{$C(24)$} &
\multicolumn{1}{c|}{$C(70)$} &
\multicolumn{1}{c|}{$C(FIR)$} &
\multicolumn{1}{c|}{$C(1.4)$} &
\multicolumn{1}{c|}{$C(4.9) $} &
\multicolumn{1}{c|}{$C(8.5) $} &
\multicolumn{1}{c|}{$C(33) $} &
\multicolumn{1}{c|}{$C(33)^{ssp}$} &
\multicolumn{1}{c|}{$t$}\\
(1) & (2)&(3) &(4) &(5) &(6) &(7) &(8) &(9) &(10) &(11) &(12)   &(13)&(14) &(15)&(16)\\
 & & $10^{-43}$   &$10^{-42}$    &$10^{-42}$    &$10^{-42}$    &$10^{-42}$     &$10^{-43} $  &$10^{-43} $& $10^{-44} $ &$10^{-28}$  &$10^{-28}$&$10^{-28}$    &$10^{-28}$  &$10^{-28}$ &\\
 \hline
 \hline
 $\rm{M}_{up}~=~40~\rm{M}_{\sun} $ & & & & & & & & &     \\
\hline
  M100 & 5.58 & 0.85 & 54.67 & 45.22 & 18.70 & 16.04  & 4.30 & 0.91 & 3.35 & 0.70  (2) & 1.96  (5) & 3.00  (8) & 7.79  (18) & 25.71 & 12.0\\
   \hline
   NGC~6946\_1 & 0.09 & 1.88 & 44.33 & 31.96 & 15.5 & 11.34  & 11.01 & 3.02 & 12.80 & 3.72  (24) & 7.62  (44) & 9.72  (54) & 15.4  (76) & 14.77 & 8.0\\
   NGC~6946\_2 & 0.20 & 1.85 & 39.31 & 31.95 & 13.8 & 11.33  & 14.70 & 4.38 & 19.30 & 4.79  (37) & 8.64  (59) & 10.40  (68) & 14.9  (85) & 14.75 & 9.0\\
   NGC~6946\_3 & 0.09 & 1.88 & 44.22 & 31.96 & 15.5 & 11.34  & 18.01 & 2.98 & 14.20 & 3.72  (24) & 7.62  (44) & 9.72  (54) & 15.4  (76) & 14.77 & 8.0\\
   NGC~6946\_5 & 0.05 & 1.88 & 44.28 & 31.96 & 15.5 & 11.34  & 25.63 & 3.98 & 17.00 & 3.72  (50) & 7.62  (71) & 9.72  (78) & 15.3  (90) & 14.77 & 8.0\\
   NGC~6946\_6 & 0.20 & 1.58 & 40.51 & 31.99 & 14.2 & 11.35  & 6.75 & 2.34 & 9.79 & 2.47  (24) & 5.53  (44) & 7.36  (54) & 12.8  (76) & 14.79 & 12.0\\
   NGC~6946\_7 & 0.29 & 2.06 & 44.38 & 31.95 & 15.6 & 11.34  & 9.70 & 3.05 & 12.70 & 5.36  (17) & 9.70  (33) & 11.70  (42) & 16.8  (66) & 14.76 & 7.0\\
   NGC~6946\_8 & 0.20 & 2.06 & 44.46 & 31.95 & 15.6 & 11.34  & 14.37 & 3.55 & 15.80 & 5.36  (37) & 9.70  (59) & 11.70  (68) & 16.8  (85) & 14.76 & 7.0\\
   NGC~6946\_9 & 0.20 & 1.88 & 44.06 & 31.96 & 15.5 & 11.34  & 9.81 & 3.09 & 12.90 & 3.72  (37) & 7.62  (59) & 9.72  (68) & 15.4  (85) & 14.77 & 8.0\\
  $<NGC~6946>$&0.20&  1.88 &44.25& 31.96& 15.5 & 11.34       & 12.69 & 3.07 & 13.55 & 3.72 (30) & 7.62 (52) & 9.72 (61) & 15.4 (80) & 14.77& 8.0  \\
 \hline
\hline 
  $\rm{M}_{up}~=~120~\rm{M}_{\sun}  $ & & & & & & & & &     \\
 \hline
M100 & 4.80 & 0.66 & 18.43 & 15.92 & 6.32 & 5.65  & 4.00 & 0.70 & 2.74 & 0.65  (24) & 1.65  (44) & 2.36  (54) & 4.88  (76) & 8.90 &12.0\\
 \hline
   NGC~6946\_1 & 0.03 & 0.80 & 16.44 & 11.46 & 5.80 & 4.07  & 3.61 & 0.90 & 4.30 & 1.34  (9) & 2.73  (19) & 3.48  (26) & 5.48  (48) & 5.27 & 11.0\\
   NGC~6946\_2 & 0.07 & 1.02 & 15.22 & 11.27 & 5.33 & 4.00  & 5.35 & 1.61 & 7.25 & 1.63  (26) & 2.98  (46) & 3.62  (56) & 5.20  (77) & 5.13 & 11.0\\
   NGC~6946\_3 & 0.03 & 0.90 & 14.62 & 11.46 & 5.11 & 4.07  & 4.78 & 0.89 & 4.27 & 1.38  (37) & 2.64  (59) & 3.27  (68) & 4.88  (85) & 5.26 & 8.0\\
   NGC~6946\_5 & 0.02 & 0.92 & 15.75 & 11.32 & 5.48 & 4.02  & 5.84 & 1.28 & 5.23 & 1.07  (32) & 2.28  (53) & 2.95  (63) & 4.79  (82) & 5.16 & 18.0\\
   NGC~6946\_6 & 0.09 & 0.80 & 16.59 & 11.46 & 5.80 & 4.07  & 2.98 & 0.92 & 4.19 & 1.34  (37) & 2.73  (59) & 3.48  (68) & 5.48  (85) & 5.27 & 11.0\\
   NGC~6946\_7 & 0.08 & 0.90 & 14.48 & 11.46 & 5.11 & 4.07  & 3.23 & 0.92 & 4.15 & 1.38  (23) & 2.64  (42) & 3.27  (52) & 4.88  (74) & 5.26 & 8.0\\
   NGC~6946\_8 & 0.06 & 0.90 & 14.51 & 11.46 & 5.11 & 4.07  & 3.83 & 0.94 & 4.48 & 1.38  (26) & 2.64  (46) & 3.27  (56) & 4.88  (77) & 5.26 & 8.0\\
  NGC~6946\_9 & 0.07 & 0.82 & 16.63 & 11.46 & 5.85 & 4.07  & 3.11 & 0.93 & 4.29 & 1.51  (32) & 2.97  (53) & 3.72  (63) & 5.68  (82) & 5.26 & 9.0\\
 $< NGC~6946 >$ &0.06& 0.90   & 15.49& 11.46&  5.41& 4.07 & 3.72 & 0.93 & 4.29 & 1.38 (29) & 2.69 (50) & 3.38 (60) & 5.04 (80)      & 5.26&10.0  \\
  \hline
  \hline
  Literature & & & & & & & & &     \\
  \hline
 &&  0.44$^{a}$ & 75.8$^{h}$&& 7.94$^{b}$ &&  2.46$^{d}$ & 0.97$^{g}$ & 4.55$^{b}$ &  0.64$^{a}$ &  1.4$^{e}$ &  2.0$^{e}$ & 6.53$^{a}$ &&  \\
 &&   &&& 5.30$^{c}$ && 4.80$^{h}$ &0.92$^{h}$  & 3.88$^{a}$ &  0.62$^{e}$ &   &  & &&  \\
 &&   &&& 6.62$^{h}$ &&   &  0.58$^{f}$  & 3.99$^{h}$ &   &   &  & &&  \\
\hline
\hline
\end{tabular}
 Col.(1): ID.
 Col.(2): average SFR (in $M_{\sun}\;  yr^{-1}$)   derived  by considering the last 100 Myr time interval for the normal star-forming  galaxy M100 and the age of the burst  in the star-forming regions of NGC~6946
 Cols.(3-15): SFR calibrations obtained using the above averaged SFR and the  luminosities (intrinsic values for
  FUV, H$\alpha$ and  H$\beta$ ) given  in Table~\ref{tab:dale_gra_derived_1_unsub}.
  In Cols.5, 7 and 15, we rather present the  calibrations  obtained from simple SSPs models, Equations~\ref{eq_sfr_hb_obi_fit} , \ref{eq_sfr_ha_obi_fit}  and \ref{eq_sfr_lff_const_fit_b}  respectively for  
  H$\beta $, H$\alpha$  and 33GHz. The later refers to free-free emission. For the radio -based SFR calibration, we  enclosed in parenthesis the percentage contribution of the thermal radio component  to the total radio emission.
 Col.(16): age of the galaxy in Gyr for M100  or the age of the burst in Myr,  for NGC~6946 extranuclear regions.
 The last row in the first and second panels (with ID, $< NGC~6946 >$) gives the median values of all quantities given for NGC~6946 star-bursting regions.
The values of the SFR calibrations given in the 3rd panel were taken from  the literature.
The superscript on these values indicates the reference as  described below:
$^a$\citet{Murphy2011}, $^b$\citet{Kennicutt1998}, $^c$\citet{Calzetti2007},  $^d$\citet{Zhu2008},  $^e$\citet{Schmitt2006},  $^f$\citet{Li2010},  $^g$\citet{Lawton2010},
 $^h$\citet{Panuzzo2003}. These authors adopted different IMF and evolutionary synthesis models. 
}
\end{table}
\end{landscape}
\begin{landscape}
\begin{table}
\caption{ \textit{\small{GRASIL}} Best-fit derived quantities related to dust attenuation.
 \label{tab:dale_olga_attens}
 }
\large{
\centering
\begin{tabular}{lcccccccccccc c}
\hline
   \multicolumn{1}{|c|}{ID} &
   \multicolumn{1}{c|}{$A_{H\alpha}$} &
   \multicolumn{1}{c|}{$A_{H\beta}$} &
   \multicolumn{1}{c|}{$A_{FUV}$} &
   \multicolumn{1}{c|}{$A_{NUV}$} &
   \multicolumn{1}{c|}{$A_{B}$} &
   \multicolumn{1}{c|}{$A_{48}$} & 
  \multicolumn{1}{c|}{$A_{V}$} &
  \multicolumn{1}{c|}{$A_{65}$} & 
  \multicolumn{1}{c|}{$\tau_{1}$} &
  \multicolumn{1}{c|}{$R_{V}$} &
  \multicolumn{1}{c|}{$E(H\beta-H\alpha)^c$} &
  \multicolumn{1}{c|}{$E(H\beta-H\alpha)^l$} &
  \multicolumn{1}{c|}{$t_{esc}$} \\
(1) & (2)&(3) &(4) &(5) &(6) &(7) &(8) &(9) &(10) &(11) &(12)  &(13) &(14)  \\
 \hline
   &$M_{up}~=~40~\rm{M}_{\sun}  $ & & & & & & &      \\
\hline
 M100 & 1.13 & 1.20 & 0.78 & 0.83 & 0.30 & 0.28 & 0.24 & 0.23 &               17.79 & 4.17 & 0.055 & 0.093 & 2.5 \\
    \hline
  NGC~6946\_1 & 0.70 & 0.76 & 0.87 & 0.91 & 0.60 & 0.63 & 0.54 & 0.57 &         12.66 & 8.08 & 0.060 & 0.059 & 1.0 \\
  NGC~6946\_2 & 0.36 & 0.40 & 0.50 & 0.53 & 0.31 & 0.33 & 0.27 & 0.29 &       15.38 & 6.03 & 0.041 & 0.037 & 0.5 \\
  NGC~6946\_3 & 0.63 & 0.69 & 0.78 & 0.82 & 0.53 & 0.56 & 0.47 & 0.51 &       26.84 & 7.34 & 0.056 & 0.055 & 1.0 \\
  NGC~6946\_5 & 0.40 & 0.45 & 0.61 & 0.65 & 0.43 & 0.42 & 0.39 & 0.37 &       8.43 & 9.14 & 0.053 & 0.054 & 0.3 \\
  NGC~6946\_6 & 0.81 & 0.88 & 0.95 & 0.99 & 0.67 & 0.71 & 0.60 & 0.66 &         8.51 & 8.62 & 0.056 & 0.067 & 1.2 \\
  NGC~6946\_7 & 0.82 & 0.88 & 0.96 & 0.99 & 0.67 & 0.73 & 0.59 & 0.66 &       17.67 & 7.71 & 0.062 & 0.059 & 1.0 \\
  NGC~6946\_8 & 0.53 & 0.59 & 0.75 & 0.79 & 0.53 & 0.54 & 0.48 & 0.48 &       9.63 & 9.48 & 0.059 & 0.053 & 0.6 \\
  NGC~6946\_9 & 0.70 & 0.76 & 0.87 & 0.91 & 0.60 & 0.63 & 0.54 & 0.57 &         8.79 & 8.08 & 0.060 & 0.060 & 1.0 \\
 \hline
   \hline
   &$M_{up}~=~120~\rm{M}_{\sun}  $ & & & & & & &      \\
  \hline
  M100 & 0.68 & 0.75 & 0.71 & 0.77 & 0.30 & 0.30 & 0.24 & 0.23 &           20.16 & 4.18 & 0.070 & 0.087 & 0.9 \\
 \hline
  NGC~6946\_1 & 1.07 & 1.16 & 1.18 & 1.24 & 0.82 & 0.95 & 0.75 & 0.78 &	 	 9.28 & 9.48 & 0.173 & 0.088 & 0.6 \\
  NGC~6946\_2 & 0.79 & 0.85 & 0.78 & 0.84 & 0.53 & 0.77 & 0.45 & 0.58 &	  	19.10 & 5.28 & 0.188 & 0.055 & 1.0 \\
  NGC~6946\_3 & 0.96 & 1.04 & 1.20 & 1.26 & 0.86 & 0.93 & 0.78 & 0.79 &	  	16.11 & 10.38 & 0.136 & 0.087 & 0.9 \\
  NGC~6946\_5 & 0.99 & 1.06 & 1.01 & 1.08 & 0.73 & 0.96 & 0.65 & 0.77 &	  	21.31 & 7.51 & 0.190 & 0.062 & 1.0 \\
  NGC~6946\_6 & 1.21 & 1.30 & 1.21 & 1.28 & 0.84 & 1.01 & 0.75 & 0.81 &	  	9.05 & 8.92 & 0.201 & 0.084 & 0.7 \\
  NGC~6946\_7 & 1.13 & 1.22 & 1.30 & 1.36 & 0.91 & 1.02 & 0.82 & 0.87 & 	  	9.08 & 9.71 & 0.152 & 0.086 & 1.0 \\
  NGC~6946\_8 & 0.92 & 1.01 & 1.15 & 1.21 & 0.82 & 0.89 & 0.74 & 0.76 & 	  	9.26 & 10.03 & 0.135 & 0.085 & 0.9 \\
  NGC~6946\_9 & 1.16 & 1.25 & 1.20 & 1.27 & 0.83 & 1.00 & 0.75 & 0.80 &   	9.11 & 9.10 & 0.194 & 0.087 & 0.6 \\
\hline
\hline
\end{tabular}

Col.(1) :ID.
Cols.(2 - 9)  give the  attenuations in H$\alpha$,  H$\beta$,  FUV (0.16$\micron$), NUV (0.20~$\micron$), \textit{B}-band(0.45~$\micron$),
the interpolated continum at  H$\beta$, \textit{V}-band(0.55~$\micron$) and
 the interpolated continum at  H$\alpha$.
They were derived  using the ratio of the intrinsic unattenuated
 and attenuated (observed) fluxes.
Col.(10): optical depth of the molecular cloud at 1~$\micron$.
Col.(11): $R_{V}$.
Col.(12):The reddening given simply  by A$_{48}$ - A$_{65}$.
Col.(13):The reddening derived using  the balmer decrement method as given by Equation~\ref{eq:ebv_lin}.
Col.(14): escape time of young stars from their birth cloud in Myr.
 }
\end{table}
\end{landscape}

\bsp	
\label{lastpage}
\end{document}